\documentclass[aps,prd,showpacs,amsmath,amssymb,nofootinbib,twocolumn]{revtex4}
\usepackage{graphics}
\usepackage{xcolor}
\usepackage{amssymb}
\usepackage{amsmath}
\usepackage{cases}
\usepackage{epsfig,subfigure}
\usepackage{graphicx}
\usepackage{epstopdf}
\usepackage{hyperref}
\usepackage{url}
\usepackage{ulem}

\begin{document}

\title{Multi-kink brane in Gauss-Bonnet gravity and its stability}

\author{Na Xu}
\affiliation{Lanzhou Center for Theoretical Physics, Key Laboratory of Theoretical Physics of Gansu Province, School of Physical Science and Technology, Lanzhou University, Lanzhou 730000, China\\
Institute of Theoretical Physics \& Research Center of Gravitation, Lanzhou University, Lanzhou 730000, China}
\author{Jing Chen}
\affiliation{Lanzhou Center for Theoretical Physics, Key Laboratory of Theoretical Physics of Gansu Province, School of Physical Science and Technology, Lanzhou University, Lanzhou 730000, China\\
Institute of Theoretical Physics \& Research Center of Gravitation, Lanzhou University, Lanzhou 730000, China}
\author{Yu-Peng Zhang}
\affiliation{Lanzhou Center for Theoretical Physics, Key Laboratory of Theoretical Physics of Gansu Province, School of Physical Science and Technology, Lanzhou University, Lanzhou 730000, China\\
Institute of Theoretical Physics \& Research Center of Gravitation, Lanzhou University, Lanzhou 730000, China}
\author{Yu-Xiao Liu}
	\email{liuyx@lzu.edu.cn}
\affiliation{Lanzhou Center for Theoretical Physics, Key Laboratory of Theoretical Physics of Gansu Province, School of Physical Science and Technology, Lanzhou University, Lanzhou 730000, China\\
Institute of Theoretical Physics \& Research Center of Gravitation, Lanzhou University, Lanzhou 730000, China}


\begin{abstract}
Einstein-Gauss-Bonnet gravity in high dimensional spacetime is intriguing. Here, the properties of thick branes generated by a bulk scalar field in the five-dimensional Einstein-Gauss-Bonnet gravity were studied. With the help of the superpotential method, we obtain a series of multi-kink brane solutions. We also analyze the linear stability of the brane system under tensor perturbations and prove that they are stable. The massless graviton is shown to be localized near the brane and hence the four-dimensional Newtonian potential can be recovered. By comparing the properties of these thick branes under different superpotentials we find with some specific choice of superpotential the Gauss-Bonnet term can determine the scalar field are multi-kink or single kink.
\end{abstract}

\maketitle

\section{Introduction}  \label{sec1}

Among several interesting descriptions of our universe, an attractive one may be that our universe acts as a four-dimensional hypersurface called 3-brane embedded in higher-dimensional  spacetime~\cite{Akama:1982jy,Rubakov:1983kea,Kaluza:1921tu,Klein:1991,Arkani:1998,Randall:1999}. Especially at the end of the 20th century, the {Arkani-Hamed-Dimopoulos-Dvali} (ADD) model~\cite{Arkani:1998} and the Randall-Sundrum (RS) model~\cite{Randall:1999} provided a new way to solve the hierarchical problem in the Standard Model of particle physics.  With the development of higher-dimensional gravity theories, more and more works focused on the nature of extra dimensions~\cite{Randall2:1999,Hsin:2010,Rizzo:2010,Maartens:2010,Liu:2017gcn} and various potential observable effects of extra dimensions were proposed. This attracts the attention of physicists once again and opens a new era for studying extra dimensions.

In brane world scenarios, there are thin or thick branes in terms of their width along the extra {dimensions. Both the branes are thin in ADD and RS models} because the thickness of the brane has been neglected. Note that, in the thin brane model, the scalar curvature comes to be singular at the core of the brane because of the zero width~\cite{Merab1998}. To deal with this singularity, the Israel-Lanczos junction condition~\cite{Israel1966} should be introduced. Physically, a brane should have thickness and it emerges as an alternative to thin brane configuration. That is to say, a brane should have nontrivial width along extra dimensions and there is {no singularity problem} as appeared in the thin braneworld {models}.

One way to obtain {such} nonsingular thick brane might be simply {replacing} the singular source term in the thin brane scenarios with a nonsingular source term. {In 1983, Rubakov and Shaposhnikov firstly proposed the picture of thick brane described by domain wall along the extra dimensional in five-dimensional flat spacetime \cite{Rubakov:1983kea}. They obtained the solution of a scalar field with a kink-like configuration. Such solution connects the two vacua of the scalar potential and has a non-trivial topology. Inspired by the idea of the domain wall in five-dimensional spacetime, lots of literatures focused on the properties of thick brane including gravity ~\cite{DeWolfe:1999cp,Bronnikov:2003gg, Bazeia:2008zx,Toharia:2010ex,Bazeia:2003aw,Novikov:2015iph,Bazeia:2015owa,Xie:2021ayr,Wan:2020smy}}. There were some models based on the nonminimal coupling between gravity and the scalar field~\cite{Nozari:2009zr,Liu:2012gv,Guo:2011wr}. Those nontrivial sources induced many new phenomena and abundant brane configurations. Furthermore, thick brane models~\cite{Xu:2014jda,Zhong:2015pta,Gu:2016nyo,Zhou:2017xaq,Zhong:2017ffr,Gu:2018lub,Cui:2020fiz,Chen:2020zzs,Yu:2015wma,Rosa:2022fhl,Rosa:2021tei,Wang:2019igp,Dzhunushaliev:2019wvv,Nozari:2019shm}. and thin brane models~\cite{Banerjee:2017lxi,Elizalde:2018rmz,Banerjee:2020uil} were investigated in the modified gravity theories such as $f(R)$ gravity. Especially, some brane solutions with rich structure were obtained in~\cite{Cui:2020fiz,Chen:2020zzs}.

When the spacetime dimension is higher than four, the Einstein-Hilbert action can be supplemented with higher order curvature corrections which do not generate three or higher order terms of equations of motion~\cite{Lovelock:1971yv}.  The gravity theory including the Gauss-Bonnet (GB) invariant term is a theory that satisfies the above-mentioned {property}, where the GB invariant arises as a correction in string theory~\cite{Wheeler:1985nh,Boulware:1985wk,Zwiebach:1985kea,Gross:1986mw} and is defined as follows
\begin{equation}
{\cal R}_{\rm GB} = R^{ A B C D} R_{A B C D} - 4 R^{A B} R_{A B} + R^2.
\label{gb}
\end{equation}
The {letters} $A,B,C,D$ in this paper are the indexes of the whole spacetime.  In four-dimensional spacetime, the GB term is a topological term  and acts as the boundary term that does not have influence on the classical field equations. When the spacetime dimension satisfies $D\geq 5$, the GB term is no longer topological invariant and its influence will exit~\cite{Zwiebach:1985kea}. Recently, the GB term was applied to the investigation of inflation after the GW170817 event~\cite{Odintsov:2020sqy,Oikonomou:2021kql,Odintsov:2020xji,Oikonomou:2020oil,Elizalde:2020zcb},~cosmology~\cite{Chirkov:2021epn,Odintsov:2021nim,Shamir:2021ptw}, as well as black hole physics~\cite{Yerra:2022alz,Li:2021wqa,Chen:2018nbh,Hu:2013cia}. In addition, the entanglement wedge cross section was investigated in a five-dimensional AdS-Vaidya  spacetime with GB corrections~\cite{Li:2021rff}. It is also interesting to consider branes in the GB gravity.
Thin brane models in the GB gravity were discussed in  Refs.~\cite{Kim:2000ym,Kim:2000pz,Neupane:2001kd,Meissner:2001xg,Aoyanagi:2004zz}.
Besides the thin brane model, the thick brane models in the GB gravity with a bulk scalar field were widely investigated~\cite{Odintsov:2021nim,Corradini:2000sw,Neupane:2000wt,Andrianov:2013vqa,German:2013sk,Giovannini:2001ta,Farakos:2006sr,HerreraAguilar:2011jm,Dias:2015gga},  and the brane model in the GB gravity was also applied to cosmology~\cite{Charmousis:2002rc,Brown:2006mh,abdesselam2002brane,Alberghi:2005vq,Nozari:2013hra,Konya:2006wr,Okada:2014eva,Herrera:2010vv,Fomin:2018typ,Nojiri:2001ae,Nojiri:2002hz,Lidsey:2002zw,Diaz:2020iwx}. The domain wall solutions constructed by two scalar fields combining either in a kink-antikink or a trapping bag configuration were found in five-dimensional GB gravity with one warped extra-dimension~\cite{Giovannini:2006rj}. Note that  there is only one defect for the domain walls with kink or antikink, when the multi kinks are considered, the gravitating multidefects can be obtained. It was originally investigated in Refs.~\cite{Giovannini:2006ye,Giovannini:2006sw}. {Same as the domain wall obtained in Ref. \cite{Rubakov:1983kea}, the multi-kink-like configurations of a background scalar field will connect the corresponding multi vacua of the scalar potential. Such novel multi-kink configurations will lead the thick brane possess a more richer inner structure. The corresponding inner structures will induce different effective potentials, and various novel KK resonances will exist \cite{Zhong:2018fdq}. It has been proved that the dynamics of the KK resonances is closely related on the effective potentials \cite{Tan:2022uex}. Therefore, construction of thick branes described by multi-kink solutions with inner structures is very important. It is well known that the effects of GB invariant is absent in four-dimensional GB gravity, but one can investigate them in a higher-dimensional spacetime. In this paper} we would like to construct multi-kink brane solution in the higher-dimensional GB gravity and study the possible effects of the GB term on the brane structure. We also analyze the linear stability of the system under tensor perturbation and the localization of gravity.


The paper is organized as follows. In Sec.~\ref{sec2}, we introduce the method to solve the thick brane solutions in GB gravity. The system {can be} reduced to the first-order formulas by introducing {a superpotential}. In Sec.~\ref{sec3}, we construct thick branes with some superpotentials and a polynomial warp factor, respectively, and study the influences of the GB term on the thick brane.
In Sec.~\ref{sec4}, the linear stability of the brane system under the tensor perturbations and {localization of gravity} are analysed. Finally, a brief summary is given in Sec.~\ref{sec5}.

\section{Brane model in GB gravity} \label{sec2}
In this section, we will introduce a new method to solve the thick brane in GB gravity. The brane model in $D$-dimensional GB gravity is described by the following action
\begin{equation}
S = \int d^D x \sqrt{-g}  \left[ \frac{ R}{ 2 \kappa} + \alpha {\cal
	R_{\rm GB}}+{\cal L}_m\right] ,
\label{ac}
\end{equation}
where $ \kappa = 8 \pi G_{D} = M_{*}^{2-D}$ with $G_{D}$ the $D$-dimensional
gravitational constant and $M_{*}$ the $D$-dimensional fundamental mass scale, and $ \alpha $ is the GB coupling constant with mass dimension $D-4$. In this paper, we use the units $ \kappa=c=\hbar=1 $. The Lagrangian density of the scalar field is given by
\begin{equation}
{\cal L}_m=- \frac{1}{2} g^{A B} \partial_{A} \phi \partial_{B} \phi
- V( \phi) ,
\label{M}
\end{equation}
where $V( \phi)$ is the scalar potential. Varying the action~\eqref{ac} with the metric and scalar field {respectively}, we can get the equations of motion as follows
\begin{eqnarray}
\label{eq of mov1}
G_{AB} - 2 \alpha \kappa Q_{AB}&=& \kappa T_{AB },\\
\label{eq of mov2}
g^{AB} \nabla _A\nabla _B\phi -\frac{\partial V(\phi)}{\partial \phi }&=&0,
\end{eqnarray}
where $G_{AB}=R_{AB}-\frac{1}{2}R g_{AB}$ is the Einstein tensor and
\begin{align}
Q_{AB}&=\frac{1}{2} g_{AB} {\cal R_{\rm GB}} -2
R R_{AB}+4 R_{AC} R^{C}{}_{B}\nonumber \\
&+4 R_{ACBD} R^{CD}-2 R_{ACDE} R_{B}^{~CDE}
\end{align}
is the Lanczos tensor~\cite{Lovelock:1972vz}. The energy-momentum tensor of the scalar field reads
\begin{equation}
\label{END}
T_{\text{AB}}=g_{AB} {\cal L}_m +\partial_{A} \phi  \partial_{B} \phi.
\end{equation}

In this paper, we focus on the flat thick brane with $Z_2$ symmetry in a five-dimensional spacetime ($D=5$). The metric is written as~\cite{Randall2:1999}
\begin{equation}
ds^2 = e^{2A(y)} \eta_{\mu \nu} dx^{\mu} dx^{\nu} + d\,y^2 ,
\label{metric5}
\end{equation}
where the warp factor $ A(y)$ is an even function of the extra dimensional coordinate $y$,  $\eta_{\mu \nu}$ is the {four-dimensional}  Minkowski metric.  The ordinary  four-dimensional coordinate indexes  $\mu, \nu$ are from 0 to 3.  The scalar curvature is
\begin{equation}
R=-4 \left(5 A'^2+2 A''\right),  \label{R}
\end{equation}
where the prime denotes the derivative with respect to the extra dimensional coordinate $y $. The {explicit} equations of motion are

\begin{subequations}
	\begin{align}
	\label{equu}
&	6 \left(A''+2 A'^2\right)
	-48 \alpha \kappa A'^2 \left(A''+A'^2\right)
	+ \kappa\left(\phi '^2+2 V\right)
	 =0,\\
	\label{eq55}
&	12A'^2
	-48  \alpha \kappa  A'^4
	- \kappa \left( \phi '^2 -2 V\right) =0,\\
	\label{eqsf}
&	4 A' \phi ' +\phi'' -\frac{\partial V}{\partial \phi } =0.
	\end{align}		
\end{subequations}
It can be shown that there are only two independent equations in~Eqs.~(\ref{equu})-(\ref{eqsf}) for the three functions $V(\phi)$, $\phi(y)$, and $A(y)$. The scalar potential $V(\phi)$ contains {the contribution of} the cosmological constant. In addation, the scalar field $\phi$ is assumed to be an odd function of the extra dimensional coordinate $y$ to localize the fermion zero mode on the brane \cite{Liu:2009ve}.

To solve Eqs.~\eqref{equu}-\eqref{eq55},  we introduce the so called superpotential used in supergravity to reduce the second-order field equations (\ref{equu})-(\ref{eq55}) to the first-order ones. Such method has been used successfully in Refs.~\cite{Bazeia:2003cv,Brito:2001hd,Afonso:2006gi,1999Gravitational,2004Fake,1993GReGr,Julia:1999tk}. We first introduce a  superpotential $W(\phi)$. The relation between the warp factor $A(y)$ and the superpotential $W(\phi)$ is given by
\begin{equation}
A'(y)= -\frac{1}{3} \kappa  W(\phi ),   \label{spp}
\end{equation}
where the superpotential $W(\phi )$ should be an odd function of $\phi$ since $A(y)$ and $\phi(y)$ are assumed to be even and odd, respectively.  From~Eq.~(\ref{spp}), we have
\begin{equation}
A''(y)= -\frac{1}{3} \kappa W_\phi\phi'(y) .  \label{A''y}
\end{equation}
Substituting the above two~ equations~(\ref{spp}) and (\ref{A''y}) into~Eqs.~(\ref{equu})-(\ref{eqsf}), we have
\begin{align}
\label{equu2}
\frac{4}{3} \kappa ^2 W^2-2 \kappa W_\phi \phi '
+\frac{16}{9} \alpha \kappa ^4 W^2 W_\phi \phi '  \nonumber \\
-\frac{16}{27} \alpha \kappa ^5 W^4
+\kappa \Big( \phi '^2+2V \Big) =0,\\
\label{eq552}
\frac{4}{3} \kappa ^2 W^2
-\frac{16}{27} \alpha \kappa ^5 W^4
-\kappa \Big( \phi '^2-2 V \Big)=0,\\
\label{eqsf2}
\phi ''-\frac{4}{3} \kappa W \phi '-V_\phi=0,
\end{align}	
where $W_{\phi}= \frac{d W(\phi)}{d \phi}$ and $V_\phi=\frac{d V(\phi)}{d \phi}$. After replacing the warp factor in terms of the relations~\eqref{spp} and~\eqref{A''y}, one can obtain a relation between the scalar field $\phi$ and the superpotential $W(\phi)$ by subtracting~Eq.~(\ref{equu2}) from~Eq.~(\ref{eq552}) as follows
\begin{equation}\label{Scl}
\phi'=\frac{1}{9} \left(9  -8 \alpha \kappa ^3 W ^2 \right)W_{\phi}.
\end{equation}
One can further obtain the scalar potential by substituting~Eq.~(\ref{Scl}) into~Eq.~(\ref{eq552}):
\begin{equation}
V=\frac{1}{162} \left(W_{\phi}^2 \big(9-8 \alpha \kappa ^3 W^2\big)^2+12 \kappa W^2 \big(4 \alpha \kappa ^3 W^2-9\big)\right).    \label{ptl}
\end{equation}
In our thick brane model, the background scalar field has a kink-like configuration and it will approach to a constant $\phi_\infty$ when the extra dimensional coordinate $y\to\infty$. Thus the spacetime is an asymptotical AdS$_5$ spacetime, which is in accord with the RS brane model. The corresponding naked cosmological constant can be calculated from the scalar potential:
\begin{equation}
\label{cosmological const}
\Lambda_5 = \lim_{y \rightarrow \infty}2\kappa V(\phi(y))=\lim_{\phi \rightarrow \phi_\infty}2\kappa V(\phi).
\end{equation}
Now, the original field equations~(\ref{equu})-(\ref{eqsf})  have been replaced by the first-order formulas (\ref{spp}), (\ref{Scl}), and (\ref{ptl}). Once the superpotential $W(\phi)$ is given, we can solve all the functions for the brane solution with the help of {the above} first-order formulas.

{Note that the solutions of non-linear coupled differential equations are not unique, which will lead to two or more different background configurations with the same Lagrangian parameters and boundary conditions. In our setup, we force the scalar field $\phi(y)$  to be an odd function along the extra dimensional coordinate $y$. Besides this assumption, we still add another condition and let the scalar field $\phi(y)$ to be a monotonic function of $y$. Thus, for such an assumption of the scalar field $\phi(y)$, it has an inverse function $y(\phi)$ and our solution is unique. With the help of the above assumption and boundary conditions, we can obtain the corresponding background solutions from the given superpotentials conveniently}. We can write the inverse function $y(\phi)$ from~Eq.~(\ref{Scl}) as follows
\begin{equation}
y(\phi) =\int \frac{9}{\left(9  -8 \alpha \kappa ^3 W ^2 \right)W_{\phi}} \, d\phi.  \label{extradim and phi}
\end{equation}
The warp factor can also be obtained as
\begin{equation}
{A}(y(\phi))=\int\frac{3 \kappa W }{ \left(8 \alpha \kappa ^3 W ^2 -9 \right)W_{\phi}} \, d\phi. \label{wf and phi}
\end{equation}
Furthermore, we introduce the conditions: $y(\phi=0)=0$ and $A(y=0)=0$.
By specifying the suitable superpotential, one can obtain the thick brane solution from Eqs. (\ref{ptl}), (\ref{extradim and phi}), and (\ref{wf and phi}). The distribution of the thick brane can be described by the effective energy density along the extra dimension with respect to the static observer $u^{A}=(e^{A},0,0,0,0)$:
\begin{equation}
\label{energy densigy}
\rho =\tau_{MN}^{(\text{eff})} u^{M}u^{N}
=-g^{00} \left(T_{00}+ 2 \alpha Q_{00} \right), 
\end{equation}
where $\tau_{MN}^{(\text{eff})} = T_{MN}+ 2 \alpha Q_{MN}$ is the effective energy-momentum tensor. It can also be expressed in terms of the superpotential as follows
\begin{equation}
\label{rho1}
\rho =\frac{1}{81}
\left[W_{\phi} ^{2 }\left(9-8  \alpha \kappa ^3 W^2\right)^2
+6 \kappa W^2 \left(4 \alpha \kappa ^3 W^2-9\right)
\right].
\end{equation}
Note that $\rho(|y|\rightarrow \infty)$ is nonvanishing since the above expression contains the contribution from the effective cosmological constant coming from the naked cosmological constant $\Lambda_5$ in (\ref{cosmological const}) and the Lanczos tensor. In order to describe the shape of the brane better, we subtract the contribution of the effective cosmological constant, i.e., we make the following replacement
\begin{equation}
\rho(y) \rightarrow \rho(y) -\rho(|y|\rightarrow \infty).  \label{rho}
\end{equation}

Actually, there is a direct way to solve Eqs.~(\ref{equu})-(\ref{eqsf}). First, giving the potential $V(\phi)$ as a function of the scalar field $\phi$, then the scalar field $\phi (y)$ and the warp factor $A(y)$ can be solved as functions of extra dimensional coordinate $y$. The scalar field $\phi (y)$ and the warp factor $A(y)$ contain the GB coupling $\alpha$. In this way, the parameters in the potential $V(\phi)$ and the GB coupling $\alpha$ are independent.  However, it is difficult to obtain the brane solution with this method  as higher order terms of the warp factor induced from the Lanczos tensor are contained in the equations of motion. Using superpotential method the equations of motion can be downgraded the order, so that one can simplify the calculation.

In the superpotential method, it is the superpotential $W(\phi)$ but not the scalar potential $V(\phi)$ that is given independently. So the parameters in the superpotential $W(\phi)$ and the GB coupling $\alpha$ are independent.  The scalar potential $V(\phi)$ is decided by the GB term and the superpotential $W(\phi)$ from Einstein equations. Therefore, the parameters in the potential $V(\phi)$ are no longer independent. In this case, one can study how the GB coupling $\alpha$ and the parameters in the superpotential $W(\phi)$ independently effect the solutions of the warp factor and scalar field. If the superpotential $W(\phi)$ was set as $W(\phi)=\phi$, the brane world solution can be solved analytically, and the potential $V(\phi)$ would be a $\phi^4$ potential. This case has been studied in Ref.~\cite{Low:2000pq}. Here, we study more general superpotential, the analytical solutions would be more difficult to get. So we will use numerical method to solve the equations of motion.

Besides the superpotential method for solving thick brane system, one can also obtain the thick brane solutions by using the relations between the scalar field as well as scalar potential and the warp factor.  This can be done conveniently with the following five-dimensional conformally flat metric
\begin{equation}
ds^2 = e^{2A(z)} [ \eta_{\mu \nu} dx^\mu dx^\nu + d\,z^2], \label{metric_conf}
\end{equation}
where the warp factor $e^{2A(z)}$ is a function of the extra dimensional coordinate $z$ and the relation between $z$ and $y$ is given by $e^{A}dz=dy$. With a known warp factor, one can derive the expressions for the scalar potential and scalar field:

\begin{align}
V(\phi(z))= &12 \alpha  e^{-4 A} \left(\partial_{z} A\right)^{2}\left[\left(\partial_{z} A\right)^{2}+\partial_{z}^{2} A\right] \nonumber \\
&-\frac{3}{2\kappa } e^{-2 A}\left[3\left(\partial_{z} A\right)^{2}+\partial_{z}^{2} A\right], \label{V_Az}\\
\phi(z) =& \int dz \sqrt{\frac{3}{\kappa}}
\sqrt{\left[1-8  \alpha \kappa \left(\partial_{z} e^{-A}\right)^{2}\right]
	\left[\left(\partial_{z} A\right)^{2}-\partial_{z}^{2} A \right]}. \label{phi_Az}
\end{align}

Therefore, we can obtain the thick brane solution by directly specifying the form of the warp factor. In Ref.~\cite{Giovannini:2001ta}, one analytical solution has been obtained, and we would like to extend this solution using other warp factors.

\section{Brane solutions}  \label{sec3}

In this section, we first consider various superpotentials for solving multi-kink branes. Then, we also construct a thick brane solution with a polynomial warp factor in the conformal coordinate. We will set $M_*=1$ in the numerical calculations or plots in this paper.

\subsection{Solutions with superpotential method}\label{sec3.1}
Once a superpotential is given, the warp factor, the scalar field, the energy density, and the scalar potential can be obtained numerically, just as discussed in Sec.~\ref{sec2}, and the property of the brane can be figured out.  For some concrete examples, we consider following three types of superpotentials:
\begin{align}
W_1(\phi) =& M_*^4\left(a~\tilde\phi + b\, \tilde\phi~\text{sech}({\tilde\phi})\right) , \label{supp1} \\
W_2(\phi) =& M_*^4\left(a~\tilde\phi+b\, \text{sinh}({\tilde\phi})\right) , \label{supp2}\\
W_3(\phi) =& M_*^4\left(a~\tilde\phi  +  b \sin({\tilde\phi})\right) , \label{supp3}
\end{align}
where $\tilde\phi=\phi/(M_*^{3/2}v)$ is a dimensionless scalar field and $a$, $b$, and $v$ are dimensionless parameters. Inspired by the sine-Gordon scalar potentials studied in Refs. \cite{Koley:2004at,Liu:2008pi}, here we give three superpotentials containing same polynomial and different nonpolynomial contributions. The three different nonpolynomial terms in these superpotentials are a finite term, a divergent term, and a  periodic term, respectively. The different superpotentials induce the different scalar potentials. Here, with these three superpotentials the scalar potentials can have richer structures, which represent more complex interactions.   We will show that multi-kink thick brane solutions can be obtained for all these superpotentials with different parameter spaces.

\subsubsection{First superpotential}  \label{case1}

For the first superpotential (\ref{supp1}), we obtain the scalar potential from~(\ref{ptl}), the function $y(\phi)$ from~\eqref{extradim and phi}, and the warp factor from~\eqref{wf and phi}:
\begin{align}
V(\phi) = & \frac{2 }{27}  M_*^5 \tilde \phi ^2 \big(b~\text{sech}({\tilde\phi}) +a \big)^2 \nonumber \\
&\times
\left(4 \tilde \alpha~  \tilde \phi ^2\big(b~\text{sech}({\tilde\phi}) +a \big)^2-9\right)
\nonumber \\
&+\frac{1 }{162} M_*^5  v^{-2} \left(9-8 \tilde \alpha~  \tilde \phi ^2~ \big(b~\text{sech}({\tilde\phi}) +a \big)^2\right)^2 \nonumber \\
&\times \Big(b~\text{sech}({\tilde\phi})  \big(1- \tilde \phi~\tanh (\tilde \phi) \big)+a\Big)^2, \label{solution4V} \\
\nonumber \\
\nonumber \\
y(\phi)=&M_*^{-1}\int_0^{\tilde{\phi}}
 \frac{9 v^2}{b~\text{sech}(\tilde{\phi})-b~\tilde{\phi}~\tanh({\tilde{\phi}})~\text{sech}(\tilde{\phi})+a}\nonumber \\
 	&\times
 \frac{1}{9-8 \tilde{\alpha}~(b~\tilde{\phi}~\text{sech}(\tilde{\phi})+a~\tilde{\phi} )^2}d \tilde{\phi}, \label{solution1_y_phi}\\
A(y(\phi))=&\int_0^{\tilde{\phi}}
\frac{3 v^2\left(b~\tilde{\phi}~\text{sech}({\tilde\phi})+a~\tilde{\phi} \right)}{b~\text{sech}({\tilde\phi})-b~\tilde{\phi}~\tanh ({\tilde{\phi}})~ \text{sech}({\tilde\phi})+a}   \nonumber \\
	&\times
\frac{1}{ 8~\tilde{\alpha}~\Big(b~\tilde{\phi}~\text{sech}({\tilde\phi})+a~\tilde{\phi} \Big)^2-9 } d\tilde{\phi} . \label{solution1_A_y}
\end{align}
We defined a dimensionless GB coupling constant $\tilde \alpha=  \alpha / M_{*}$. Here, we also keep $M_*$ in these expressions in order to check the dimensions of the quantities. The energy density $\rho(y)$ can be obtained from Eqs.~\eqref{energy densigy}, ~\eqref{rho1}, and~\eqref{rho}. We do not list it here.

The kink-like solution for the scalar field demands that the scalar field satisfies $\phi \rightarrow \text{const}=\phi_{\infty}$ when $y \rightarrow \infty$. That is to say, the integral function in Eq.~(\ref{solution1_y_phi}) for $y(\phi)$ must approach to infinity when $\phi \rightarrow \phi_{\infty}$. Thus we get the constraints on the parameters $\tilde{\alpha}$, $a$, and $b$ that could support the existence of thick brane solutions. We list the constraints as follows
\begin{itemize}
	\item when $\tilde{\alpha} > 0$, the restriction is $a \neq -b$;
	\item when $\tilde{\alpha} \leq 0$ and $ b \geq 0$, the restriction is $-b<a<0.2511b$;
	\item when $\tilde{\alpha} \leq 0$ and $ b < 0$, the restriction is $0.2511b<a<-b$.
\end{itemize}

\begin{figure}[!h]
	\centering
	\begin{center}
		\subfigure[Warp factor]
		{\includegraphics [width=4.0cm] {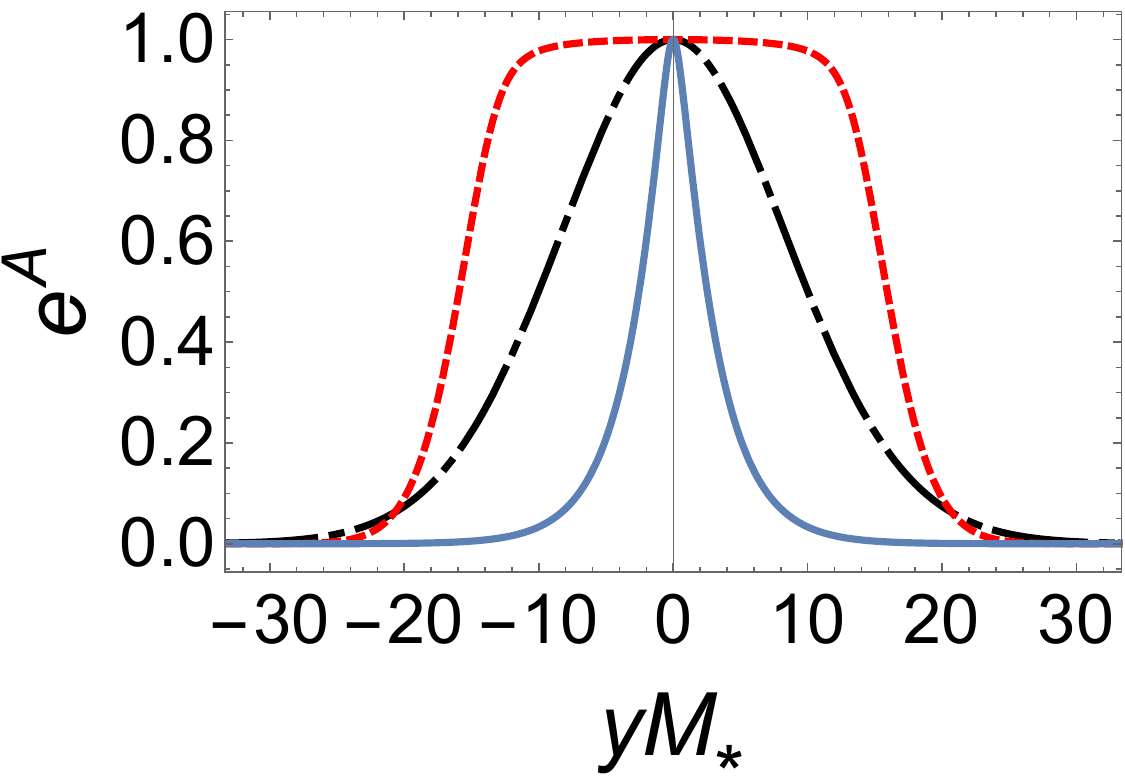}}
		\subfigure[ Scalar field ]
		{\includegraphics [width=4.0cm] {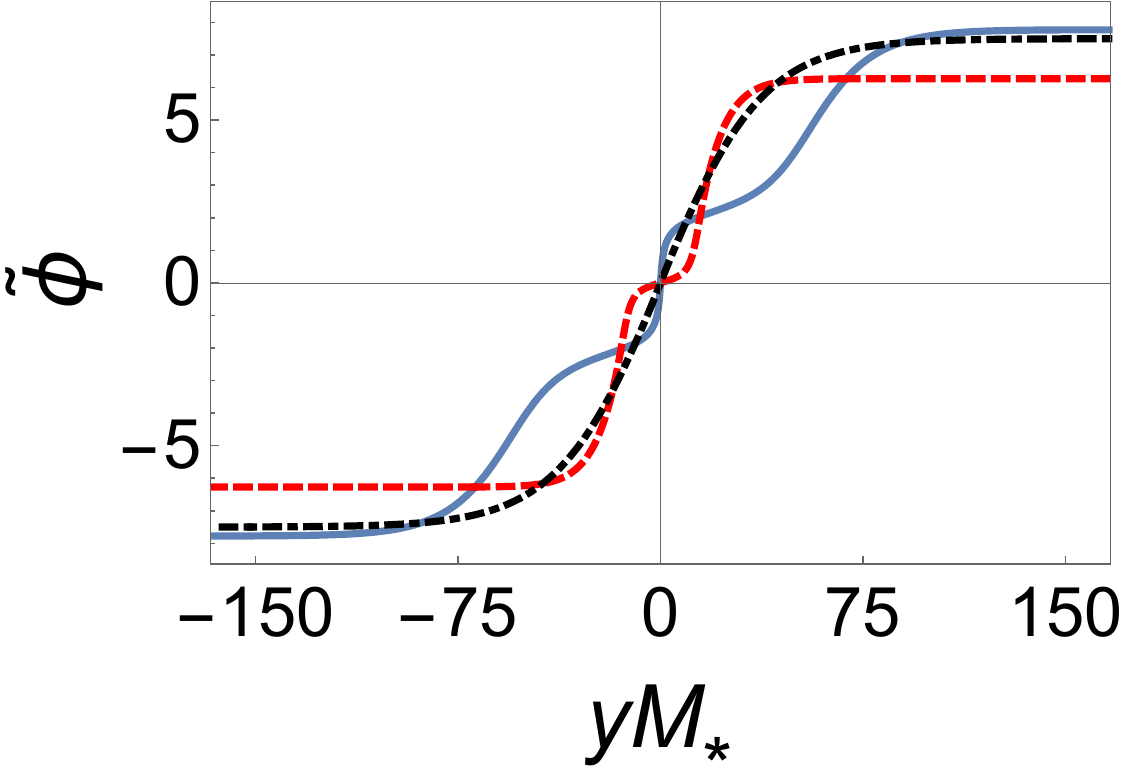}}
		\subfigure[Energy density]
		{\includegraphics [width=4.0cm] {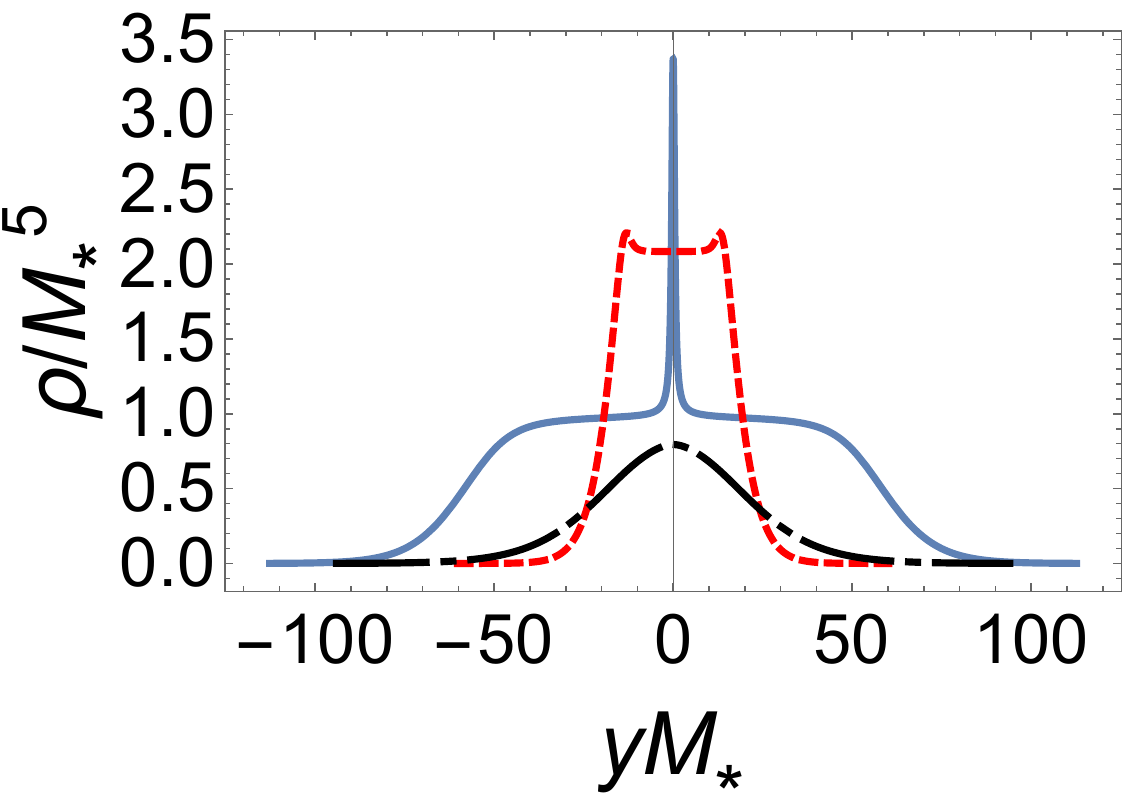}}
		\subfigure[Scalar potential]
		{\includegraphics [width=4.0cm] {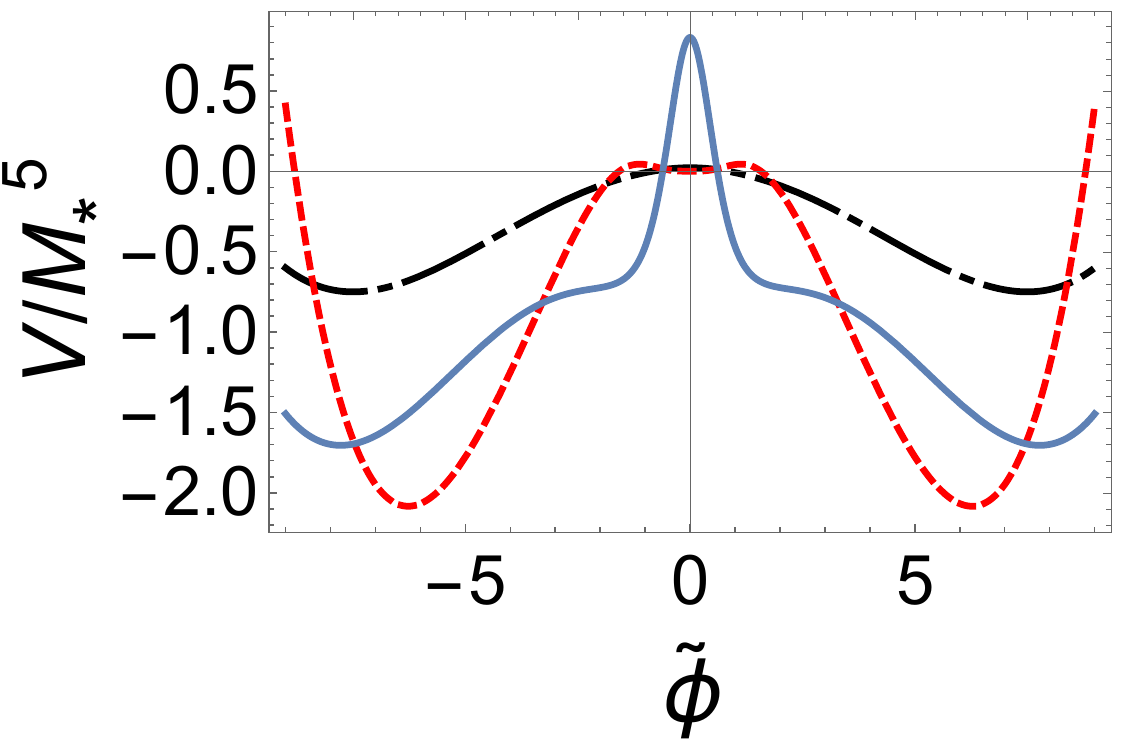}}
	\end{center}
	\caption{The brane solutions for the first superpotential (\ref{supp1}). The black dot dashed lines, red dashed lines, and dark blue lines correspond to the single-kink, double-kink, and triple-kink scalar field solutions, respectively. The values of the parameters are listed in Table~\ref{tab_case1}.\label{fig_case1}}
\end{figure}

Then, specifying the suitable values of parameters $a$ and $b$, we can obtain the single-kink, double-kink, and triple-kink thick brane solutions. See the corresponding results in Fig.~\ref{fig_case1} with the values of the parameters given in Table~\ref{tab_case1}. In this subsection, we set $v=1$ for the scalar potential. From Fig.~\ref{fig_case1} we can see that the triple-kink scalar field configuration has a step-like energy density which is a new feature of the multi-kink scalar field.

\begin{table}[!htb]
	\begin{center}
		\caption{Parameters used in Fig.~\ref{fig_case1}}
		\begin{tabular}{|c|c|c|c|c|}
			\hline
			solution &Lines &$\tilde \alpha$&$ a $&$ b $   \\
			\hline	\hline
			solution\_a1 &Black dot dashed lines & 0.500 & 0.200 & 0.010\\
			\hline
			solution\_a2 &Red dashed lines & 0.180   & 0.400 & -0.375 \\
			\hline
			solution\_a3 &Dark blue lines & 0.220   & 0.290 & 1.000  \\
			\hline
		\end{tabular}
		\label{tab_case1}
	\end{center}
\end{table}

\subsubsection{Second superpotential}   \label{case2}

Next, we use the second superpotential (\ref{supp2}) to construct thick brane solutions. The corresponding expression of the scalar potential is
\begin{align}
& V(\phi)=
\frac{2 M_*^5}{27} \left(a \tilde{\phi }+b \sinh (\tilde\phi)\right)^2 \left(4 \tilde{\alpha } \left(a \tilde{\phi }+b \sinh (\tilde\phi)\right)^2-9\right) \nonumber \\
&+\frac{M_*^5}{162}  v^{-2} \left(b \cosh (\tilde\phi)+a\right)^2 \left(9-8 \tilde{\alpha } \left(a \tilde{\phi }+b \sinh (\tilde\phi)\right)^2\right)^2.
\label{solution5V}
\end{align}
Substituting the superpotential~\eqref{supp2} into Eqs.~\eqref{extradim and phi}, ~\eqref{wf and phi}, and~\eqref{rho}, the function $y(\phi)$, the warp factor $A(y)$, and the energy density $\rho(y)$ can be obtained.
We also need to restrict the parameters $\tilde{\alpha}$, $a$, and $b$ with Eq.~\eqref{extradim and phi} that could support the existence of thick brane solutions just like we did in Sec.~\ref{case1}. The result is
\begin{itemize}
	\item when $\tilde{\alpha} >0$, the restriction is $b\neq -a$;
	\item when $\tilde{\alpha} \leq 0$, the restriction is $b^2 <-ab$.
\end{itemize}

\begin{figure}[!ht]
	\centering
	\begin{center}
		\subfigure[Warp factor]
		{\includegraphics [width=4.0cm] {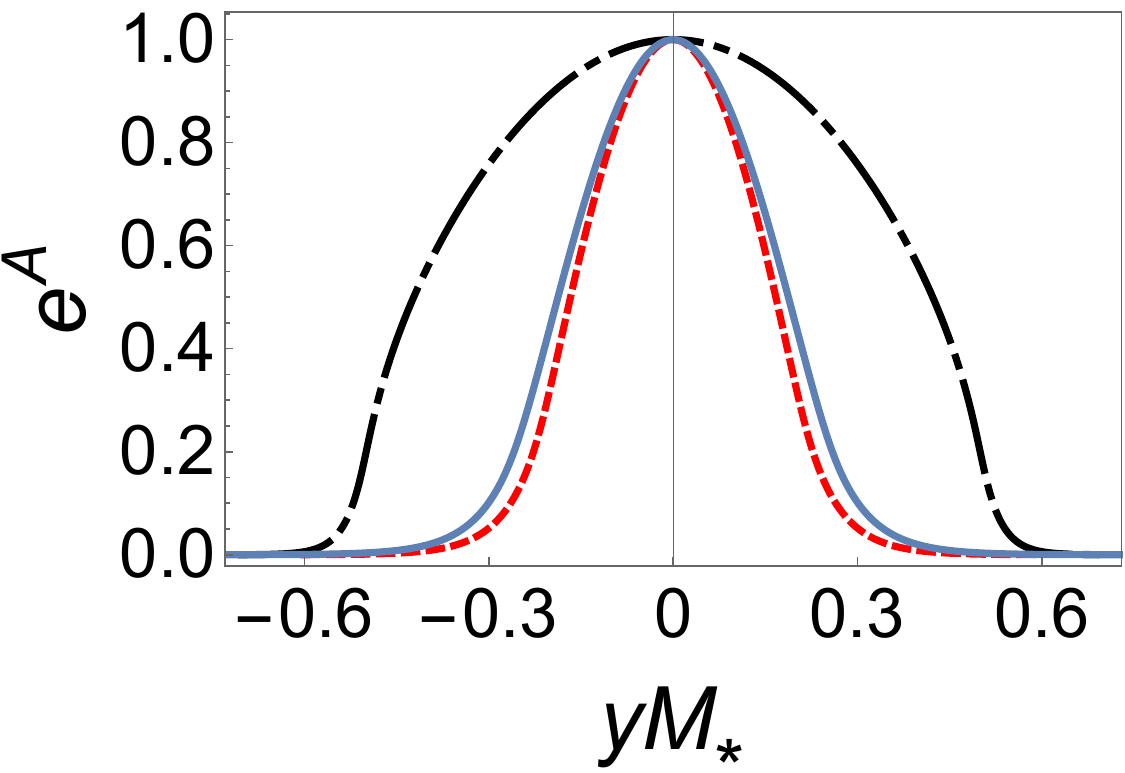}}
		\subfigure[Scalar field]
		{\includegraphics [width=4.0cm] {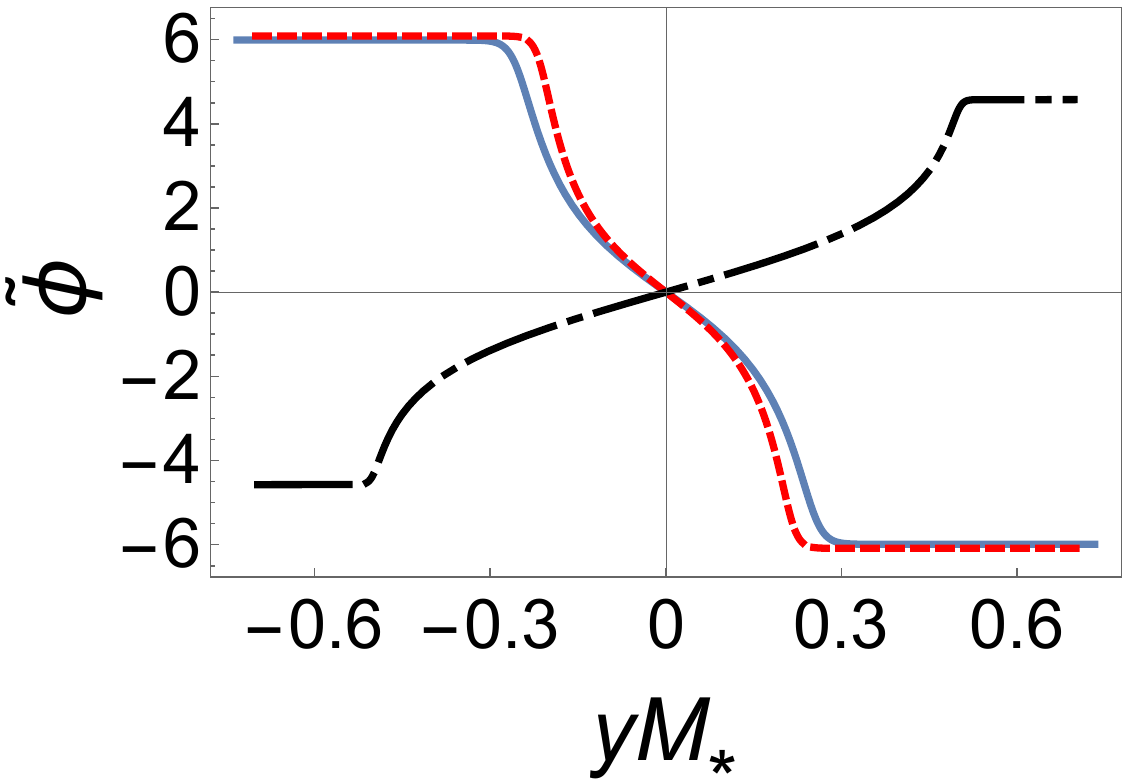}}
		\subfigure[Energy density]
		{\includegraphics [width=4.0cm] {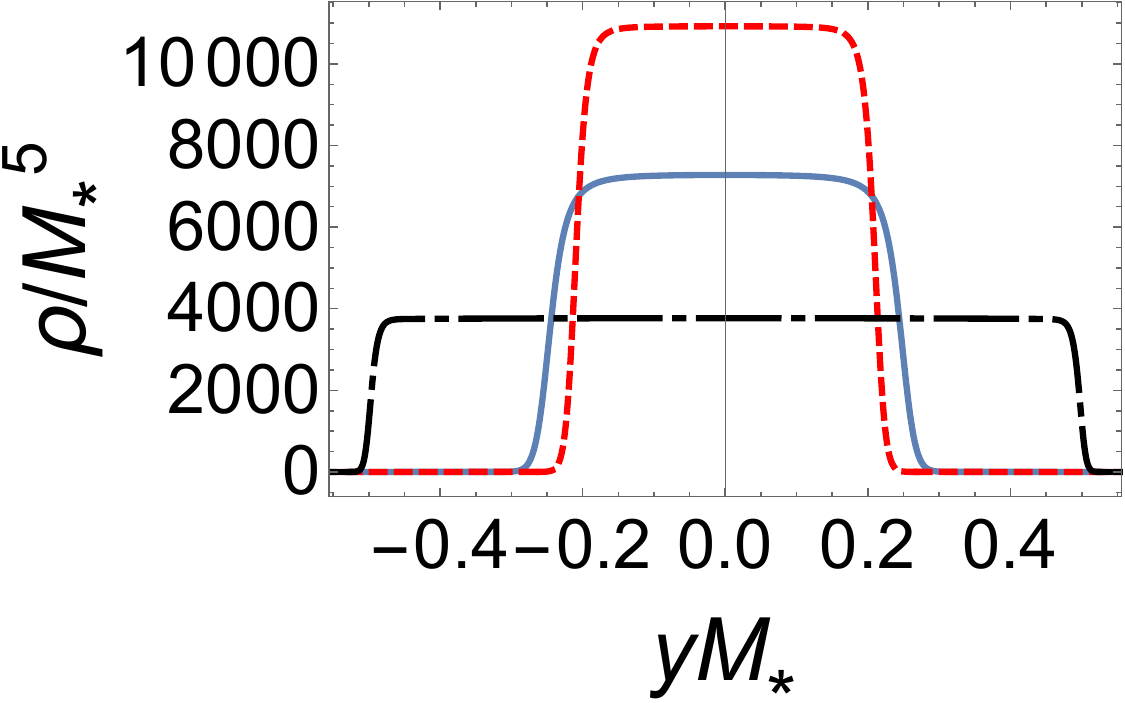}}
		\subfigure[Scalar potential]
		{\includegraphics [width=4.0cm] {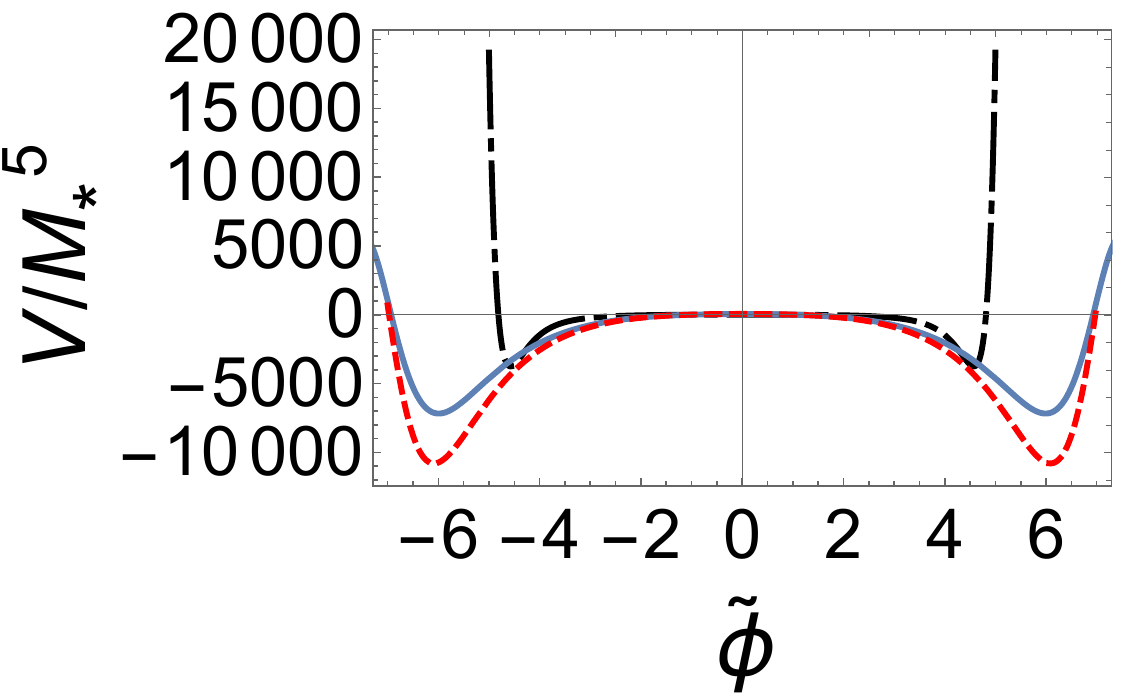}}
	\end{center}
	\caption{The brane solutions for the second superpotential (\ref{supp2}). All three lines correspond to the double-kink solutions. The values of the parameters are listed in Table~\ref{tab_case2}.}
	\label{fig_case2}
\end{figure}

With this superpotential, we can also find thick brane solutions with single-kink and double-kink scalar field configurations. We show three different solutions with double-kink configurations in Fig.~\ref{fig_case2}, for which the parameters are given in Table~\ref{tab_case2}.
\begin{table}[!htb]		
	\begin{center}
		\caption{Parameters used in Fig.~\ref{fig_case2}}
		\begin{tabular}{|c|c|c|c|c|}
			\hline
			solution&Lines &$\tilde \alpha$&$ a $&$ b $   \\
			\hline	\hline
			solution\_b1  & Black dot dashed lines & {0.0001}  & {2.00} & {2.00} \\
			\hline
			solution\_b2  &Red dashed lines & -0.003  & -11.0 & 0.050 \\
			\hline
			solution\_b3  &Dark blue lines & {-0.003}  & {-10.0} & {0.05}  \\
			\hline
		\end{tabular}
		\label{tab_case2}
	\end{center}
\end{table}

\subsubsection{Third superpotential} \label{case3}
\begin{figure}[!ht]\centering
	\begin{center}
		\subfigure[Warp factor]
		{\includegraphics[width=4.0cm] {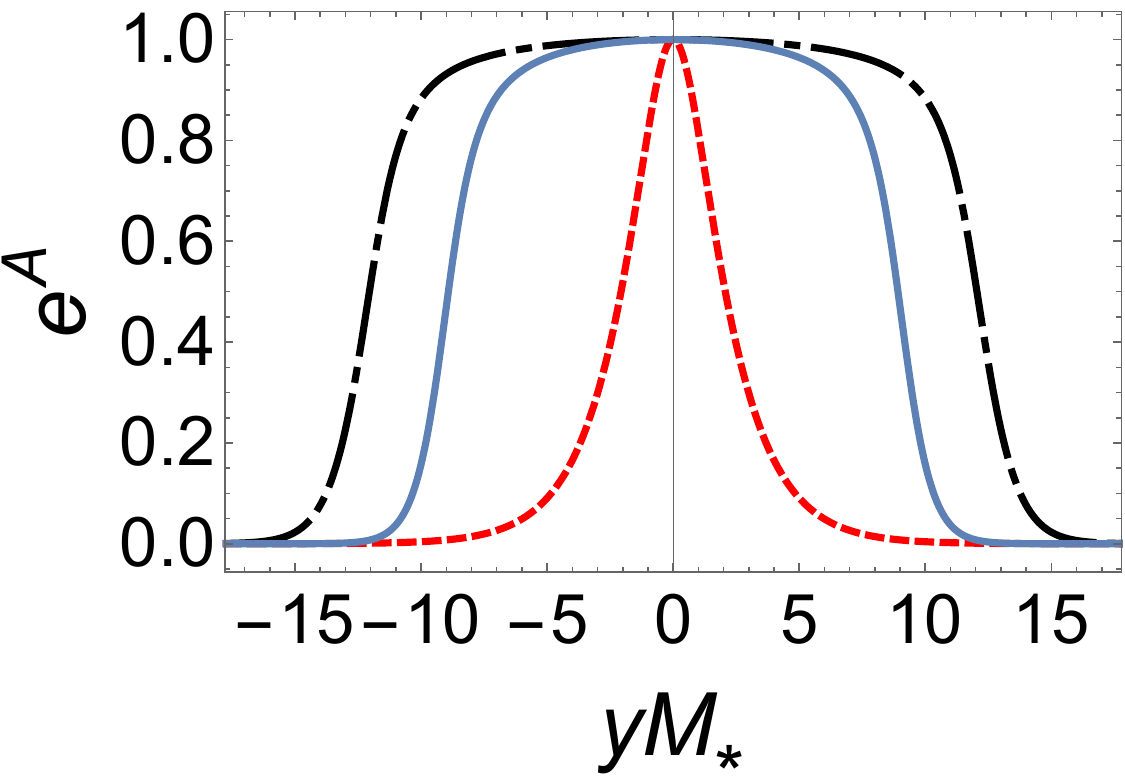}}
		\subfigure[Scalar field ]
		{\includegraphics[width=4.0cm] {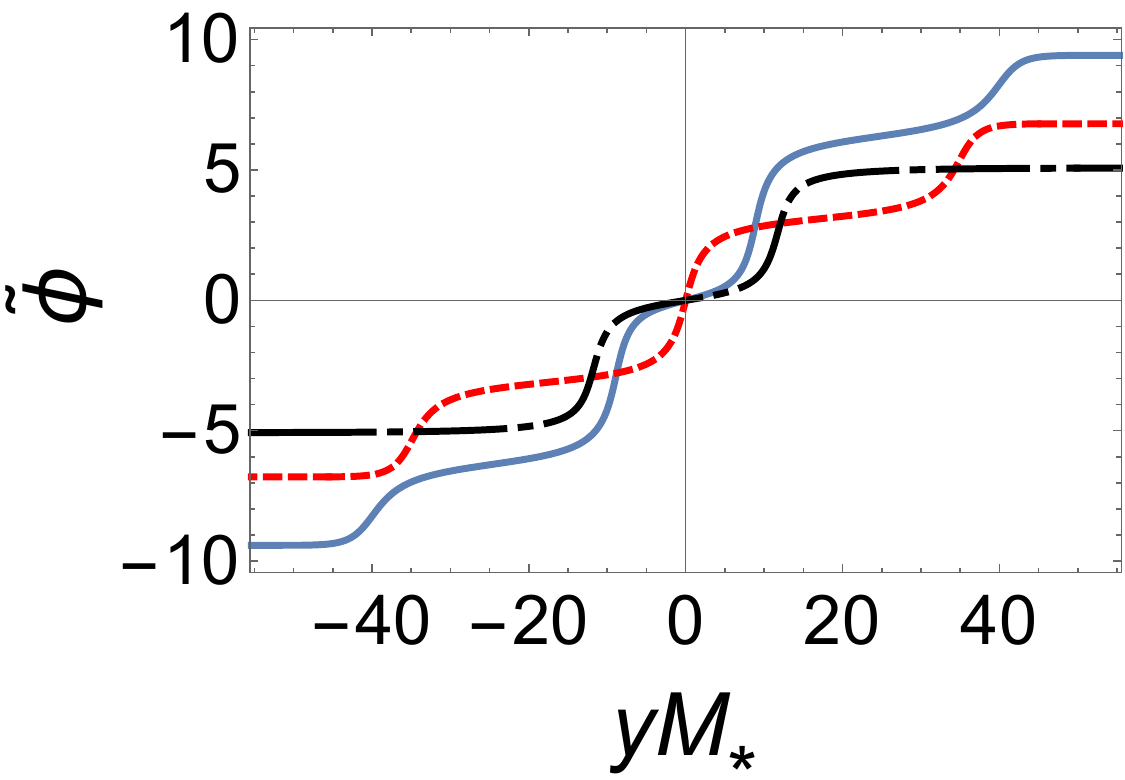}}
		\subfigure[Energy density]
		{\includegraphics[width=4.0cm] {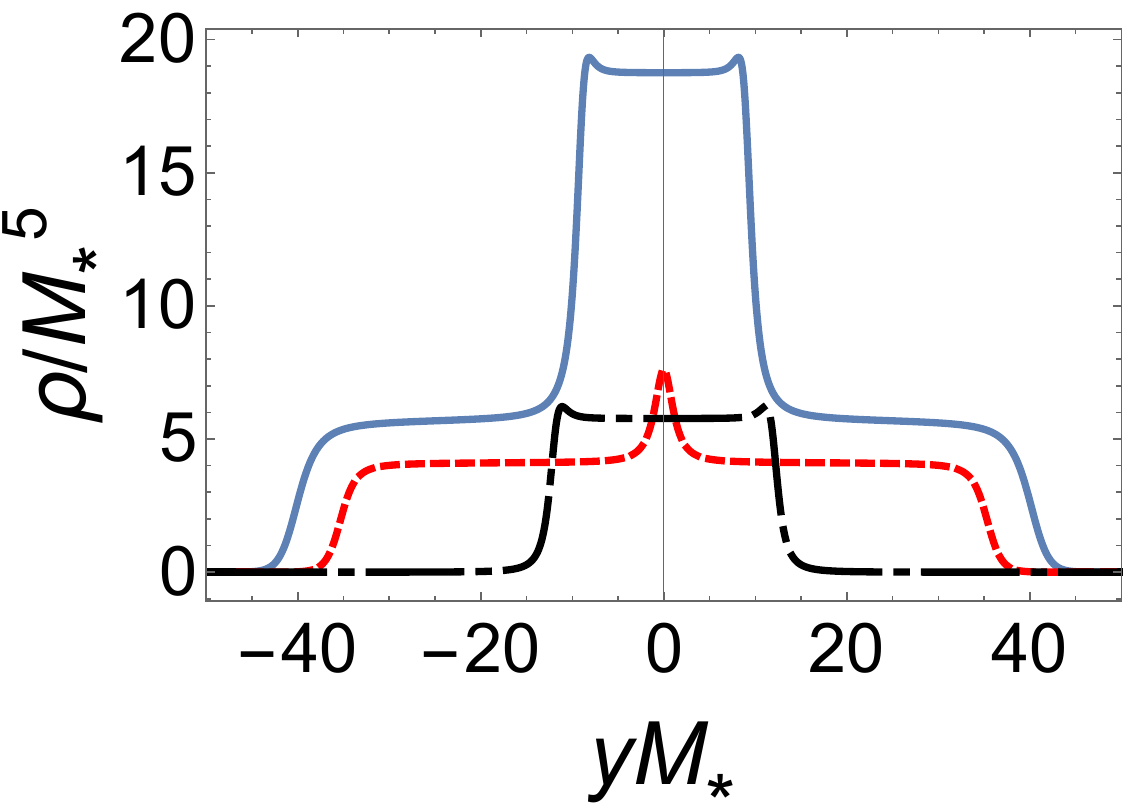}}
		\subfigure[Scalar potential]
		{\includegraphics[width=4.0cm] {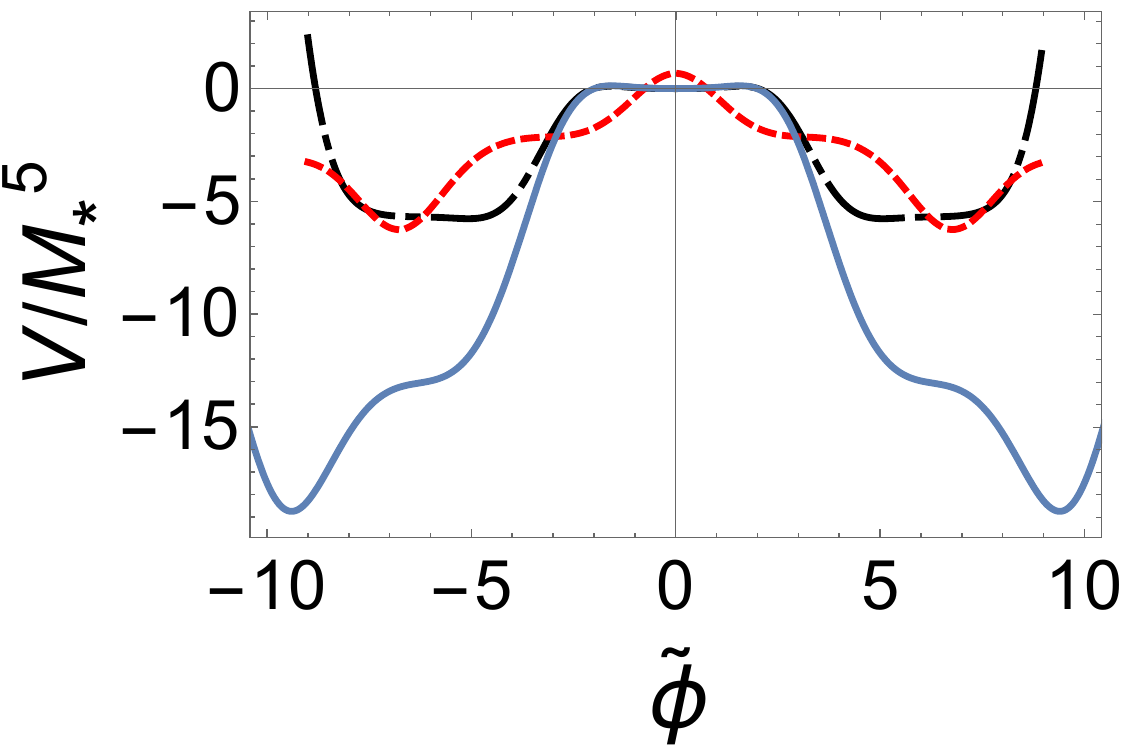}}
	\end{center}
	\caption{The brane solutions for the third superpotential (\ref{supp3}). The black dot dashed lines, red dashed lines, and dark blue lines correspond to the double-kink, triple-kink, and 4-kink scalar field solutions, respectively. The values of the parameters are listed in Table~\ref{tab_case3}.}\label{fig_case3}
\end{figure}

At last, we use the third superpotential (\ref{supp3})
to derive the thick brane solutions. With the above superpotential, we get the scalar potential as follows
\begin{align}
V(\phi)&= \frac{2  M_*^5 }{27} \big(a~\tilde\phi +b \sin (\tilde\phi) \big)^2 \big(4 \tilde{\alpha } \big(a~\tilde\phi +b~\sin(\tilde\phi) \big)^2-9\big)  \nonumber \\
&+\frac{ M_*^5 }{162} v^{-2} (a+b \cos (\tilde\phi) )^2 \Big(9-8~\tilde{\alpha } (a~\tilde\phi +b \sin(\tilde\phi) )^2\Big)^2. \label{solution1V}
\end{align}
Other functions can also be obtained with the above superpotential.
For thick brane solutions, the constrains for the parameters that support the existence of thick brane solutions are
\begin{itemize}
	\item when $\tilde{\alpha}>0$, then $a \neq \pm b$;
	\item when $\tilde{\alpha} \leq 0$, then $-b^2 <ab<b^2$.
\end{itemize}
The superpotential (\ref{supp3}) has rich structures due to the combination of the linear term $\tilde{\phi}$ and the periodic function $\sin(\tilde{\phi})$. For such superpotential, we find that any number of kinks for a background scalar field can be constructed with some suitable parameters.
For instance, the number of kinks approaches to infinity when the parameters satisfy $a>-b>0$ and $\tilde{\alpha}\rightarrow +0$. This is a new result of the GB term and the superpotential (\ref{supp3}). We show some multi-kink brane solutions in Fig.~\ref{fig_case3}.

\begin{table}[!htb]
	\begin{center}
		\caption{Parameters used in Fig.~\ref{fig_case3}}
		\begin{tabular}{|c|c|c|c|c|}
			\hline
			solution&Lines &$\tilde \alpha$&$ a $&$ b $   \\
			\hline	\hline
			solution\_c1  &Black dot dashed lines & 0.065 & 0.700 &-0.650\\
			\hline
			solution\_c2  &Red dashed lines & 0.060  & 0.600 & 0.560 \\
			\hline
			solution\_c3  &Dark blue lines & 0.020  & 0.800 & -0.720  \\
			\hline
		\end{tabular}
		\label{tab_case3}
	\end{center}
\end{table}

Similarly, we can also find other brane world solutions for a given superpotential with our method given in the last section. This method is powerful in the GB brane model and can be generalized to brane models in other gravity theories.

From the above three typical examples of multi-kink brane solutions, we see that the warp factors with even kink solutions seem fatter than the ones with odd kink solutions. The energy densities and scalar potentials of multi-kink scalar fields are obviously different for different numbers of kinks. In these solutions we can not distinguish the number of kinks from the warp factor. This is because the influence of the multi-kink scalar field to the warp factor is too small. With some suitable superpotentials, the influence to the warp factor can be seen. We also find that the GB term plays an important role in finding multi-kink scalar field solutions. When $\tilde \alpha=0$ the scalar field in these three cases could not be multi-kink scalar fields, at most single kink scalar fields. When $\tilde \alpha \neq 0$ the scalar fields could be multi-kink scalar fields. So that the GB term could lead to richer brane structure. Note that the superpotentials contain polynomial and nonpolynomial contributions and can be seen as three extensions of the superpotential in Ref. \cite{Low:2000pq}. With these contributions from the nonpolynomial and the GB term, the multi-kink scalar fields can be gotten.


\subsection{Solution with a polynomial expanded warp factor}\label{sec3.2}
\begin{figure}
	\begin{center}
		\subfigure[Warp factor\label{ay4}]
		{\includegraphics[width=4.0cm] {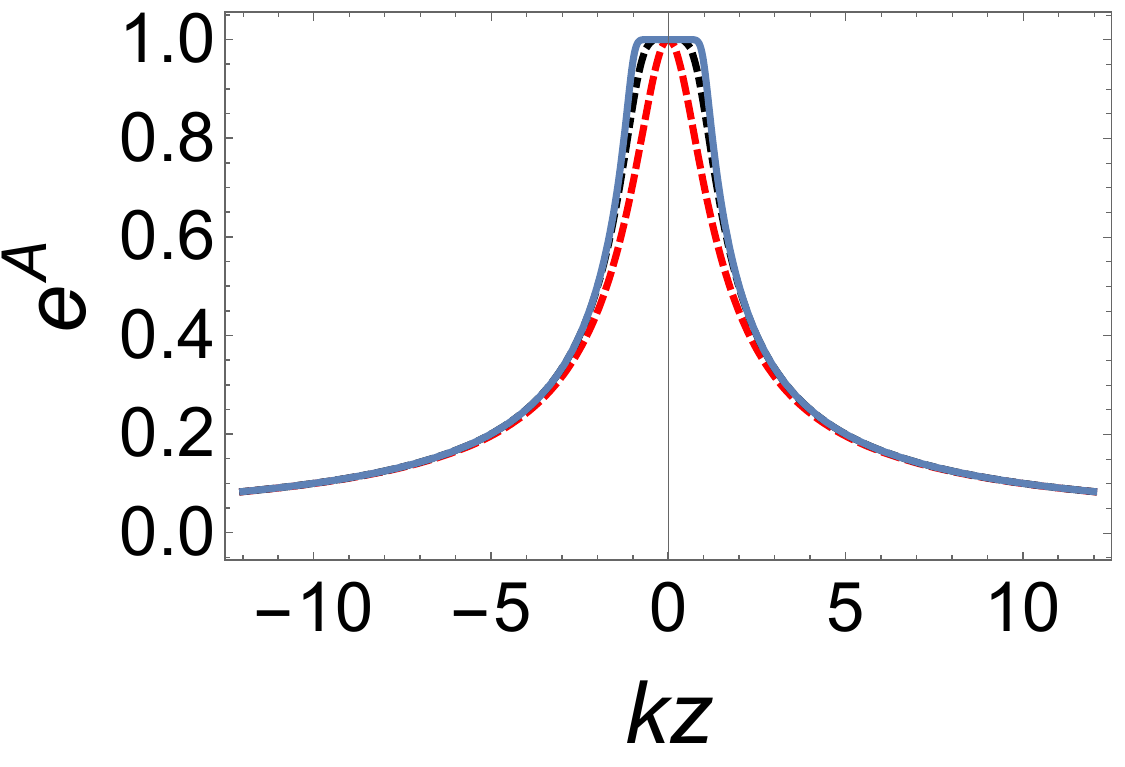}}
		\subfigure[Scalar field\label{phi4}]
		{\includegraphics[width=4.0cm] {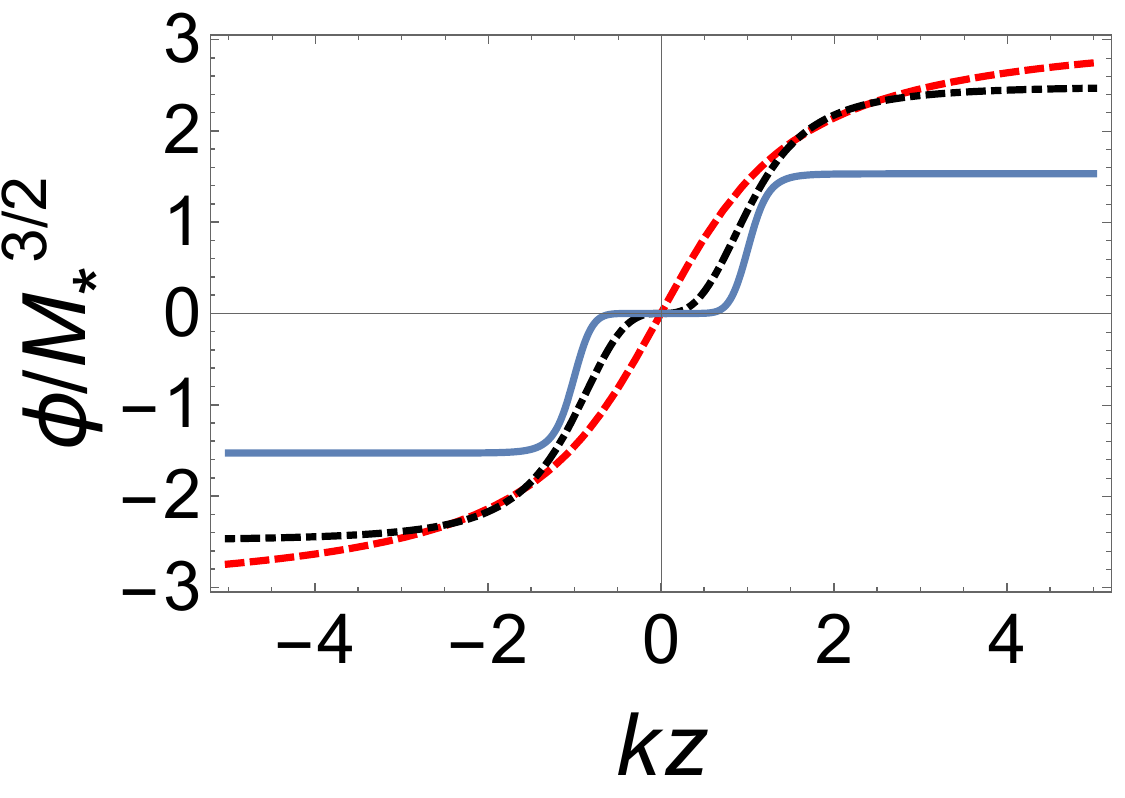}}
		\subfigure[Energy density\label{energy4}]
		{\includegraphics[width=4.0cm] {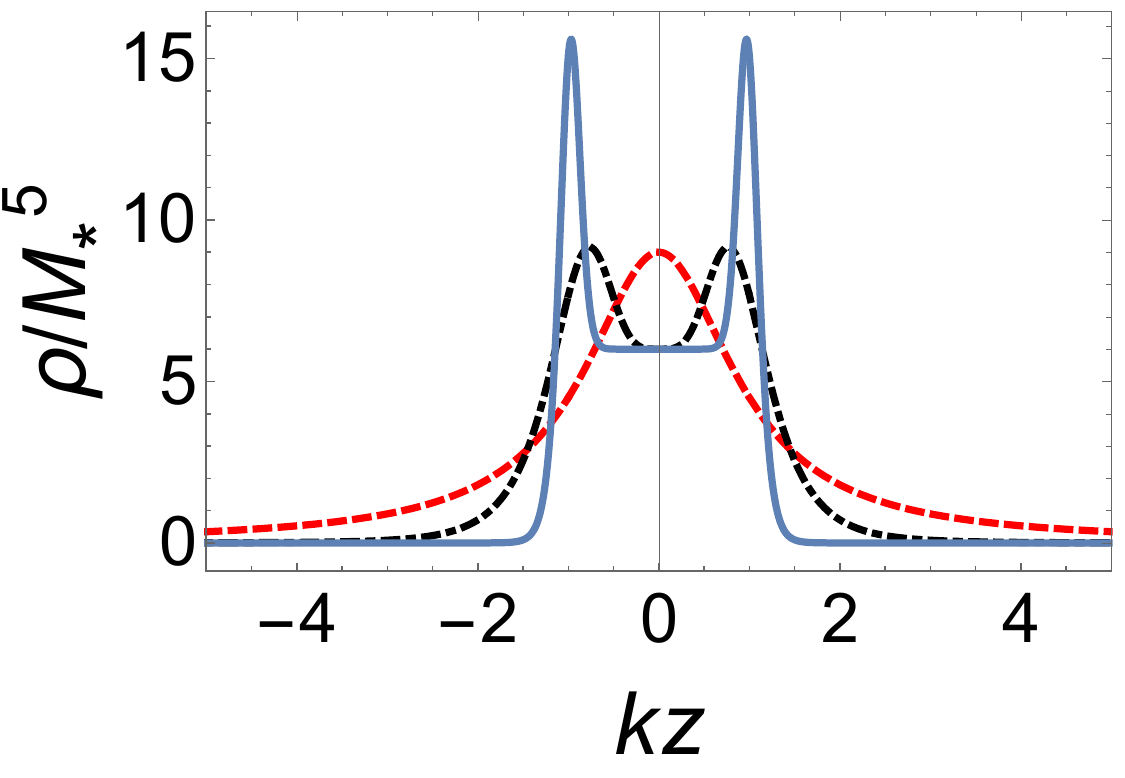}}
		\subfigure[Scalar potential\label{Vphi4} ]
		{\includegraphics[width=4.0cm] {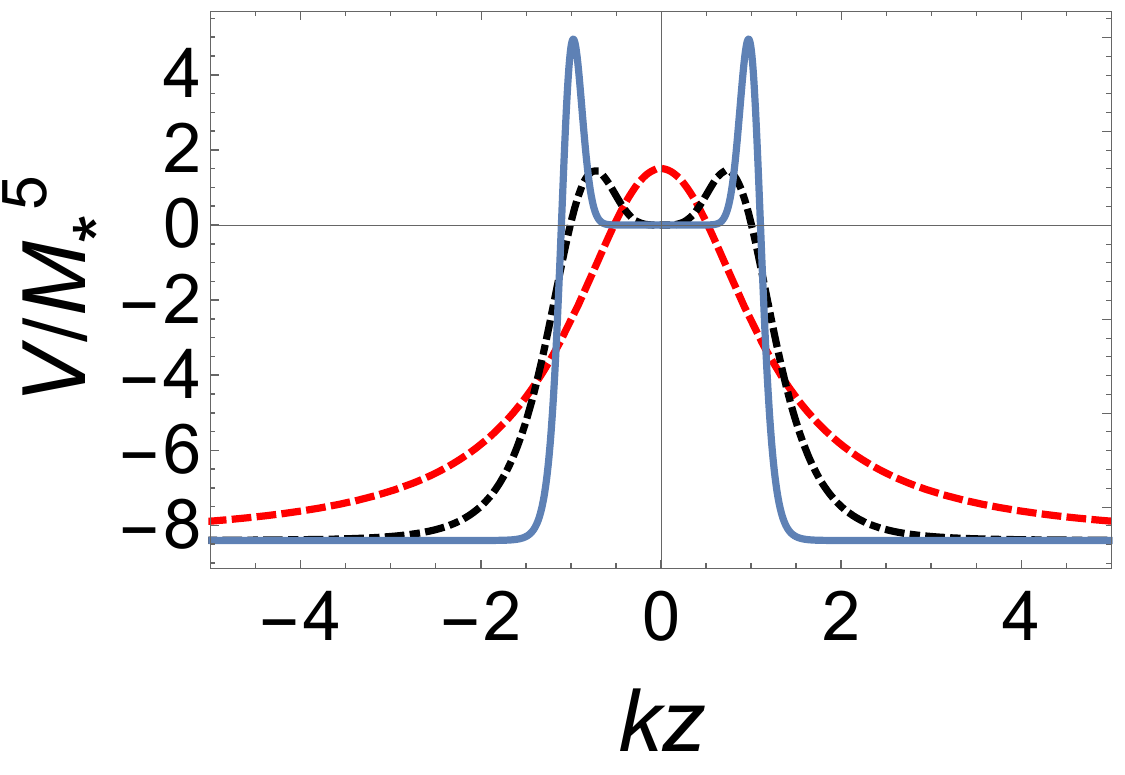}}
	\end{center}
	\caption{The brane solutions for the warp factor ~\eqref{solution6A} with different values of $n$. The red dashed lines, black dot dashed lines, and dark blue lines correspond to $n=0$, $n=2$, and $n=9$, respectively. Here we let $\tilde{\alpha}=-0.1$. \label{fig_case4}}
\end{figure}

\begin{figure}
	\begin{center}
		\subfigure[Scalar field\label{phi4_2}]
		{\includegraphics[width=4.0cm] {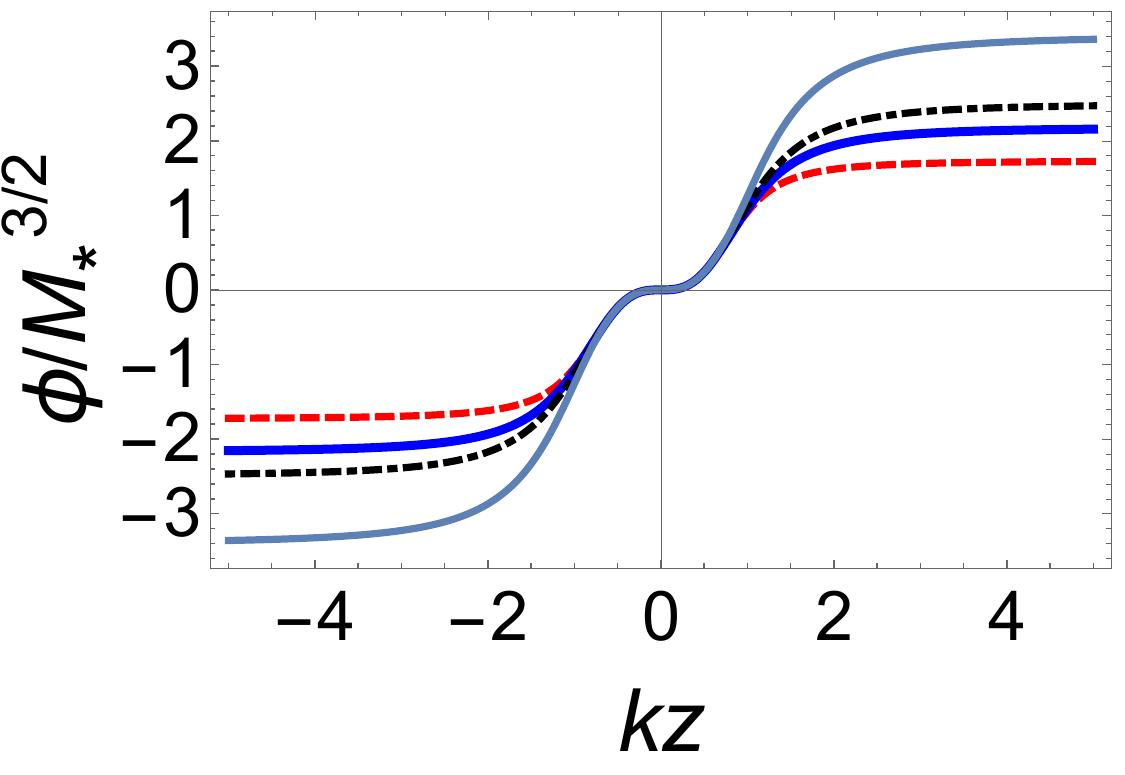}}
		\subfigure[Scalar potential\label{Vphi4_2}]
		{\includegraphics[width=4.0cm] {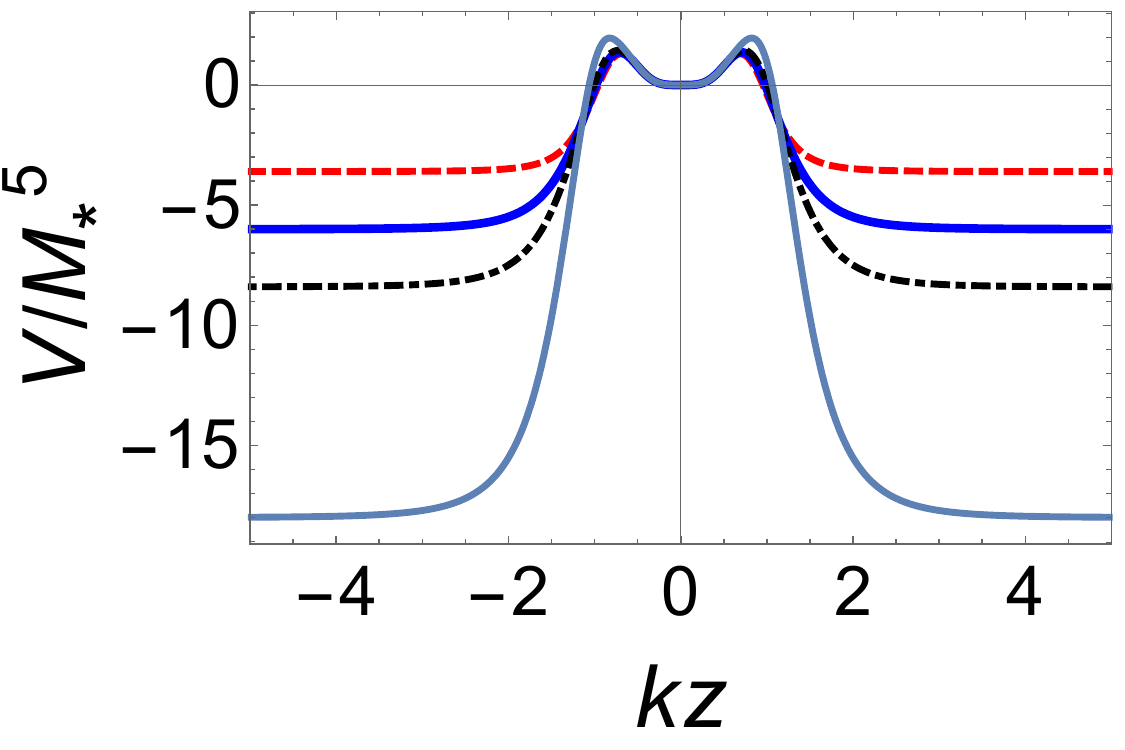}}
	\end{center}
	\caption{The brane solutions for the warp factor ~\eqref{solution6A} with different values of $\tilde{\alpha}$. The red dashed lines, blue solid lines, black dot dashed lines, and dark blue lines correspond to $\tilde{\alpha}=0.1$, $\tilde{\alpha}=0$, $\tilde{\alpha}=-0.1$, and $\tilde{\alpha}=-0.5$, respectively. Here we let $n=2$.\label{fig_case4_2}}
\end{figure}

In the last subsection, we have introduced the superpotential method to solve the thick brane system and shown that we can obtain the thick brane solutions with multi-kink background scalar fields by choosing some suitable superpotentials. In this subsection, we focus on the second method introduced in the end of Sec.~\ref{sec2}. As an example, we adopt the following polynomial expanded warp factor with respect to the extra dimensional coordinate $z$:
\begin{align}
e^{A(z)} &=\sqrt{ \frac {s_1(z)}{s_2(z)}}, \label{solution6A} \\
s_1(z) &= \sum_{i=0}^{n} (kz)^{2i}, \label{s1} \\
s_2(z) &= \sum_{i=0}^{n+1} (kz)^{2i}, \label{s2}
\end{align}
where $n$ is a nonnegative integer and $k$ is a positive real parameter with mass dimension. The scalar potential and the scalar field can be solved from (\ref{V_Az}) and (\ref{phi_Az}), respectively.
It can be seen that $\tilde{\alpha} \leq \frac{1}{8}\frac{M_{*}^2}{k^2}$ can guarantee the scalar field is real. We plot the brane solutions in Fig.~\ref{fig_case4} and Fig.~\ref{fig_case4_2} by choosing different values of $n$ and $\tilde{\alpha}$, respectively. Since the warp factor and energy density are independent of $\tilde{\alpha}$, we do not repeat their plots in Fig.~\ref{fig_case4_2}. With the increase of the parameter $n$, the warp factor shown in Fig.~\ref{ay4} becomes much wider with a platform around $z=0$, and the scalar field shown in Fig.~\ref{phi4} becomes double-kink from single-kink when $n\geq 1$. 
The energy density in Fig.~\ref{energy4} shows that the brane splits into two sub-branes. When the parameter $n$ increases to infinity, the warp factor, the scalar field, the energy density, and the scalar potential still have the similar shapes as the case of a finite $n$. From Fig.~\ref{fig_case4_2} we see that the scalar field and scalar potential are stretched with the decrease of the parameter $\tilde{\alpha}$.
\begin{table}[!htb]
	\begin{center}
		\caption{Parameters used in Fig.~\ref{fig_case4}}
		\begin{tabular}{|c||c|c|c|}
			\hline
			solution&Lines &~~~~$\tilde \alpha$~~~~&~~~~$n$~~~~   \\
			\hline	\hline
			solution\_d1  &Red dashed lines & -0.1  &0\\
			\hline
			solution\_d2  &Black dot dashed lines & -0.1  & 1 \\
			\hline
			solution\_d3  &Dark blue lines & -0.1  &  2  \\
			\hline
		\end{tabular}
		\label{tab_case4}
	\end{center}
\end{table}
\begin{table}[!htb]
	\begin{center}
		\caption{Parameters used in Fig.~\ref{fig_case4_2}}
		\begin{tabular}{|c||c|c|c|}
	\hline
	solution&Lines &~~~~$\tilde \alpha$~~~~&~~~~$n$~~~~   \\
	\hline	\hline
	solution\_e1  &Red dashed lines & 0.1  &2\\
	\hline
	solution\_e2  &Blue solid lines & 0  & 2 \\
	\hline
	solution\_e3  &Black dot dashed lines & -0.1  &  2  \\
	\hline
	solution\_e4  &Dark blue lines & -0.5  &  2  \\
    \hline
		\end{tabular}
		\label{tab_case5}
	\end{center}
\end{table}

\section{Stability of the system and localization of gravity}  \label{sec4}
So far, we have obtained four kinds of thick brane solutions. {It has been shown that the four-dimensional Newtonian potential can be recovered. Note that the four-dimensional Newtonian potential is generated by the massless KK graviton and an additional correction will be induced by the massive KK gravitons. Therefore, to check that whether the four-dimensional Newtonian potential can be recovered, we need to study the properties of the brane system under four-dimensional tensor perturbations of the background metric}. The total metric is given by
\begin{equation}
\label{metric perturbation}
ds^2 = e^{2 A(z)} \big[(\eta_{\mu \nu}+h_{\mu \nu}) dx^\mu dx^\nu + dz^2\big],
\end{equation}
where $h_{\mu \nu}=h_{\mu \nu}(x^{\mu},z)$ is a transverse and traceless tensor perturbation of the background metric (\ref{metric_conf}). The transverse and traceless gauge conditions are given by $\partial^{\mu}h_{\mu \nu}=0=h^{\mu}_{\mu}$.

The linear perturbation equation is
\begin{equation}\label{tenperEineq}
\delta G_{\mu }^{\nu }-\epsilon \delta Q_{\mu }^{\nu }=\kappa \delta T_{\mu }^{\nu },
\end{equation}
where $\epsilon=2 \tilde \alpha \kappa M_* $. The above equation can be expanded as
\begin{align}
& \left(1+2 \epsilon \bar{R}\right) \delta R_{\mu }^{\nu }-\frac{1}{2}\bar{R} h_{\mu}^{\nu}-2 \epsilon
\bigg[\frac{1}{4} {\cal \bar{R}_{\rm GB}} h_{\mu}^{\nu}\nonumber\\
&+2 (\delta R_{\mu \beta } \bar{R}^{\nu \beta }+ \bar{R}_{\mu \beta }\delta R^{\nu \beta }) + 2 (\delta R_{AB } \bar{R}_{\mu }^{ A \nu B }+\bar{R}_{AB } \delta R_{\mu }^{A \nu B })
\nonumber\\
&-(\delta R_{\mu ABC} \bar{R}^{\nu ABC}+ \bar{R}_{\mu ABC}\delta R^{\nu ABC})\bigg]=\kappa \delta T_{\mu}^{\nu}.  \label{perturbedEq}
\end{align}
The quantities with a bar and with a delta in Eq.~\eqref{perturbedEq} correspond to the background and the perturbation, respectively. With the help of the equations of motion of the background, the equation for the tensor perturbation under the transverse and traceless gauges can be derived as follows~\cite{Giovannini:2001ta}
\begin{align}
\label{perturbationeq}
\left(1-4 \epsilon e^{-2 A} \partial_z^2 A\right) \Box^{(4)} h_{\mu }^{\nu }
+ \left[1-4 \epsilon e^{-2 A}(\partial_z A)^2\right]\partial_z^2 h_{\mu }^{\nu }\nonumber \\
+ \left[ 3\partial_z A -4 \epsilon e^{-2 A} \left( 2 (\partial_z A) \partial_z^2 A+ (\partial_z A)^3 \right) \right] \partial_z h_{\mu }^{\nu }
=0,
\end{align}
where $\Box^{(4)} =\eta^{\mu \nu} \partial_{\mu} \partial_{\nu} $ denotes the four-dimensional d'Alembert operator on the brane. The above equation can also be given in the physical coordinate $y$ by the coordinate transformation $ dy=e^{A} dz $. The result is
\begin{equation}\label{uncomformeq}
\square^{(4)} h_{\mu }^{\nu }+F^2(y) h_{\mu }^{\nu }{''}+ H(y) h_{\mu }^{\nu }{'}=0,
\end{equation}
where the prime denotes the derivative respect to physical coordinate $y$ and
\begin{align}
F(y)&=e^{A}\sqrt{\frac{ \left(1-4 \epsilon A'^2\right)}{1-4 \epsilon (A'^2+A'')}},\label{Fy} \\
H(y)&=e^{2 A} \frac{4(A'-2 \epsilon A' A''-4 \epsilon A'^3)}{1-4 \epsilon (A'^2+A'')}.\label{Hy}
\end{align}
Then, by considering the KK decomposition of the tensor perturbation $h_{\mu }^{\nu }(x^{\mu},y)=\chi_{\mu}^{\nu}(x^{\mu})  \varphi(y)$, we obtain the four-dimensional Klein-Gordon equation of the four-dimensional KK graviton $\chi_{\mu}^{\nu}(x^{\mu})$ with mass $m$ and the equation of the extra-dimensional part $\varphi(y)$:
\begin{align}
\left(\Box^{(4)} -m^2 \right) \chi_{\mu}^{\nu}(x^{\mu})&=0,\label{eq4dim} \\
F^2(y) \varphi''(y) +H(y) \varphi'(y) +m^2 \varphi(y)&=0 .\label{eq5dim}
\end{align}
Using another coordinate transformation $dy=F(y(w))dw$, equation~\eqref{eq5dim} can be rewritten in the coordinate $w$ as
\begin{equation}
\partial_w^2 \varphi (w)+ \mathcal{K}(w)\partial_w \varphi (w)+m^2 \varphi (w) =0,
\end{equation}
where $\mathcal{K}(w)=\frac{H-\partial_w F }{F} $. Finally, by making the field transformation $\varphi (w)=f(w)\xi(w)$ with the function $f(w)$ satisfying
\begin{equation}
\frac{\partial_w f(w)}{f(w)}=\frac{1}{2}\left(\frac{\partial_w F(w)}{F(w)}-\frac{H(w)}{F(w)}\right),
\end{equation}
we can transform the equation for $\varphi (w)$ as the following Schr\"{o}dinger-like equation
\begin{equation}
\left( -\partial_w^2 + V_{\text{eff}}(w)\right) \xi(w) =m^2 \xi(w),  \label{schEq}
\end{equation}
where the effective potential in the coordinate $w$ is given by
\begin{equation}
V_{\text{eff}} (w)=\frac{ 2 F\left(\partial_w H-\partial_w^2 F\right)-4 H \partial_w F+3 (\partial_w F)^2+H^2}{4 F^2}.  \label{Veff(w)}
\end{equation}
The Schr\"{o}dinger-like equation (\ref{schEq}) can be expressed as the following form
\begin{equation}  \label{scheqqm}
Q^{\dagger} Q \xi (w)=m^2 \xi (w),
\end{equation}
where the operators $Q$ and $Q^{\dagger}$ are given by
\begin{align}
Q= \partial_{w}+P(w), ~~~~
Q^{\dagger}= -\partial_{w}+ P(w), \\
P(w)=\frac{1}{2}\left[\frac{\partial_w F}{F}-\frac{H}{F}\right].
\end{align}
Equation \eqref{scheqqm} guarantees $m^2\geq 0$ and there is no tachyonic graviton. So it is proved that the system is stable under the linear tensor perturbations. The tensor zero mode can be obtained by solving Eq.~(\ref{scheqqm}) with $m^2=0$:
\begin{equation}
\xi_{0}(w) =N_{0} e^{\pm\int{P(w)dw}},  \label{xi0}
\end{equation}
where $N_{0}$ is the normalization constant. A localized graviton zero mode should satisfy the normalization condition $\int_{-\infty}^{\infty}\xi_{0}^{2} (w)dw=1$.

Next, we use the relation $\frac{\partial w}{\partial y}=\frac{1}{F}$ to write the expressions of the effective potential $V_{\text{eff}}$ and the tensor zero mode $\xi_{0} $ in the physical coordinate $y$:
\begin{align}\label{veff}
V_{\text{eff}}(w(y))&=\frac{1}{4} \left(\frac{H^2(y)}{F^2(y)}-\frac{4 H(y) F'(y)}{F(y)}+F'^2(y)\right)\nonumber \\
& +\frac{1}{4} \Big(2 H'(y)-2 F(y) F''(y)\Big),\\
\xi_{0}(w(y)) &=N_{0} e^{\int{ \mathcal{P}_{\pm}(y) dy}},   \label{zm}
\end{align}
where the functions $\mathcal{P}_{\pm}(y)$ are given by
\begin{equation}
\mathcal{P}_{\pm}(y)=\pm\frac{F(y) F'(y)-H(y)}{2 F(y)^2}. \label{Py}
\end{equation}
For an asymptotic AdS$_{5}$ spacetime described by the metric \eqref{metric5}, the warp factor $e^{A(y)}$ tends to $ A(y)\rightarrow -k |y| $ when $ |y|\rightarrow \infty$. Thus, the behavior of the effective potential at infinity is
\begin{equation}
\lim_{|y| \rightarrow \infty} V_{\text{eff}} \rightarrow \frac{15}{4} k^2 e^{-2 k | y|},
\end{equation}
which shows that it will tend to zero when $ |y|\rightarrow \infty$. The asymptotic behavior of the factor $\int \mathcal{P}_{\pm} dy$ of the zero mode at infinity is
\begin{equation}
\label{y_infinity}
\lim_{|y|\rightarrow  \infty} \int {\mathcal{P}_{\pm}(y)dy} = \pm \left(\frac{3 k}{2}|y|+C_{0}\right),
\end{equation}
where $C_{0}$ is an integration constant. The asymptotic behavior \eqref{y_infinity} shows that the following condition
\begin{equation}
\int_{-\infty}^{\infty}\xi_{0}(y)^2 dy =1
\label{nomilization}
\end{equation}
can be satisfied with a finite normalization constant $N_{0}$ for the case of $\mathcal{P}_{-}$ in the expression~(\ref{y_infinity}). Therefore, it is proved that the corresponding zero mode $\xi_{0}(w(y)) =N_{0} e^{\int{ \mathcal{P}_{-}(y) dy}}$ is localized on the brane and the four-dimensional Newtonian potential can be recovered.

Next, we focus on the profiles of the effective potentials and the corresponding configurations of the tensor zero modes on the thick brane solutions obtained in Sec.~\ref{sec3}.

Note that, the effective potential (\ref{veff}) would diverge when the function $F(y)$ given in Eq.~\eqref{Fy} goes to zero or the function $H(y)$ given in Eq.~\eqref{Hy} diverges at certain finite extra dimension coordinate $y=y_0$, which is equivalent to
\begin{equation}
\left(4 \epsilon  A'^2-1\right) \left(4 \epsilon  \left(A'^2+A''\right)-1\right)=0. \label{singularityCondition}
\end{equation}
The divergence of the effective potential means that there exists a nonsmooth tensor zero mode in the smooth thick brane.

\begin{figure}[!ht]
	\begin{center}
		\subfigure[Effective potential]
		{\includegraphics[width=4.0cm] {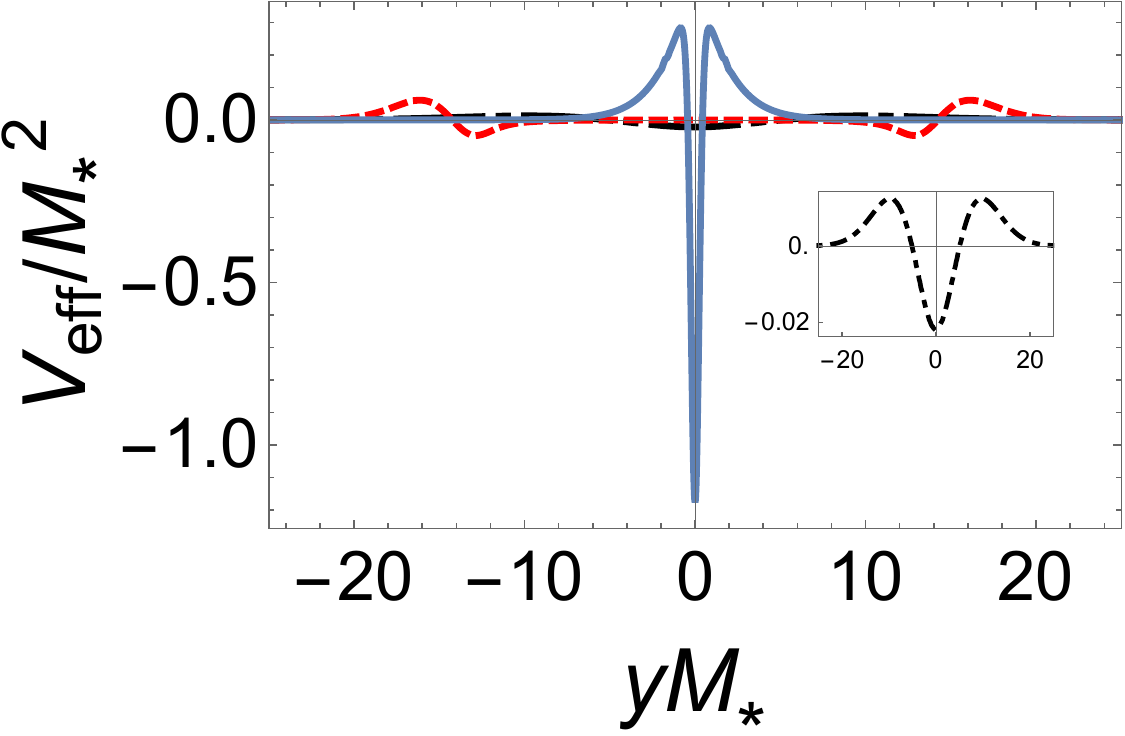}}
		\subfigure[Zero mode]
		{\includegraphics[width=4.0cm] {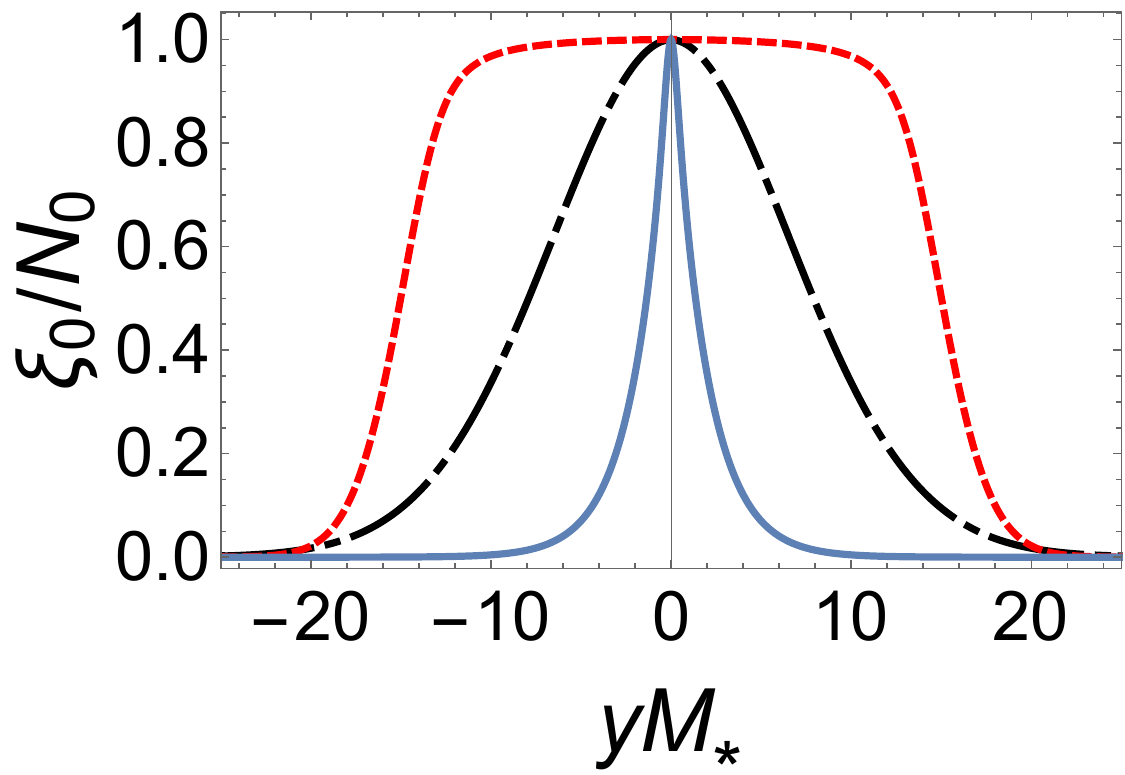}}
	\end{center}
	\caption{The effective potential and zero mode for the first brane solution given in Sec. \ref{sec3.1} with the superpotential \eqref{supp1}. The parameters are given in Table \ref{tab_case1}.
		\label{Veff_xi0_1}}
\end{figure}

\begin{figure}[!ht]
	\begin{center}
		\subfigure[Effective potential]
		{\includegraphics[width=4.0cm] {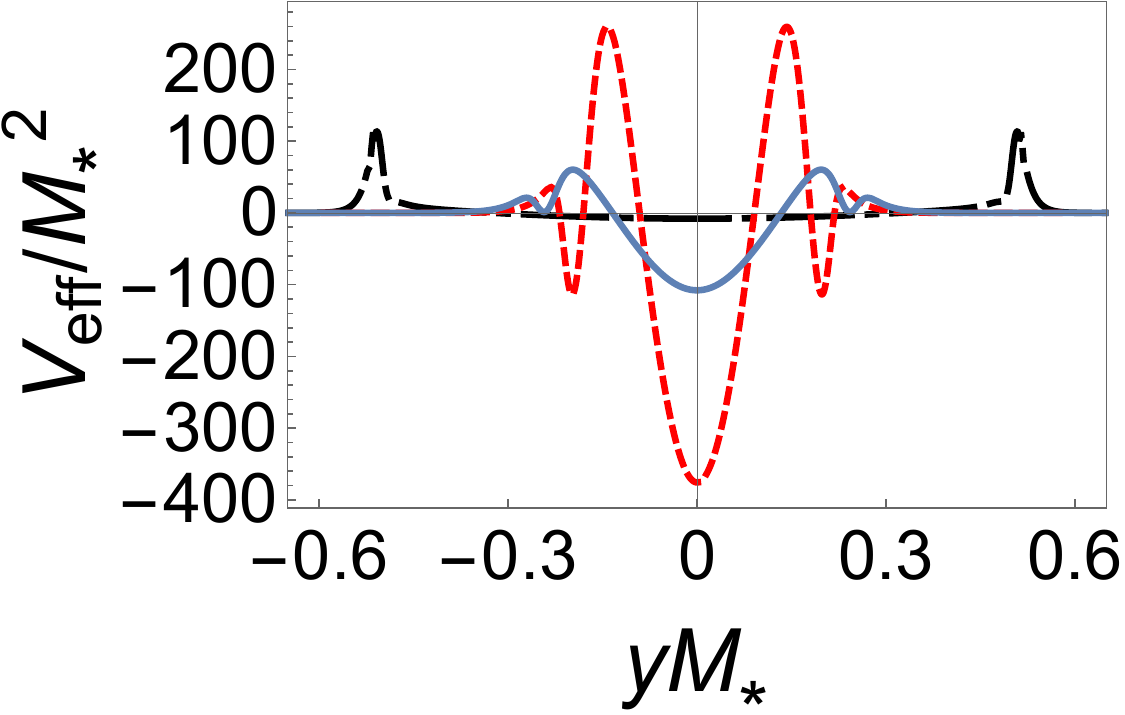}}
		\subfigure[Zero mode]
		{\includegraphics[width=4.0cm] {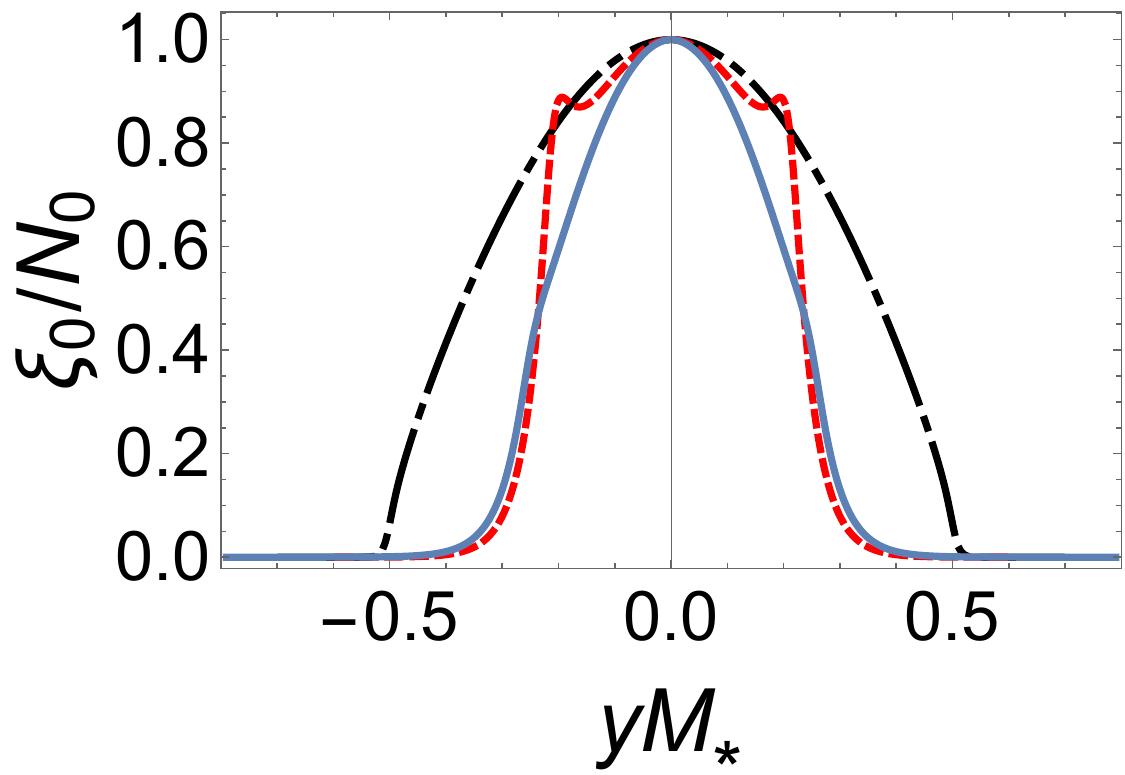}}
	\end{center}
	\caption{The effective potential and zero mode for the second brane solution in Sec. \ref{sec3.1} with the superpotential \eqref{supp2}. The parameters are given in Table \ref{tab_case2}.
		\label{Veff_xi0_2}}
\end{figure}

\begin{figure}[!ht]
	\begin{center}
		\subfigure[Effective potential]
		{\includegraphics[width=4.0cm] {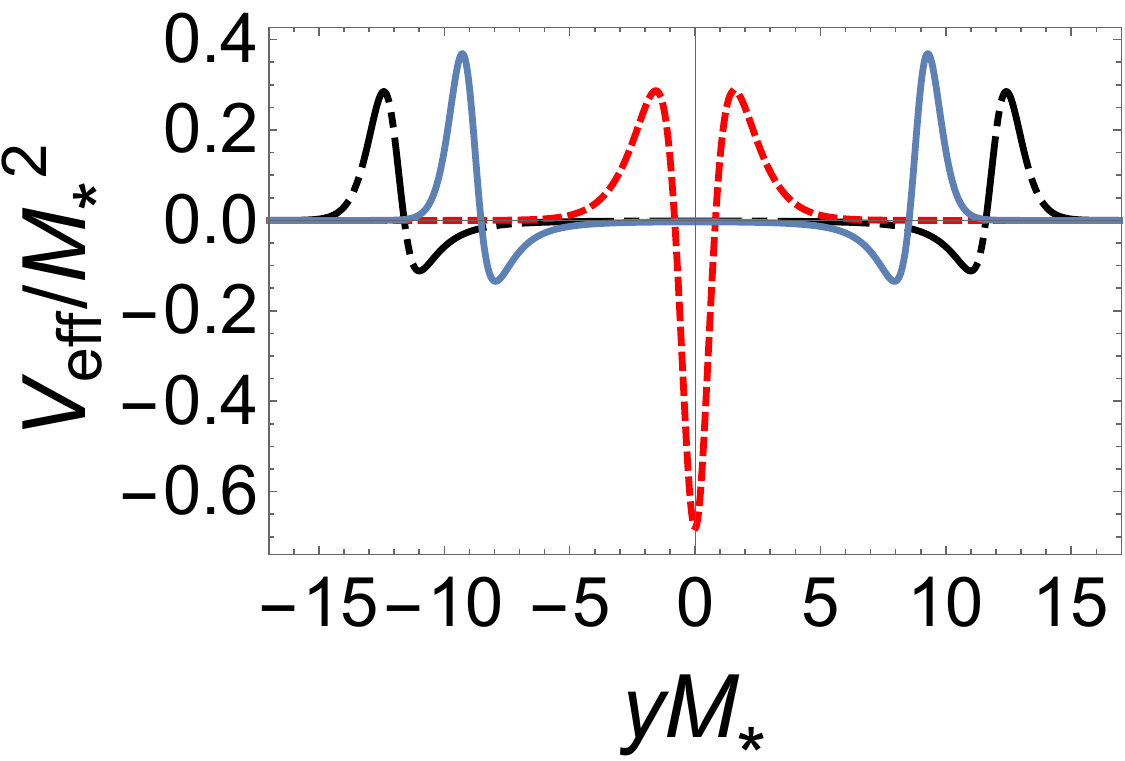}}
		\subfigure[Zero mode]
		{\includegraphics[width=4.0cm] {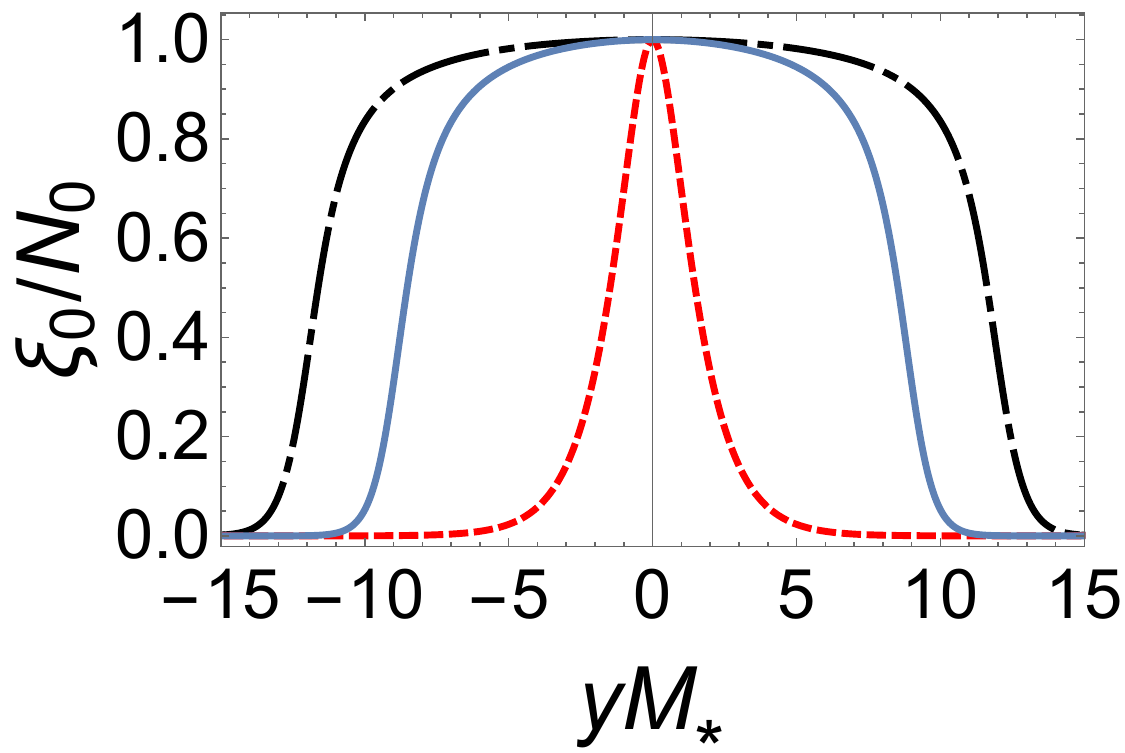}}
	\end{center}
	\caption{The effective potential and zero mode for the third brane solution in Sec. \ref{sec3.1} with the superpotential \eqref{supp3}. The parameters are given in Table \ref{tab_case3}.
		\label{Veff_xi0_3}}
\end{figure}

\begin{figure}[!ht]
	\begin{center}
		\subfigure[Effective potential]
		{\includegraphics[width=4.0cm] {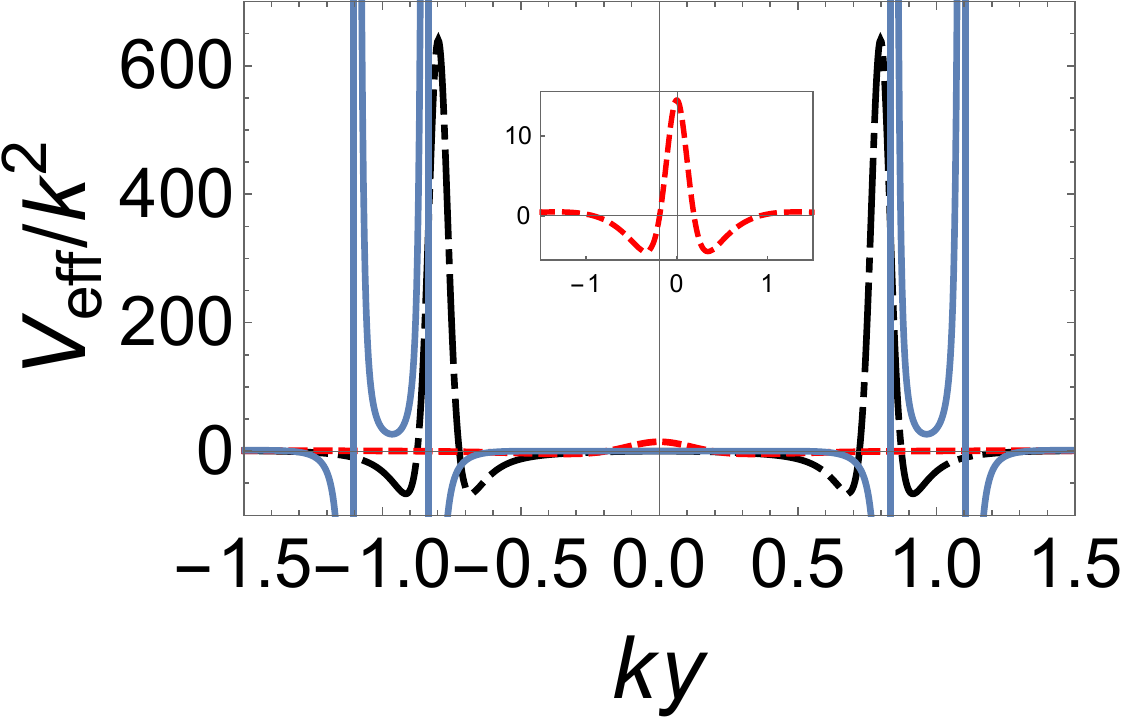}}
		\subfigure[Zero mode]
		{\includegraphics[width=4.0cm] {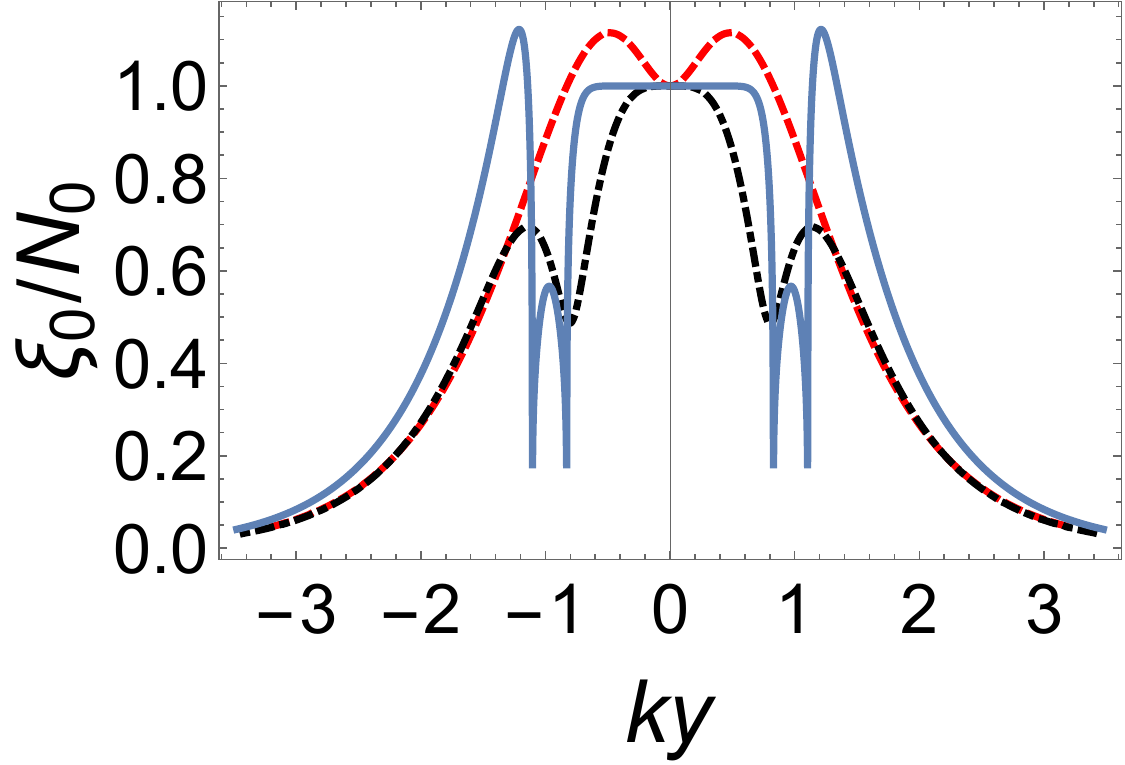}}
	\end{center}
	\caption{The effective potential and zero mode for the brane solution given in Sec. \ref{sec3.2} with the given warp factor \eqref{solution6A} for different values of $n$. The parameters are set to $\tilde{\alpha}=-0.1$ and $n=(0, 2, 9)$ for the red dashed, black dot dashed,  and dark blue lines, respectively.
		\label{Veff_xi0_4}}
\end{figure}

\begin{figure}[!ht]
	\begin{center}
		\subfigure[Effective potential]
		{\includegraphics[width=4.0cm] {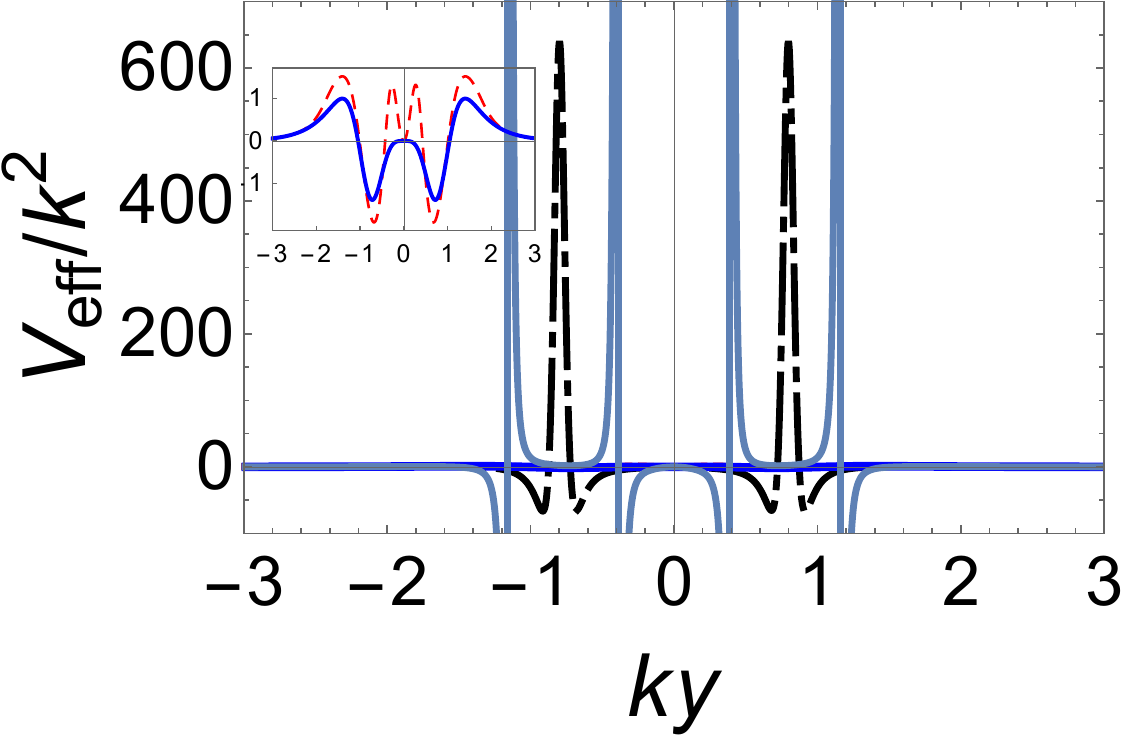}}
		\subfigure[Zero mode]
		{\includegraphics[width=4.0cm] {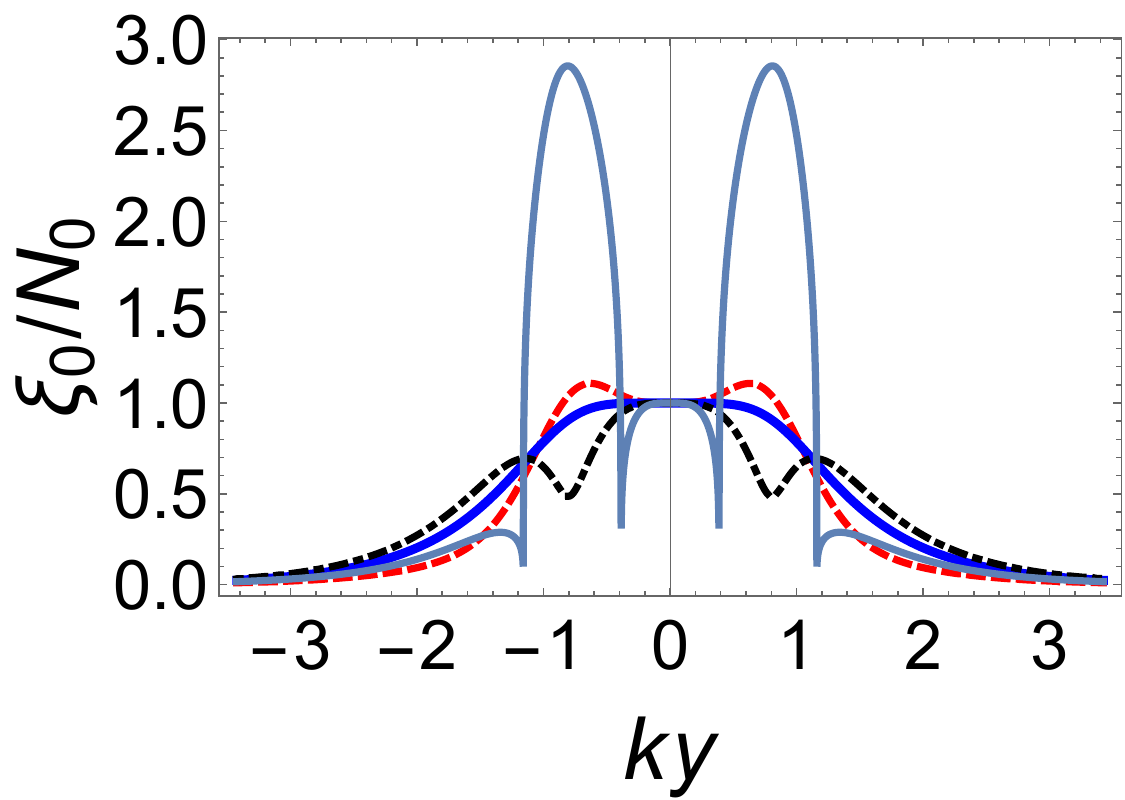}}
	\end{center}
	\caption{The effective potential and zero mode for the brane solution given in Sec. \ref{sec3.2} with the given warp factor \eqref{solution6A} for different values of $\tilde{\alpha}$. The parameters are set to $n=2$ and $\tilde{\alpha}=(0.1, 0, -0.1, -0.5)$ for the red dashed, blue solid, black dot dashed, and dark blue lines, respectively.
		\label{Veff_xi0_4_2}}
\end{figure}

Finally, we give the shapes of the effective potentials and tensor zero modes of our four kinds of thick brane solutions. Figures~\ref{Veff_xi0_1},~\ref{Veff_xi0_2}, and~\ref{Veff_xi0_3} show the effective potentials and tensor zero modes of the thick brane solutions generated by the superpotentials~\eqref{supp1},~\eqref{supp2},~and~\eqref{supp3}, respectively. It can be seen that the distribution of each effective potential is consistent with the energy density of the corresponding thick brane. With the specific superpotential $W(\phi)$, the effective potential $V_{\text{eff}}$ can have a richer structure. For instance, in Fig.~\ref{Veff_xi0_2} one can see that there are several wells in the effective potential. Figures \ref{Veff_xi0_4} and  \ref{Veff_xi0_4_2} show the effective potential and tensor zero mode of the thick brane solution \eqref{solution6A} for different values of $n$ and $\tilde{\alpha}$, respectively. Here, we obtain the discontinuous effective potential and nonsmooth tensor zero mode for a smooth thick brane when $n$ or $-\tilde{\alpha}$ has a large value. This is a new feature that did not find in general relativity. Although the profiles of these effective potentials and tensor zero modes are quite different, the localization condition \eqref{nomilization} are satisfied and so all the zero modes we obtained can be localized on the brane and the four-dimensional Newtonian potential can be recovered.

{
We have discussed the localization of the KK gravitons and the corresponding tachyon stability. Next, we further discuss the ghost problem of the tensor perturbations. It has been proved that the linear perturbation equations can be derived via the Hamiltonian variation principle in terms of the quadratic actions \cite{Zhong:2012nt,Zhong:2015pta}. Therefore, the coefficients of the kinetic term for the quadratic action of $h_{\mu \nu}$ can be recast in terms of the equation of motion \eqref{perturbationeq}. That is to say, the kinetic terms of the quadratic action will have the following forms
\begin{align}
	 & -K_1 (z) h^{\mu \nu} \partial^2_t h_{\mu \nu} +K_1 (z)h^{\mu \nu}\partial^2_{\stackrel{\rightarrow}{ x}} h_{\mu \nu} + \nonumber \\
	 &~  K_2(z) h^{\mu \nu}\partial_z^2 h_{\mu \nu} ,
\end{align}
where
\begin{align}
	K_1(z)&=1-8 \tilde \alpha \kappa e^{-2 A} \partial_z^2 A, \\
	K_2(z)&=1-8 \tilde \alpha \kappa e^{-2 A}(\partial_z A)^2.
\end{align}
Therefore, the no-ghost conditions for the KK tensor modes are
\begin{align}
	K_1 >0 \quad \mathrm{and} \quad K_2 > 0.
\end{align}
It can be seen that the above conditions are satisfied in GR with $\tilde\alpha=0$, and there is no ghost in GR. {Next we will give the constraint on the GB coupling parameter $\tilde{\alpha}$. It is noted that the value of function $e^{-2 A}(\partial_z A)^2$ is always positive definite, for which we can get the following inequality}
\begin{equation}
 \tilde\alpha <  \frac{e^{2A(z)}}{8 \kappa (\partial_z A)^2 }. \label{condition1}
\end{equation}
For the function $e^{-2 A} \partial_z^2 A$, its value can be positive or negative, therefore, we can get another two conditions
\begin{align}
	\tilde\alpha > \frac{e^{2A(z)}}{8 \kappa \partial_z^2 A},~~\text{if}~~\partial_z^2 A<0, \label{condition2a} \\
	\tilde\alpha < \frac{e^{2A(z)}}{8 \kappa \partial_z^2 A },~~\text{if}~~\partial_z^2 A>0. \label{condition2b} \\
\end{align}	
To check the ghost instability of our model, we define the two following functions
\begin{align}
	\Omega_1 &=\frac{e^{2A(z)}}{8 \kappa \partial_z^2 A },\\
	\Omega_2 &=\frac{e^{2A(z)}}{8 \kappa (\partial_z A)^2 }.
\end{align}

Then we can check whether the conditions ({\ref{condition1}), (\ref{condition2a}), and (\ref{condition2b}) are satisfied for our solutions. For the thick brane models given in case \ref{case1}, we give the result in Fig. \ref{ghost_a}. It can be seen  that the values of the functions $\Omega_1$ and $\Omega_2$ along the extra dimension will gradually tend towards the values of $\tilde{\alpha}$, for which the functions $K_1$ and $K_2$ are positive definite and there are no ghost instabilities for the tensor perturbations.
For the thick brane models that are given in case of \ref{case2}, we give the result in Fig. \ref{ghost_b} and we show that there are no ghost instabilities for the solutions in subfigures \ref{ghost_b_a}, \ref{ghost_b_b}, and \ref{ghost_b_c}. For the thick brane models that are given in case of \ref{case3}, we give the results in Figs. \ref{ghost_c} and prove that there are no ghost instabilities for these solutions.

We should note that, it is hard to figure out what are the exact effects of the GB term on the properties of thick brane for the thick brane solutions listed in Tables \ref{tab_case1}, \ref{tab_case2}, and \ref{tab_case3}. However, the solutions listed in Tables \ref{tab_case4} and \ref{tab_case5} are derived in terms of the directly given warp factors. Therefore, for a determined warp factor $A(z)$, the conditions (\ref{condition1}), (\ref{condition2a}), and (\ref{condition2b}) can directly give the constraint of the GB coupling parameter $\tilde\alpha$. As the concrete examples, we have checked the corresponding constraints obtained from the warp factors \eqref{solution6A}, we give the results in Figs. \ref{ghost_d}, \ref{ghost_e}, and \ref{ghost_f}. For the solutions listed in Table \ref{tab_case5}, the value of $\tilde{\alpha}$should satisfy
	\begin{equation}
	-0.1<\tilde\alpha<0.1
	\end{equation}
in terms of the constraints \eqref{condition1}, \eqref{condition2a}, and \eqref{condition2b}. We find that when we choose $\tilde\alpha=-0.5$, the tensor perturbation will have the ghost instability. We also choose different values of $\tilde\alpha$ as follows
\begin{equation}
\tilde\alpha=(-0.1, -0.05, -0.03, 0, 0.03, 0.05, 0.01)
\end{equation}
and give the results in Fig. \ref{ghost_f}, our results confirm that the correctness of constraints \eqref{condition1}, \eqref{condition2a}, and \eqref{condition2b}. We will give a summary about the ghost instability of the four-dimensional tensor perturbation in following Table \ref{stability}.}

\begin{figure}
	\centering
	\subfigure[Thick brane solution with $\tilde \alpha = 0.500$ , $a=0.200 $, and $b=0.010$.]{\includegraphics[width=4.0cm]{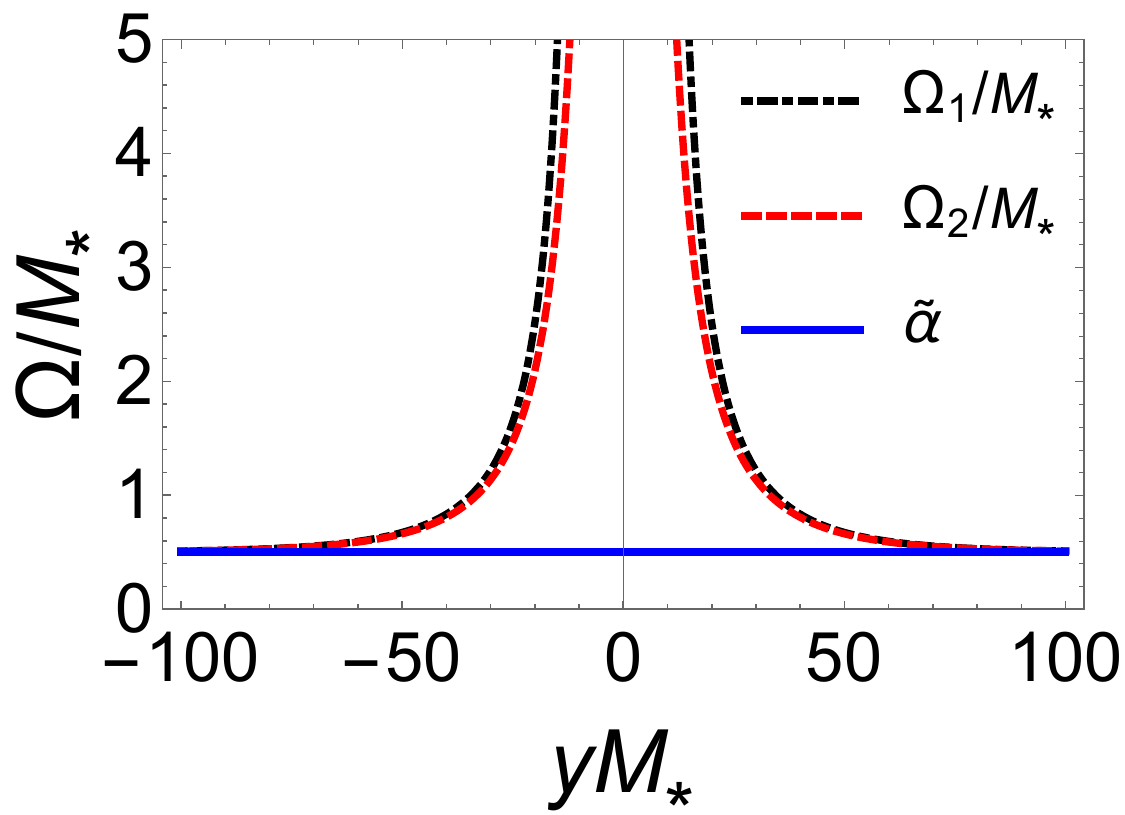}}
	\subfigure[Thick brane solution with $\tilde \alpha= 0.180$, $a=0.400$, and $b=-0.375$.]{\includegraphics[width=4.0cm]{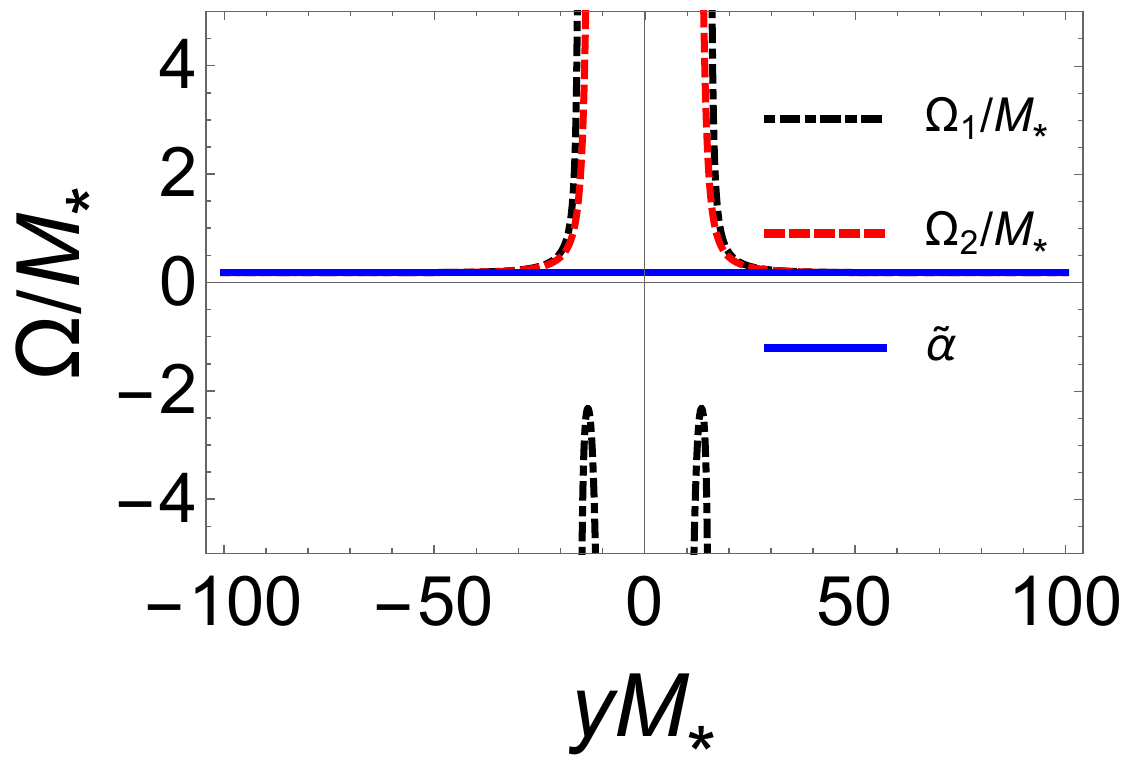}}
	\subfigure[Thick brane solution with $\tilde \alpha= 0.220$, $a=0.290$, and $b=1.000$.]{\includegraphics[width=4.0cm]{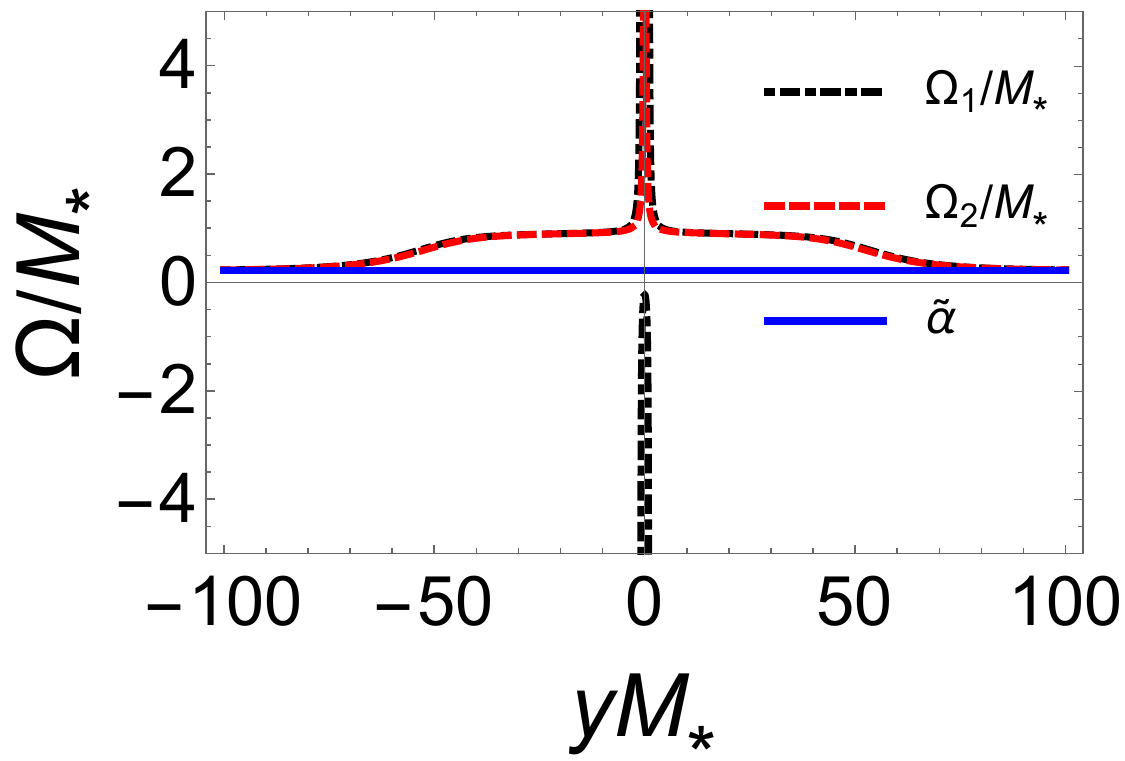}}
	\caption{Comparison between the functions $(\Omega_1, \Omega_2)$ and values of $\tilde\alpha$ for the thick brane solution of case \ref{case1}.}
	\label{ghost_a}
\end{figure}
\begin{figure}
	\centering
	\subfigure[Thick brane solution with $\tilde \alpha = 0.0001$ , $a=2.00 $, and $b=2.00$.\label{ghost_b_a}]{\includegraphics[width=4.0cm]{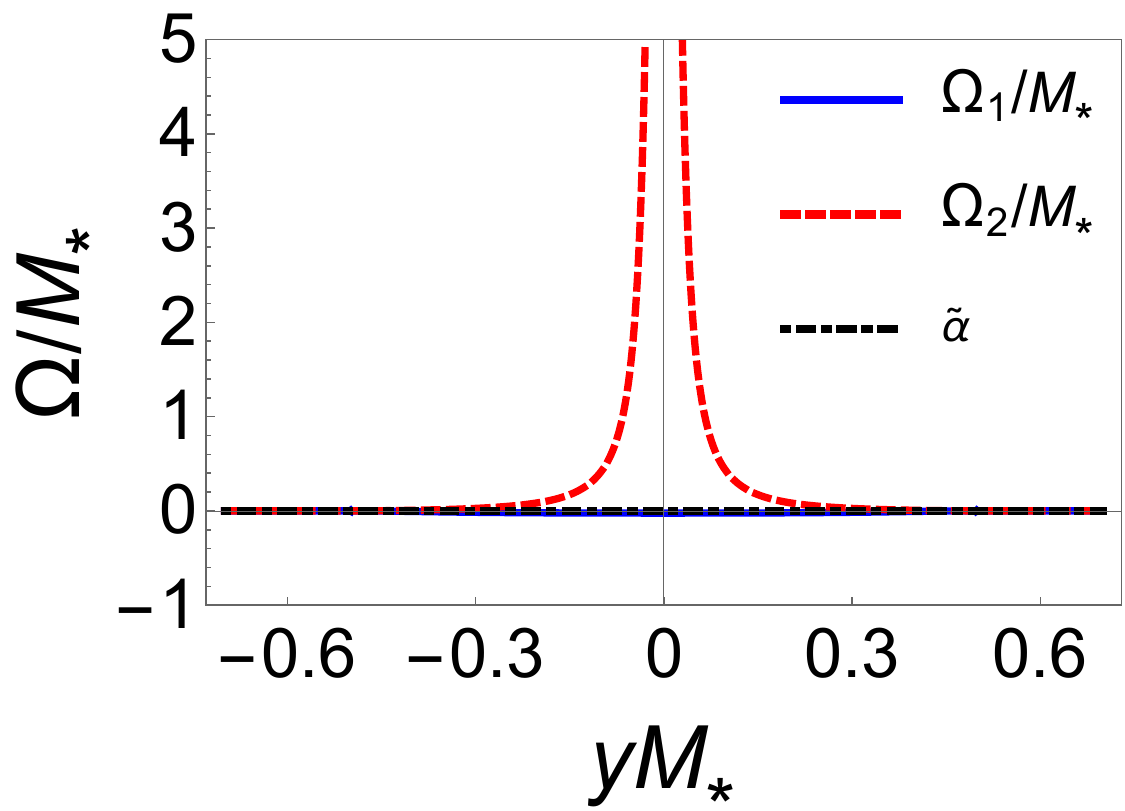}}
	\subfigure[Thick brane solution with $\tilde \alpha = -0.003$ , $a=-11.0 $, and $b=0.050$.\label{ghost_b_b}]{\includegraphics[width=4.0cm]{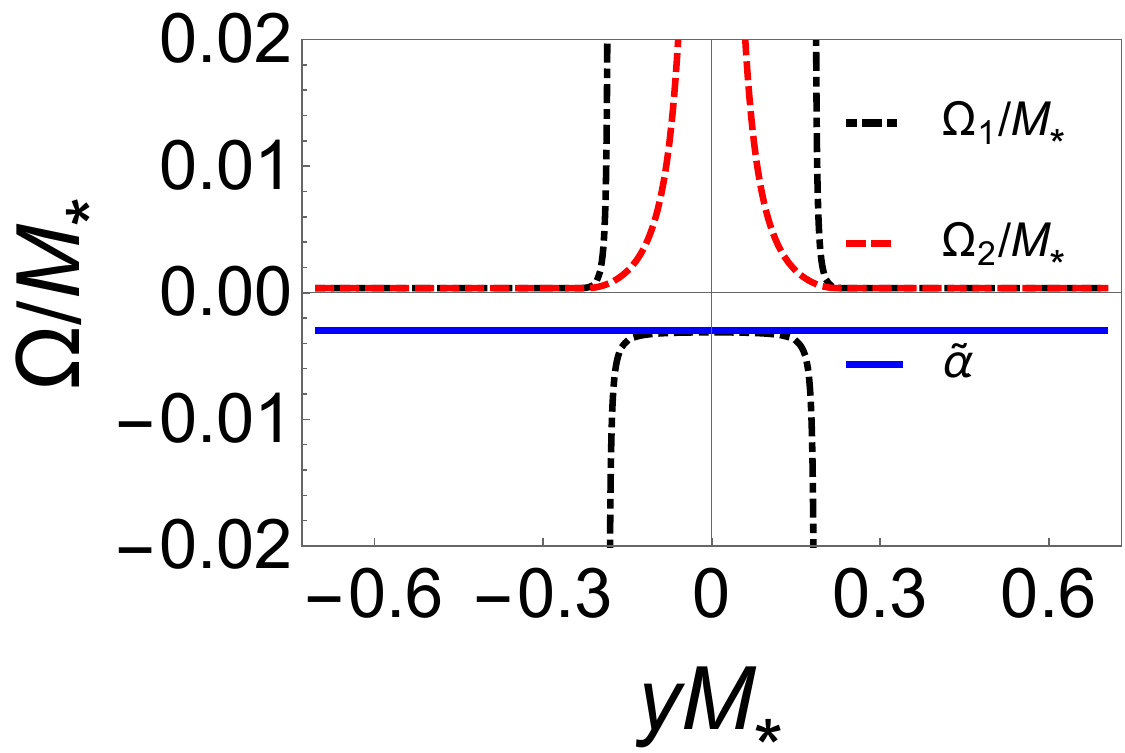}}
	\subfigure[Thick brane solution with $\tilde \alpha = -0.003$ , $a=-10 $, and $b=-0.05$.\label{ghost_b_c}]{\includegraphics[width=4.0cm]{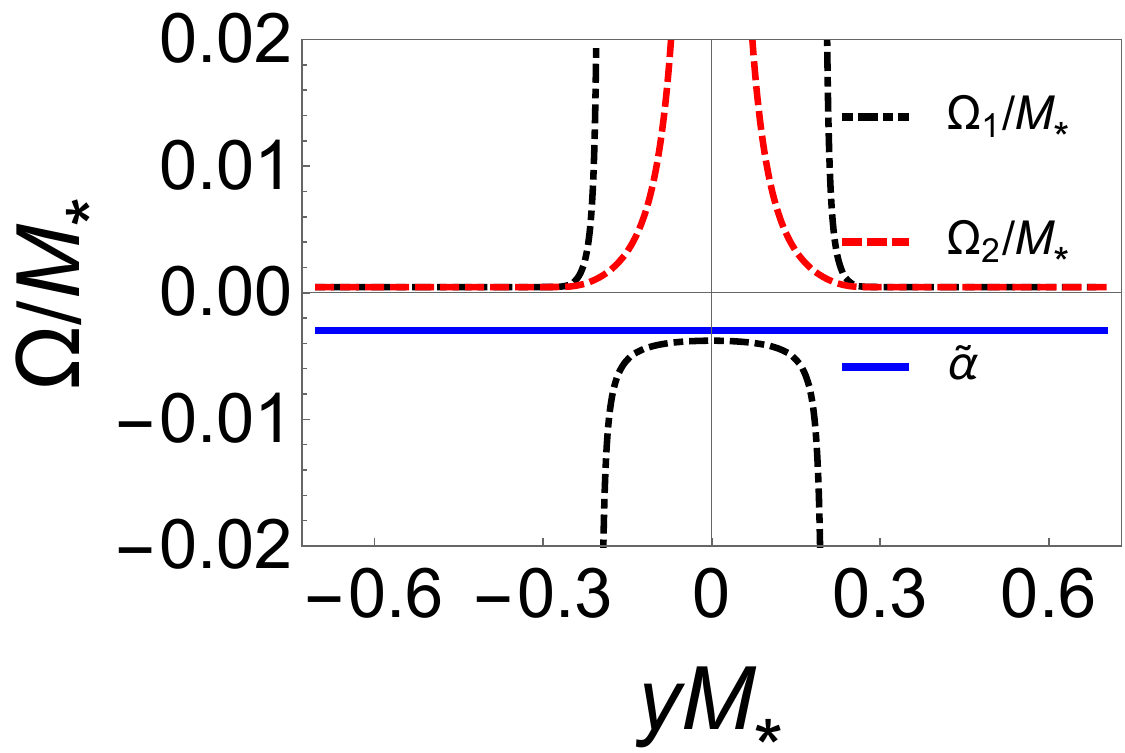}}
	\caption{Comparison between the functions $(\Omega_1, \Omega_2)$ and values of $\tilde\alpha$ for the thick brane solution of case \ref{case2}.}
	\label{ghost_b}
\end{figure}
\begin{figure}
	\centering
	\subfigure[Thick brane solution with $\tilde \alpha = 0.065$ , $a=0.700 $, and $b=-0.650$.]{\includegraphics[width=4.0cm]{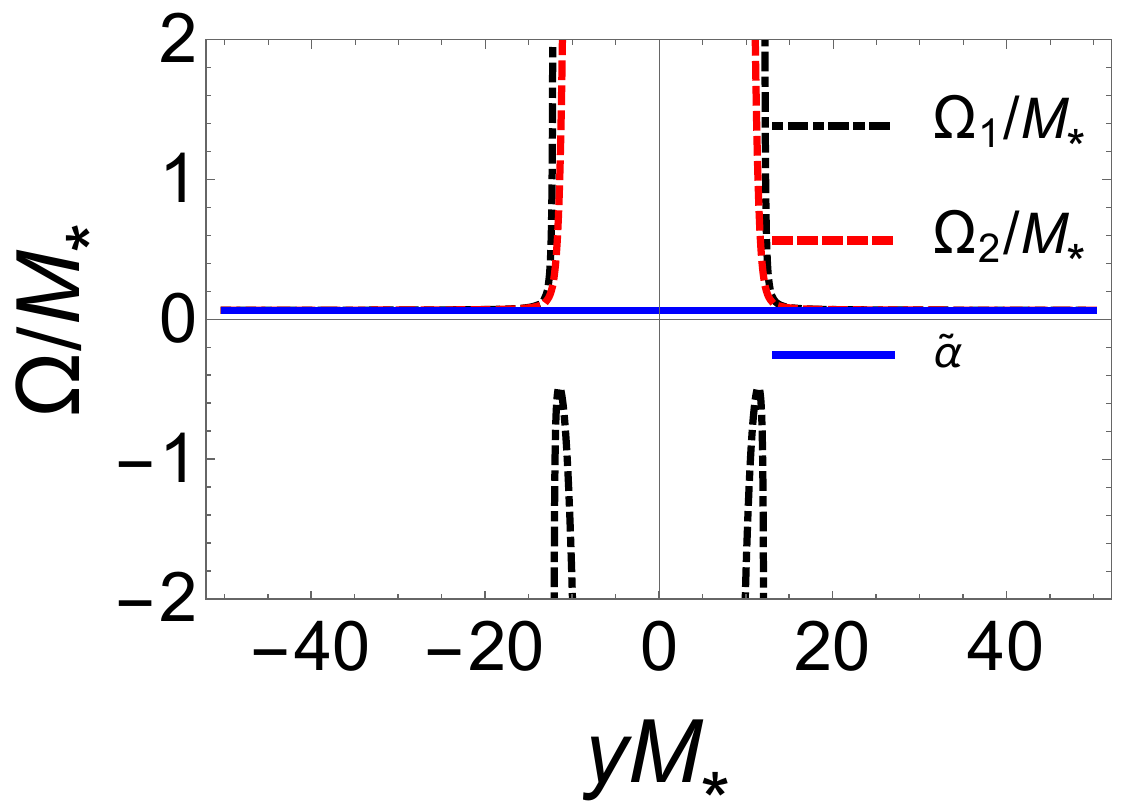}}
	\subfigure[Thick brane solution with $\tilde \alpha = 0.060$ , $a=0.600 $, and $b=0.560$.]{\includegraphics[width=4.0cm]{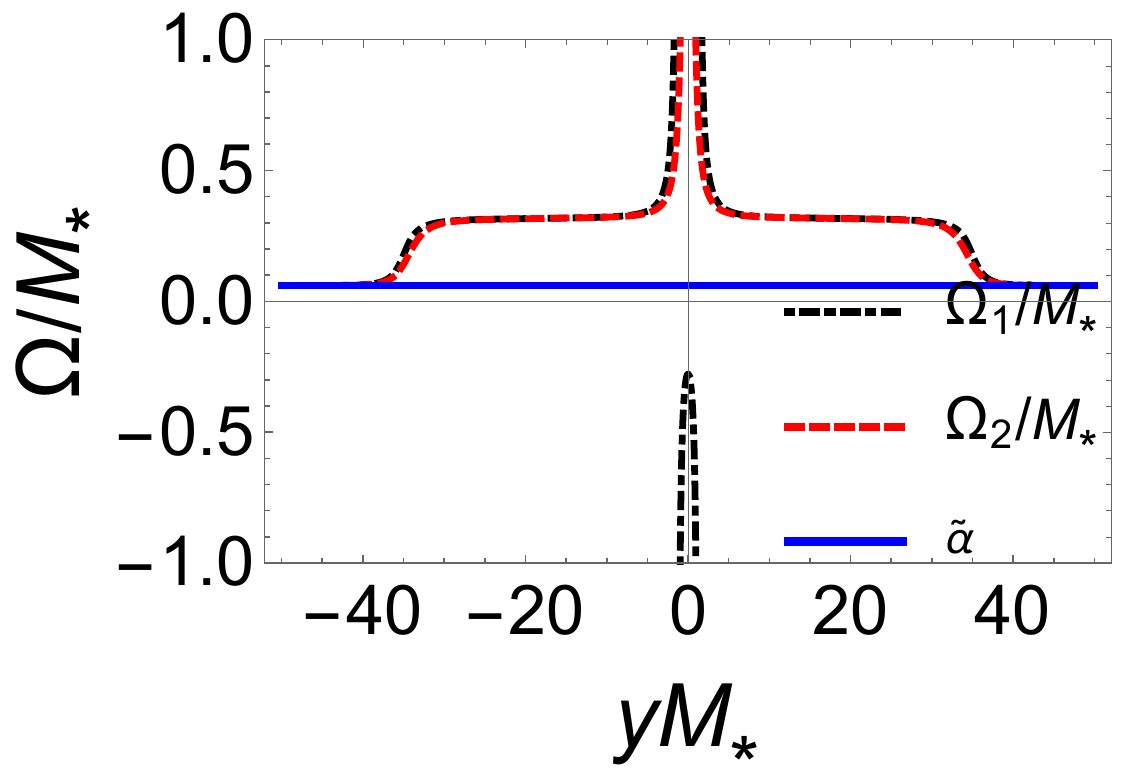}}
	\subfigure[Thick brane solution with $\tilde \alpha = 0.020$ , $a=0.800 $, and $b=-0.720$.]{\includegraphics[width=4.0cm]{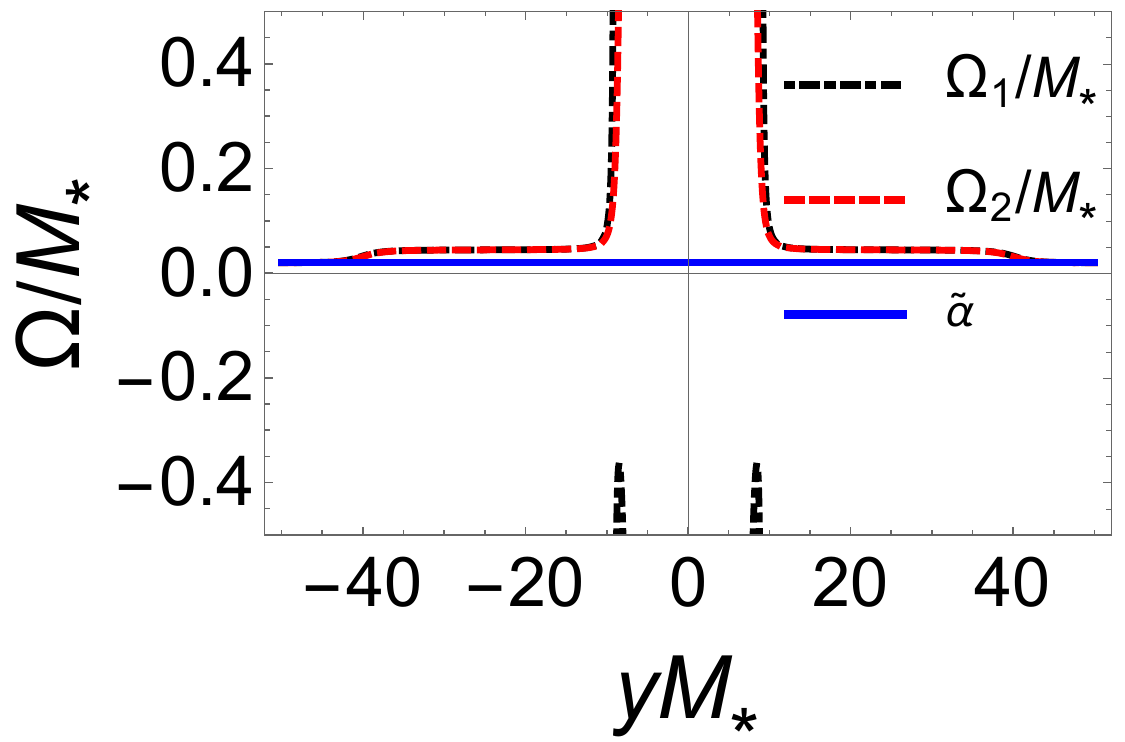}}
	\caption{Comparison between the functions $(\Omega_1, \Omega_2)$ and values of $\tilde\alpha$ for the thick brane solution of case \ref{case3}.}
	\label{ghost_c}
\end{figure}
\begin{figure}
	\centering
	\subfigure[Thick brane solution with $\tilde \alpha = -0.1$ and $n=0$.]{\includegraphics[width=4.0cm]{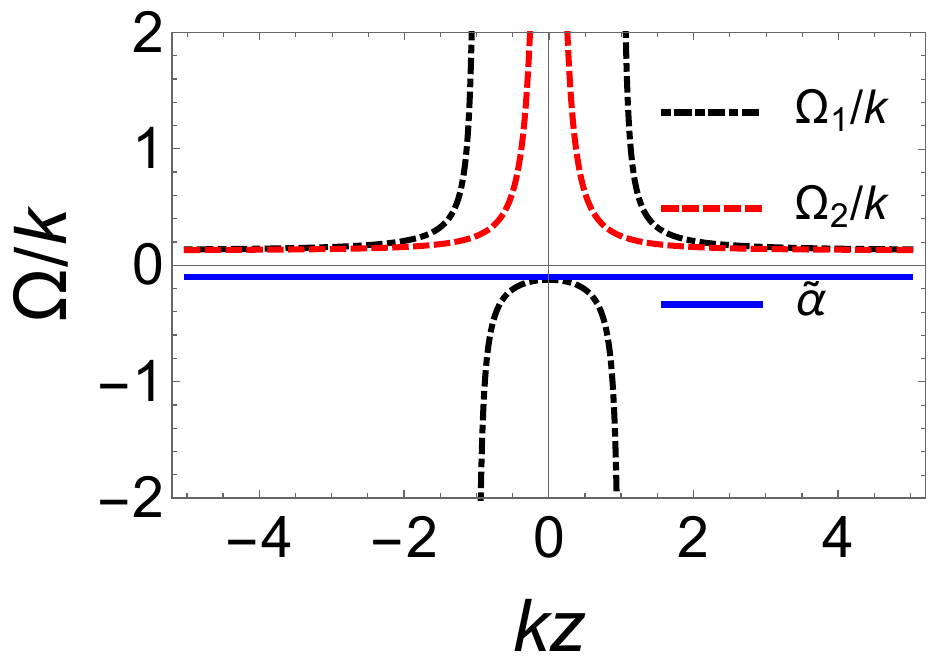}}
	\subfigure[Thick brane solution with $\tilde \alpha = -0.1$ and $n=2$.]{\includegraphics[width=4.0cm]{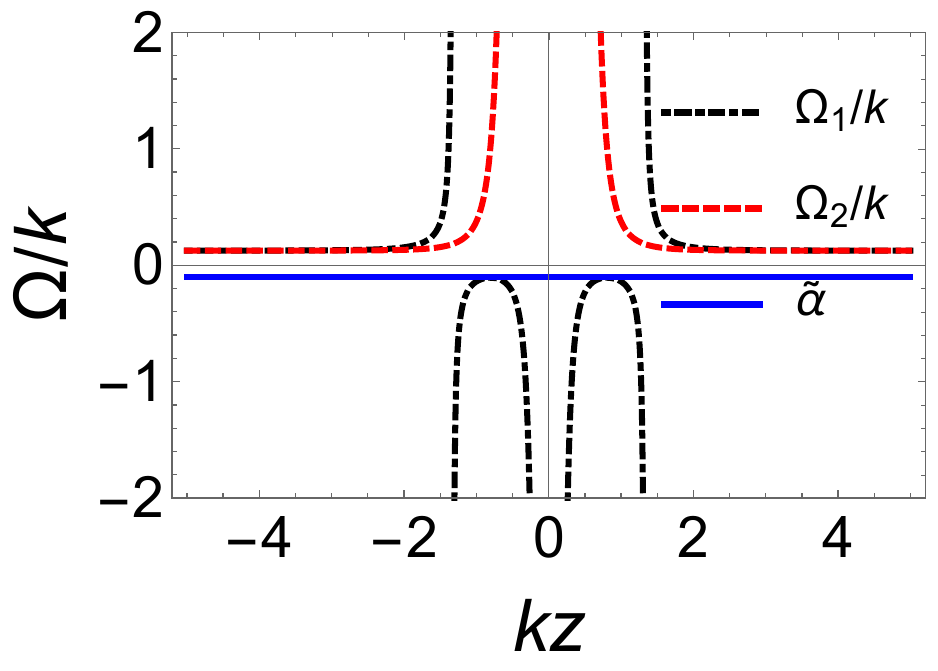}}
	\subfigure[Thick brane solution with $\tilde \alpha = -0.1$ and $n=9$.]{\includegraphics[width=4.0cm]{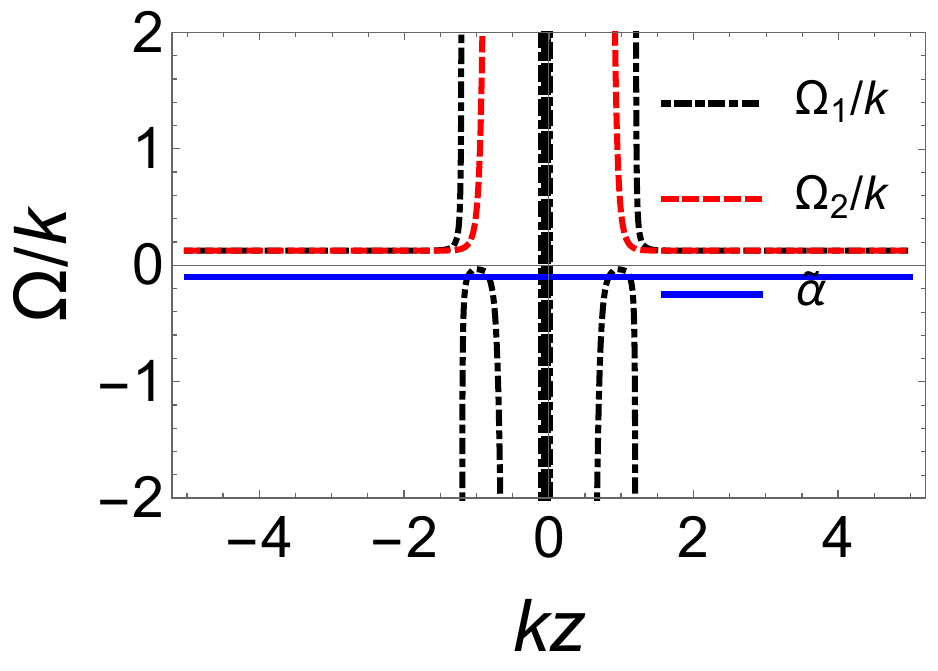}}
	\caption{Comparison between the functions $(\Omega_1, \Omega_2)$ and values of $\tilde\alpha$ for the thick brane solutions in Fig. \ref{fig_case4}.}
	\label{ghost_d}
\end{figure}
\begin{figure}
	\centering
	\subfigure[Thick brane solution with $\tilde \alpha = 0.1$ and $n=2$.\label{ghost_e_a}]{\includegraphics[width=4.0cm]{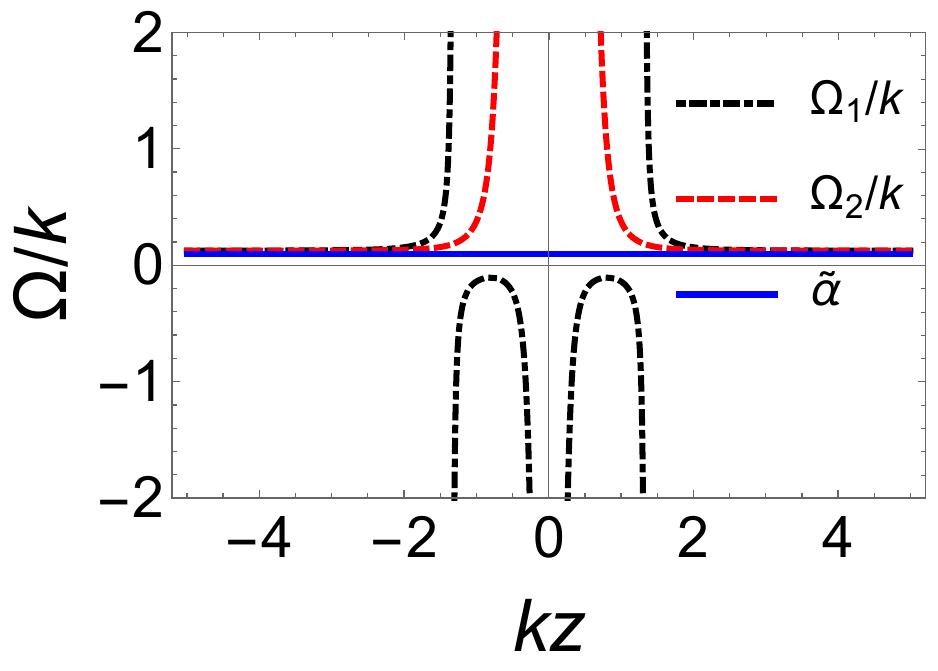}}
	\subfigure[Thick brane solution with $\tilde \alpha = 0$ and $n=2$.\label{ghost_e_b}]{\includegraphics[width=4.0cm]{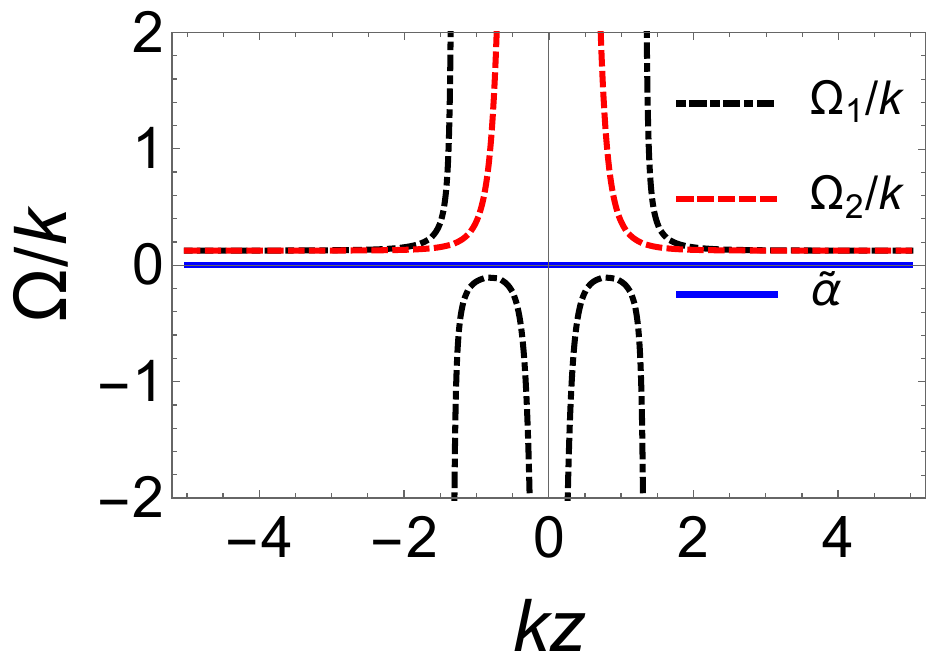}}
	\subfigure[Thick brane solution with $\tilde \alpha = -0.1$ and $n=2$.\label{ghost_e_c}]{\includegraphics[width=4.0cm]{Case4Ghost26.pdf}}
	\subfigure[Thick brane solution with $\tilde \alpha = -0.5$ and $n=2$.\label{ghost_e_d}]{\includegraphics[width=4.0cm]{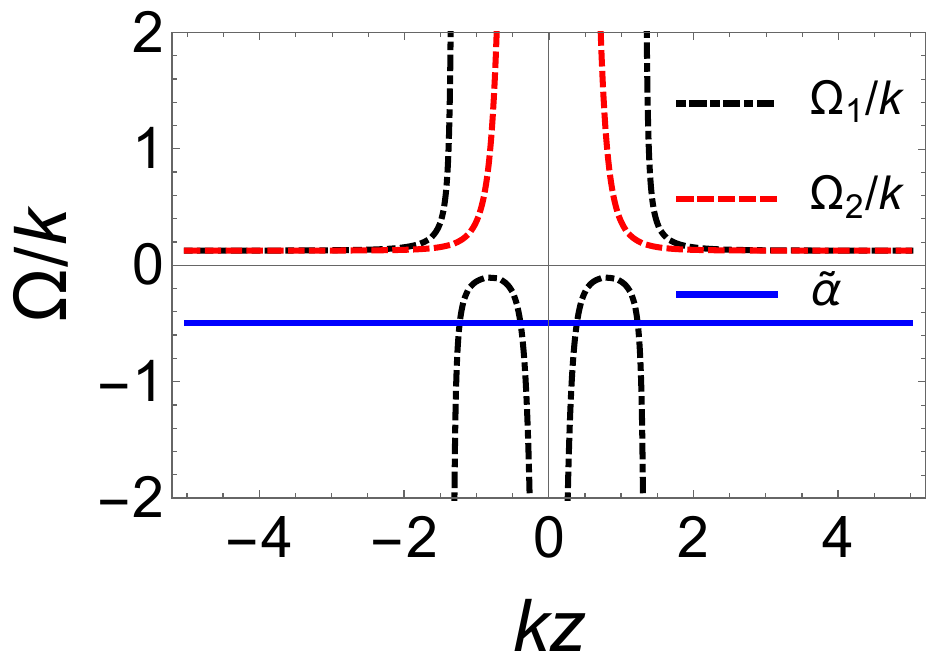}}
	\caption{Comparison between the functions $(\Omega_1, \Omega_2)$ and values of $\tilde\alpha$ for the thick brane solutions in Fig. \ref{fig_case4_2}.}
	\label{ghost_e}
\end{figure}
\begin{figure}
	\centering
	\subfigure[Thick brane solution with $\tilde \alpha = 0.1$ and $n=2$.\label{ghost_f_a}]{\includegraphics[width=4.0cm]{Case4Ghost4.pdf}}
	\subfigure[Thick brane solution with $\tilde \alpha = 0.05$ and $n=2$.\label{ghost_f_b}]{\includegraphics[width=4.0cm]{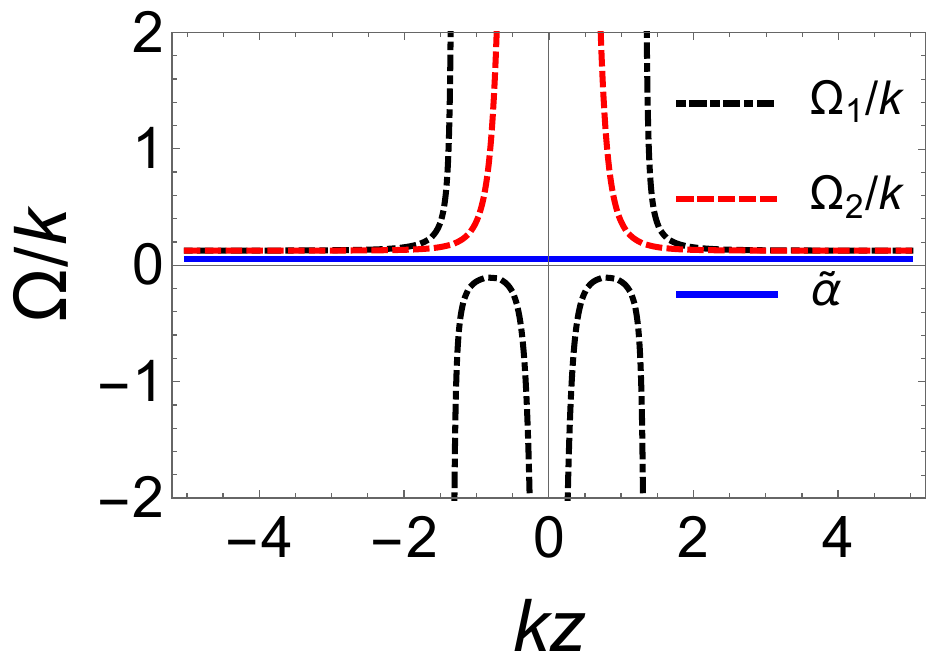}}
	\subfigure[Thick brane solution with $\tilde \alpha = 0.03$ and $n=2$.\label{ghost_f_c}]{\includegraphics[width=4.0cm]{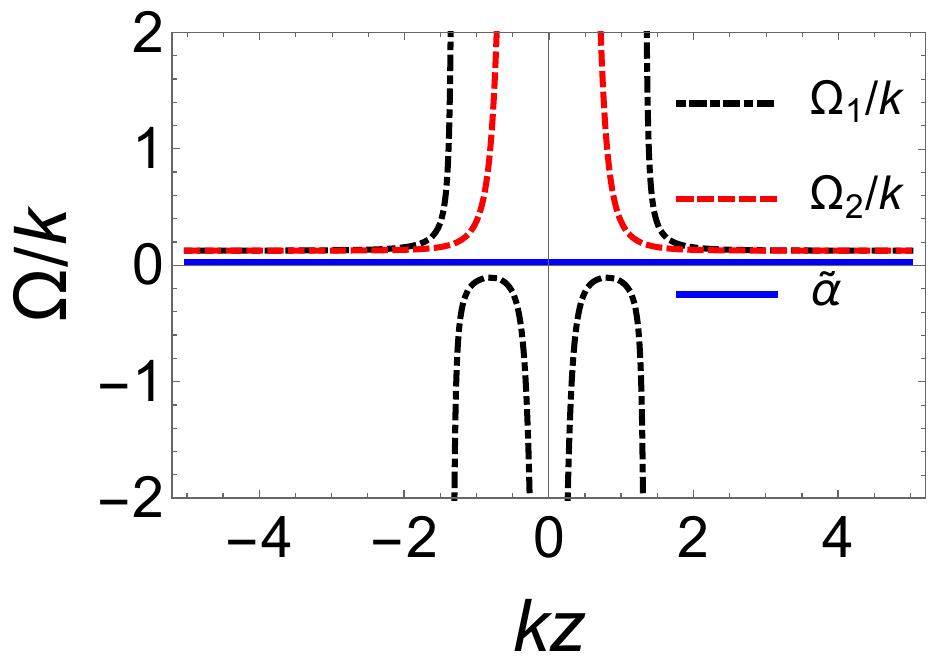}}
	\subfigure[Thick brane solution with $\tilde \alpha = 0$ and $n=2$.\label{ghost_f_d}]{\includegraphics[width=4.0cm]{Case4Ghost5.pdf}}
	\subfigure[Thick brane solution with $\tilde \alpha = -0.03$ and $n=2$.\label{ghost_f_e}]{\includegraphics[width=4.0cm]{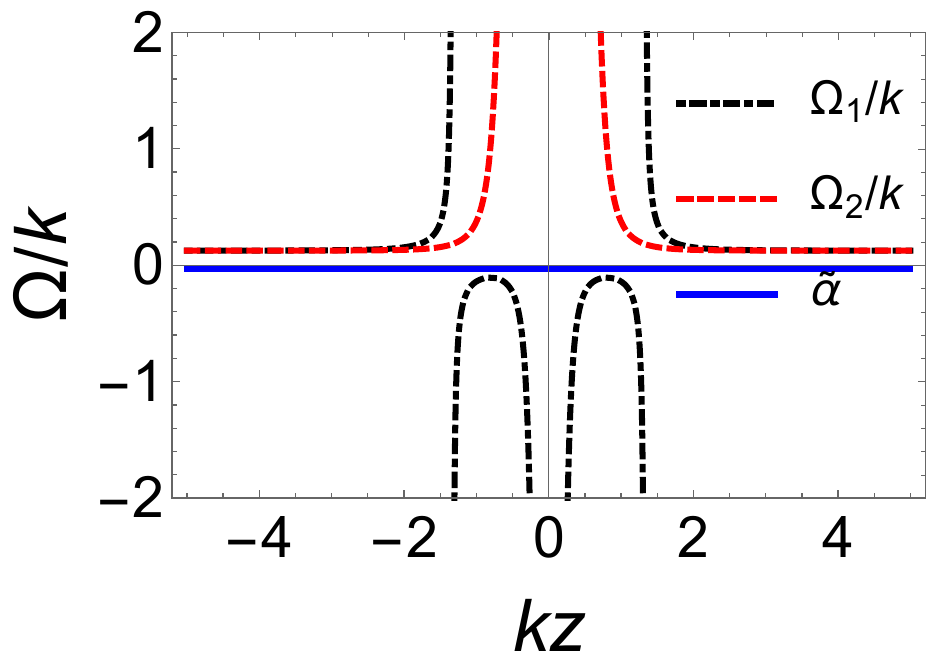}}
	\subfigure[Thick brane solution with $\tilde \alpha = -0.05$ and $n=2$.\label{ghost_f_f}]{\includegraphics[width=4.0cm]{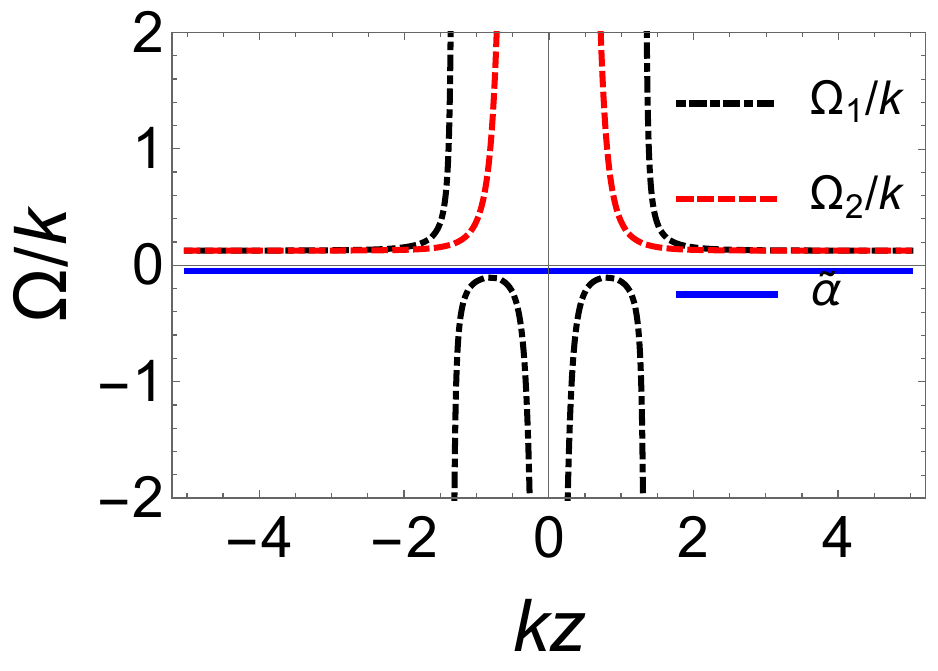}}
	\subfigure[Thick brane solution with $\tilde \alpha = -0.1$ and $n=2$.\label{ghost_f_g}]{\includegraphics[width=4.0cm]{Case4Ghost26.pdf}}
	\caption{Comparison between the functions $(\Omega_1, \Omega_2)$ and values of $\tilde\alpha$ for the thick brane solutions with an determined warp factor.}
	\label{ghost_f}
\end{figure}

\section{Scalar perturbation}
{
In this section, we investigate the properties of scalar perturbations of our model. Reference \cite{Giovannini:2001ta} has studied the full perturbations and provided the corresponding results about the localization of the tensor modes, vector modes, and scalar modes. Here, we give a brief summary about the scalar perturbations and discuss the corresponding localization and stability of our models.

The complete form of metric with the scalar perturbation is
\begin{equation}
\delta g_{M N}=e^{2A(z)}\left(\begin{array}{cc}
2 \eta_{\mu \nu} \psi+2 \partial_\mu \partial_\nu E & \partial_\mu C \\
\partial_\mu C & 2 \xi
\end{array}\right) \text {. }
\end{equation}
It can be seen that the scalar perturbations are parameterized by the four scalar functions $\psi, \xi, C, E$. However, the above scalar perturbation functions are not gauge invariant \cite{Giovannini:2001ta} under an infinitesimal transformation as follows
\begin{equation}
x^M\longrightarrow \tilde{x}^M = x^M+\epsilon^M,
\end{equation}
where $\epsilon_M=a^2(y)(\epsilon_\mu,\epsilon_z)$ and the gauge functions are $\epsilon_\mu=\partial_\mu\epsilon+\zeta_\mu$ with $\partial_\mu\zeta^\mu=0$. In Ref. \cite{Giovannini:2001ta}, the authors have constructed the gauge-invariant scalars as follows
\begin{align}
& \Psi=\psi-\partial_z A\left(\partial_z E+C\right), \\
& \Xi=\xi+\frac{1}{e^{A(z)}}\partial_z \left[e^{A(z)}\left(C+\partial_z E\right)\right],\\
& X=\chi-\partial_z \phi \left(\partial_z E+C\right).
\end{align}
The equations of motion for these gauge-invariant scalars can also be obtained, see the details in Ref. \cite{Giovannini:2001ta}.

Using the re-scaled $\Psi$ in terms of
\begin{equation}
	\Phi=\frac{e^{3 A(z) / 2} q}{\partial_z \phi} \Psi,
\end{equation}
the master equation of motion for the scalar perturbation is
\begin{equation}
	-\partial_z^2 \Phi+u\partial_z^2 \left(\frac{1}{u}\right) \Phi-\left(1-\frac{\partial_z q}{\partial_z A q}\right) \partial_\alpha \partial^\alpha \Phi=0,
	\label{final_equation}
\end{equation}
where
\begin{equation}
q=1-\frac{4 \epsilon (\partial_z A)^2}{e^{2 A(z)}}
\end{equation}
and
\begin{equation}
	u(z)=\frac{e^{3 A(z) / 2} \partial_z \phi}{\partial_z A}.
\end{equation}
It has been proved that the equation obeyed by $\Phi$ is also obeyed by the appropriately re-scaled $\Xi$ \cite{Giovannini:2001ta}.

Applying the following KK decomposition
\begin{equation}
	\Phi=\Sigma_n^{\infty}s_n(x^{\mu}) \theta_n(z)
\end{equation}
and assuming the four-dimensional part $s_n(x^{\mu})$ satisfies the four-dimensional Klein-Gordon equation as follows
\begin{equation}
\partial_{\alpha} \partial^{\alpha} s_n(x^{\mu})=m_n^2 s_n(x^{\mu}),
\label{4dkg}
\end{equation}
one can derive following equation
\begin{equation}
-\partial_z^2 \theta_n(z)+\left[u \partial_z^2 \left(\frac{1}{u}\right)\right]\theta_n(z)=m_n^2\left(1-\frac{\partial_z q }{\partial_z A q }\right)\theta_n(z).
\label{ScalarPEQ}
\end{equation}
Here, $m_n$ is the four-dimensional observed effective mass of the scalar perturbation $s_n$. It can be seen that when the coupling constant $\alpha=0$, we have $q=1$, and above equation \eqref{schordiner} reduce to GR case \cite{Kobayashi:2001jd,Giovannini:2001fh,Giovannini:2001xg}
\begin{equation}
-\partial_z^2 \theta_n(z)+u \partial_z^2 \left(\frac{1}{u}\right)\theta_n(z)=m_n^2\theta_n(z).
\label{schordiner}
\end{equation}

Note that, the localization of massless scalar perturbation has been ruled out \cite{Kobayashi:2001jd,Giovannini:2001fh,Giovannini:2001xg} and there is no tachyon instability in GR. To check that whether there are tachyon instabilities for the scalar perturbations of our thick brane solutions, we define following functions
\begin{align}
	\Theta & =1-\frac{\partial_z q}{\partial_z A q},\\
	V_s & =u \partial_{\tilde{z}}^2\left(\frac{1}{u}\right).
\end{align}
Clearly, one positive definite $\Theta$ and one positive definite $V_s$ will rule out the existence of the solution with an imaginary $m_n$ \cite{Kobayashi:2001jd}. We have checked all our solutions and give the corresponding results in Figs. \ref{sf_per_tachyon_a} and \ref{sf_per_tachyon_b}. Combining the results about ghost instabilities of tensor perturbation and tachyon instabilities of scalar perturbations, we finally give a summary about the instabilities of our thick brane solutions in Table. \ref{stability}.
\begin{figure}
	\centering
	\subfigure[The parameters are given in Table \ref{tab_case1}\label{sf_per_tachyon_a_a}]{\includegraphics[width=4cm]{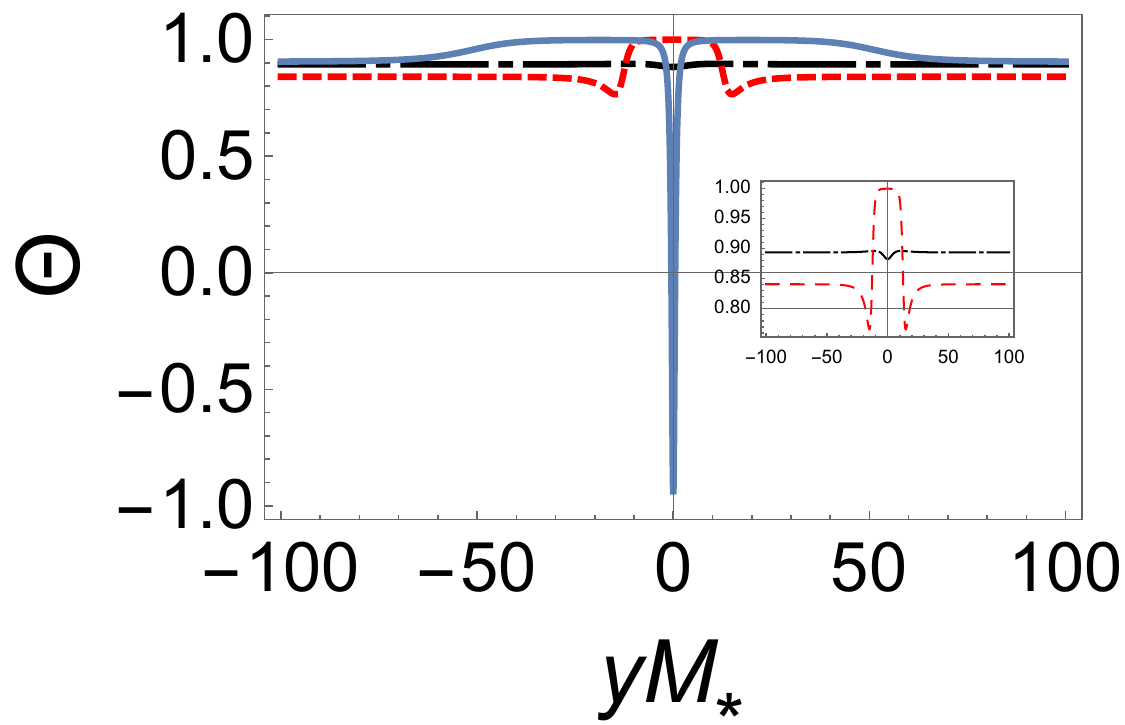}}
	\subfigure[The parameters are given in Table \ref{tab_case2}\label{sf_per_tachyon_a_b}]{\includegraphics[width=4cm]{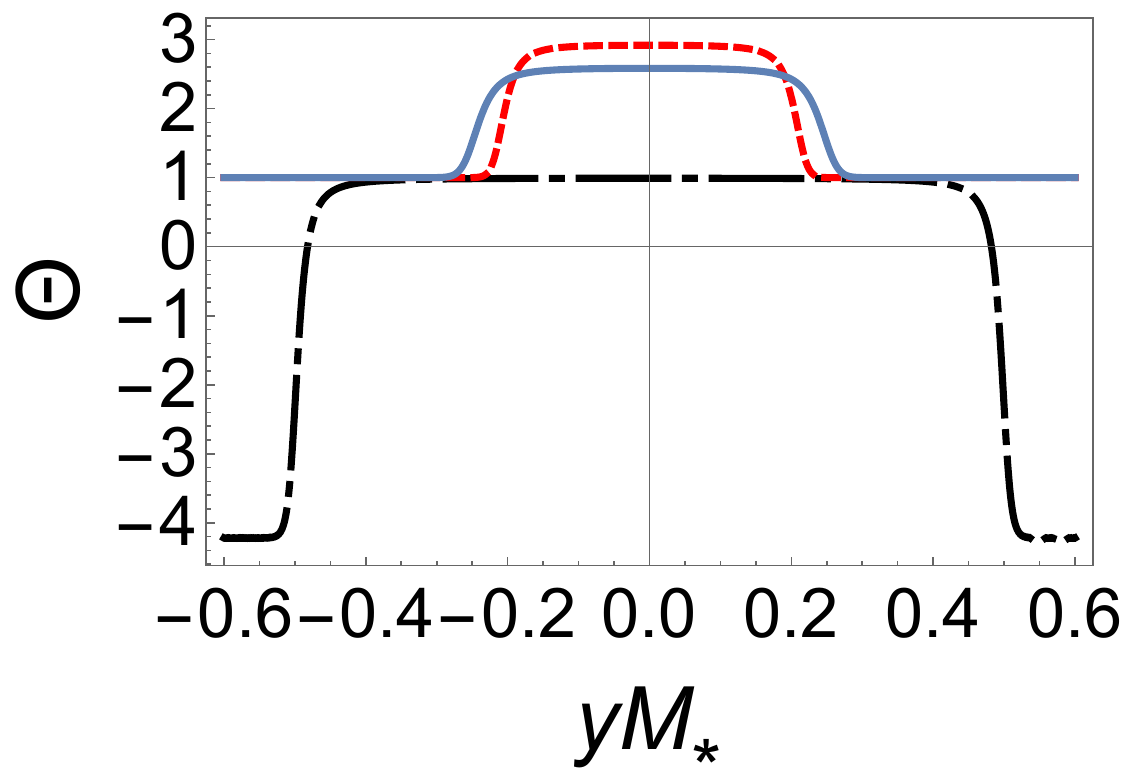}}
	\subfigure[The parameters are given in Table \ref{tab_case3}\label{sf_per_tachyon_a_c}]{\includegraphics[width=4cm]{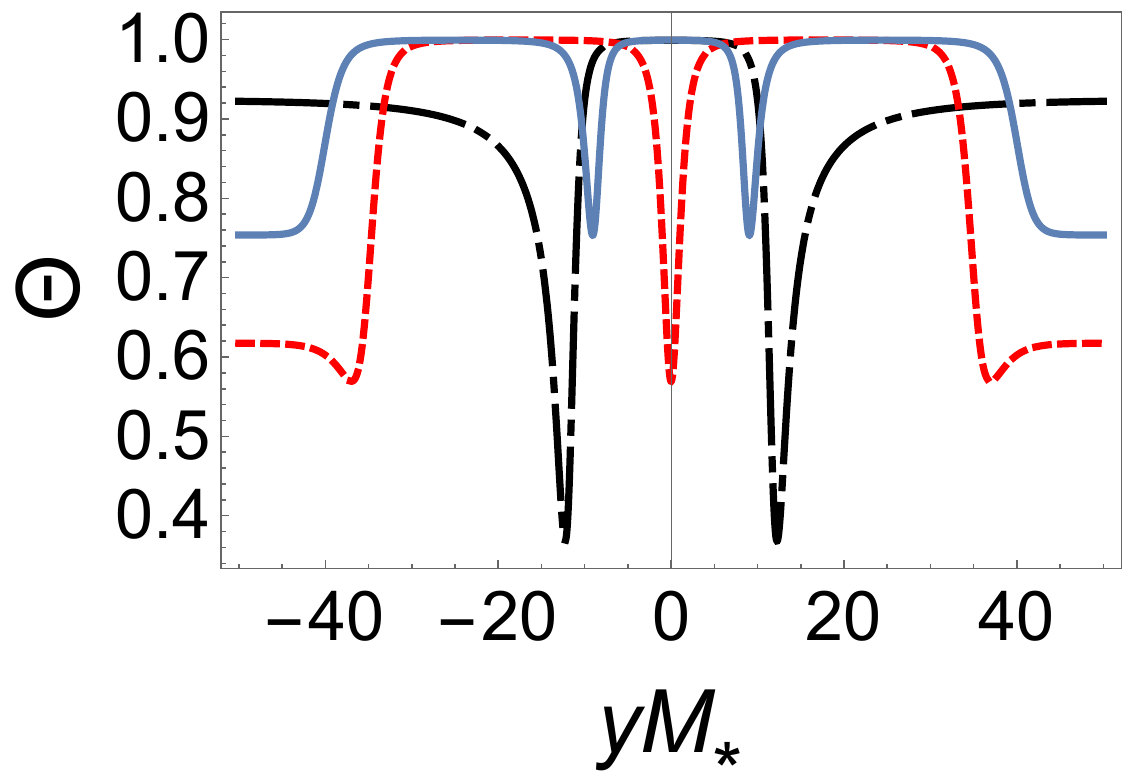}}
	\subfigure[The red dashed lines, black dot dashed lines, and dark blue lines correspond to $n=0$, $n=2$, and $n=9$, respectively. We let $\tilde{\alpha}=-0.1$.\label{sf_per_tachyon_a_d}]{\includegraphics[width=4cm]{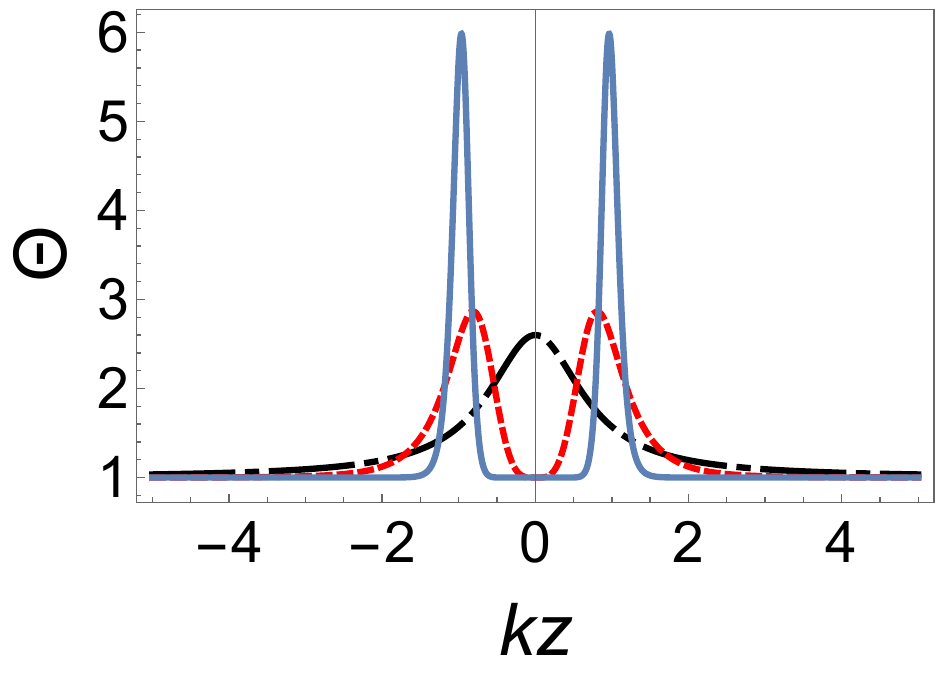}}
	\subfigure[The red dashed lines, blue solid lines, black dot dashed lines, and dark blue lines correspond to $\tilde{\alpha}=0.1$, $\tilde{\alpha}=0$, $\tilde{\alpha}=-0.1$, and $\tilde{\alpha}=-0.5$, respectively. We let $n=2$.\label{sf_per_tachyon_a_e}]{\includegraphics[width=4cm]{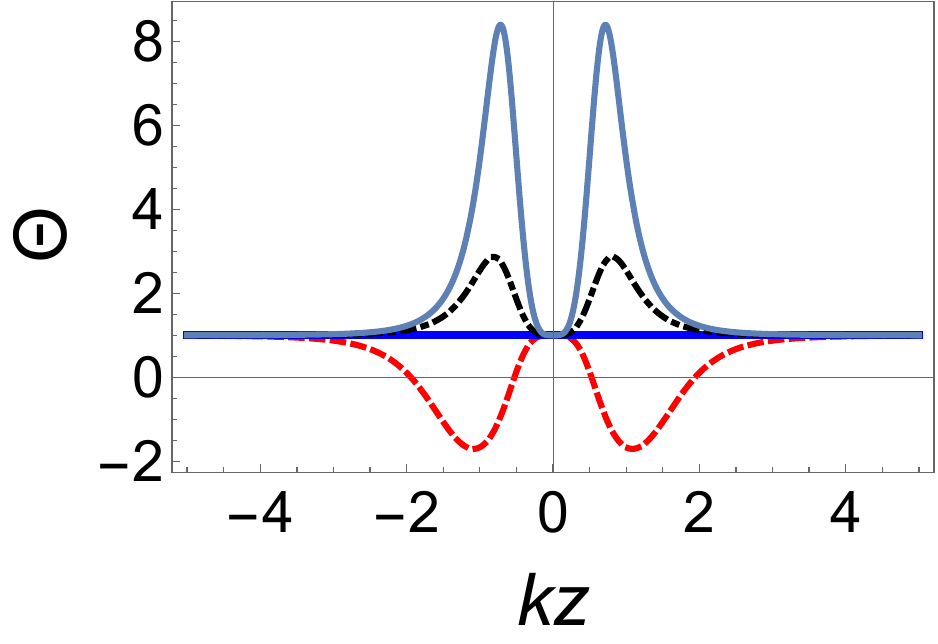}}
	\caption{Profiles of function $\Theta$ along the extra dimension for all the thick brane models.}
		\label{sf_per_tachyon_a}
\end{figure}
\begin{figure}
	\centering
	\subfigure[The parameters are given in Table \ref{tab_case1}]{\includegraphics[width=4cm]{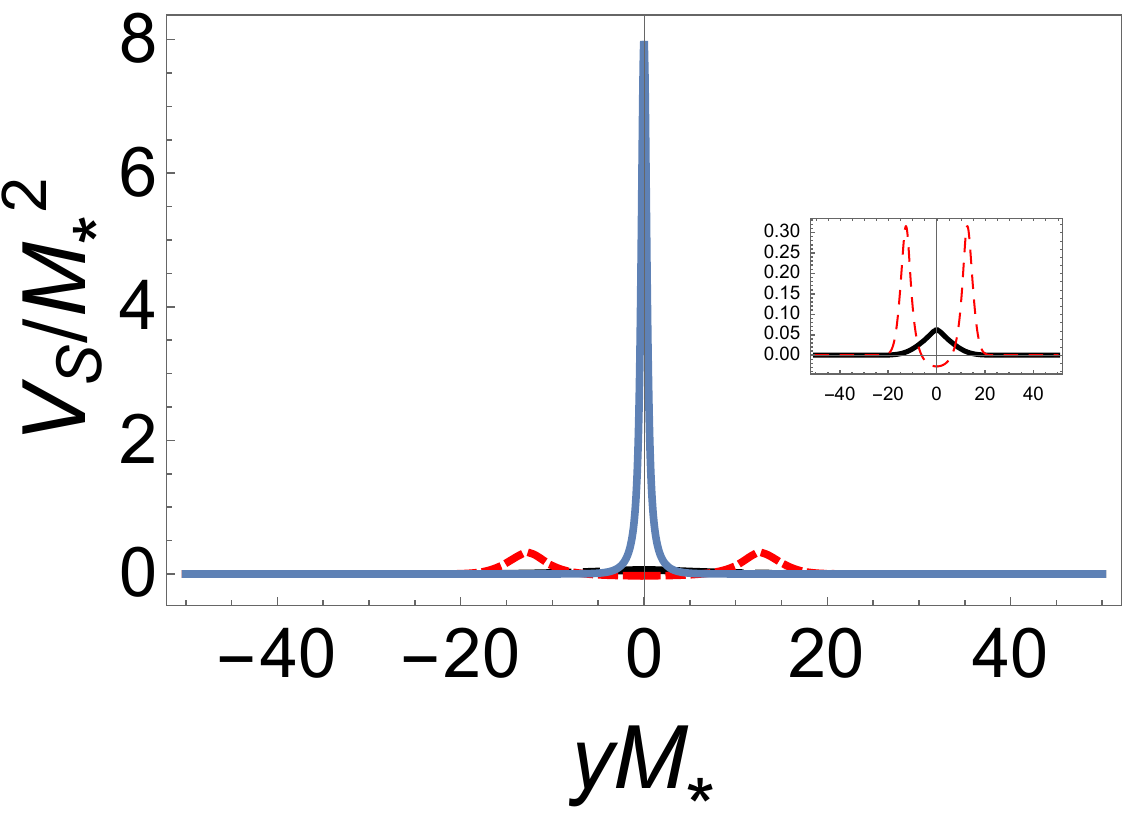}}
	\subfigure[The parameters are given in Table \ref{tab_case2}]{\includegraphics[width=4cm]{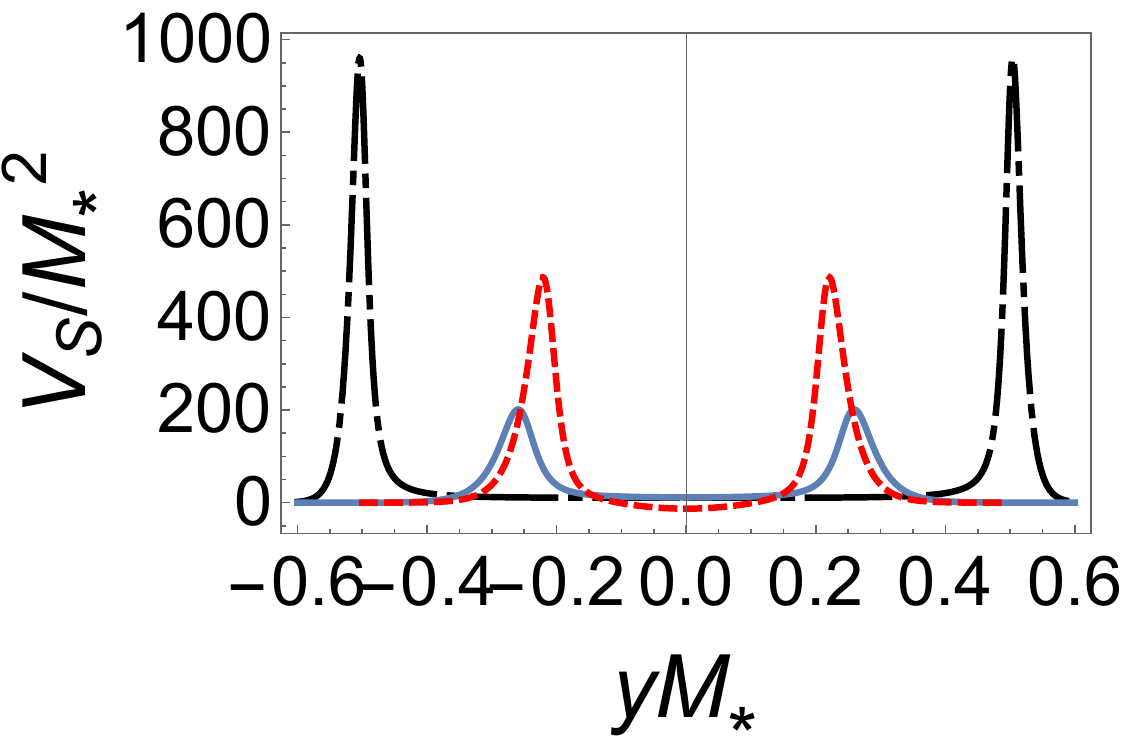}}
	\subfigure[The parameters are given in Table \ref{tab_case3}]{\includegraphics[width=4cm]{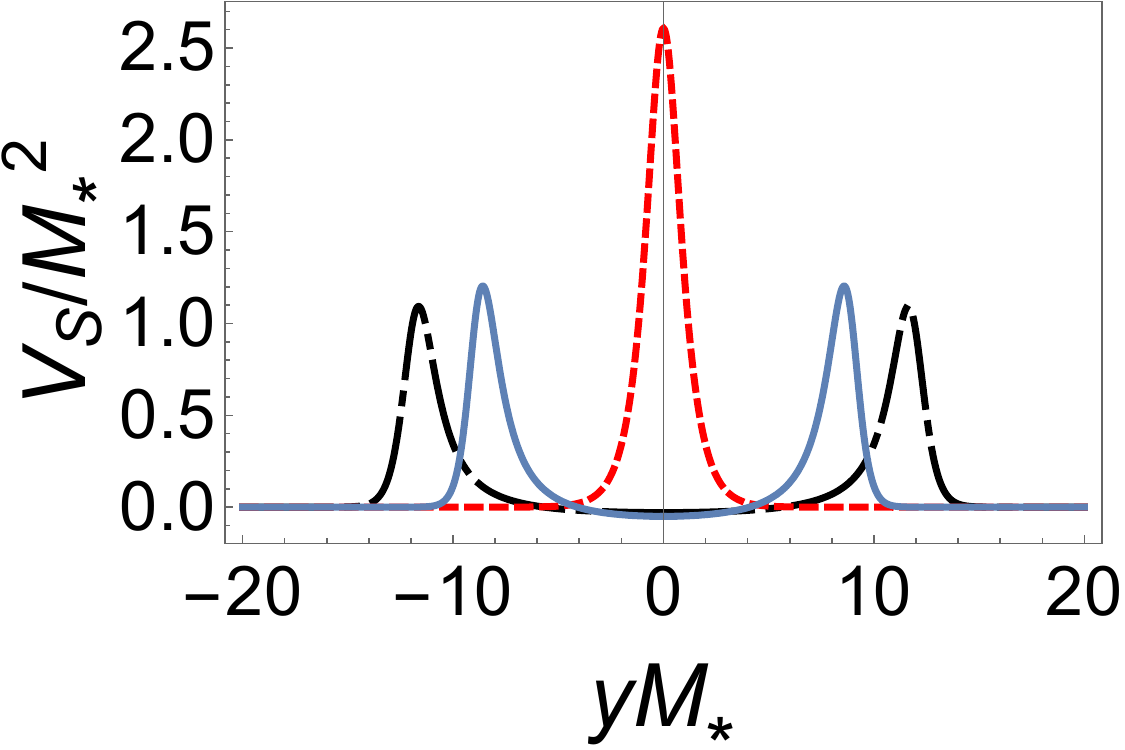}}
	\subfigure[The red dashed lines, black dot dashed lines, and dark blue lines correspond to $n=0$, $n=2$, and $n=9$, respectively. And $\tilde{\alpha}=-0.1$.]{\includegraphics[width=4cm]{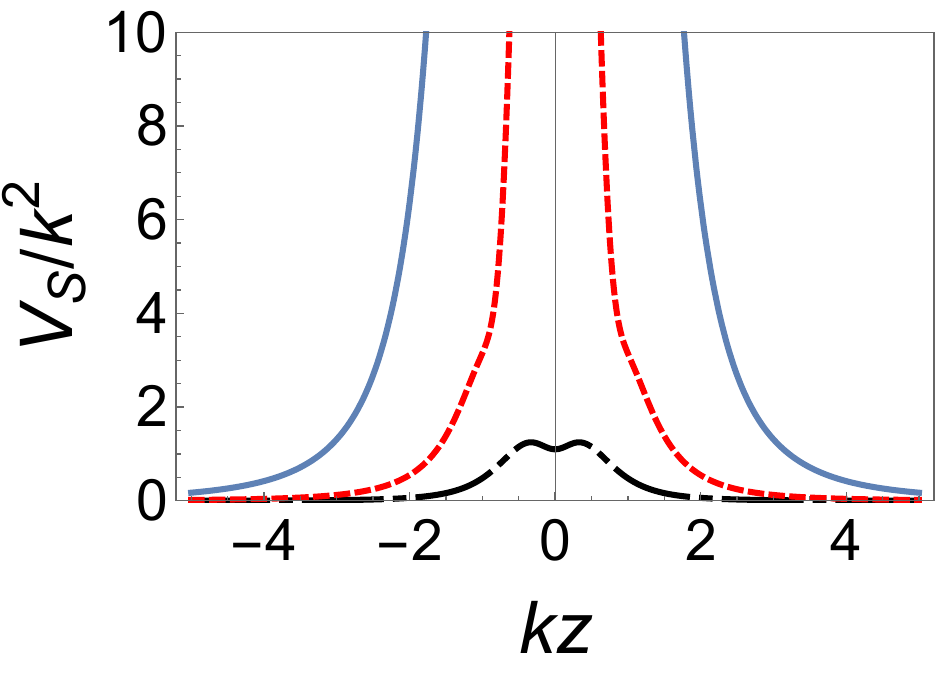}}
	\subfigure[The red dashed lines, blue solid lines, black dot dashed lines, and dark blue lines correspond to $\tilde{\alpha}=0.1$, $\tilde{\alpha}=0$, $\tilde{\alpha}=-0.1$, and $\tilde{\alpha}=-0.5$, respectively. And $n=2$]{\includegraphics[width=4cm]{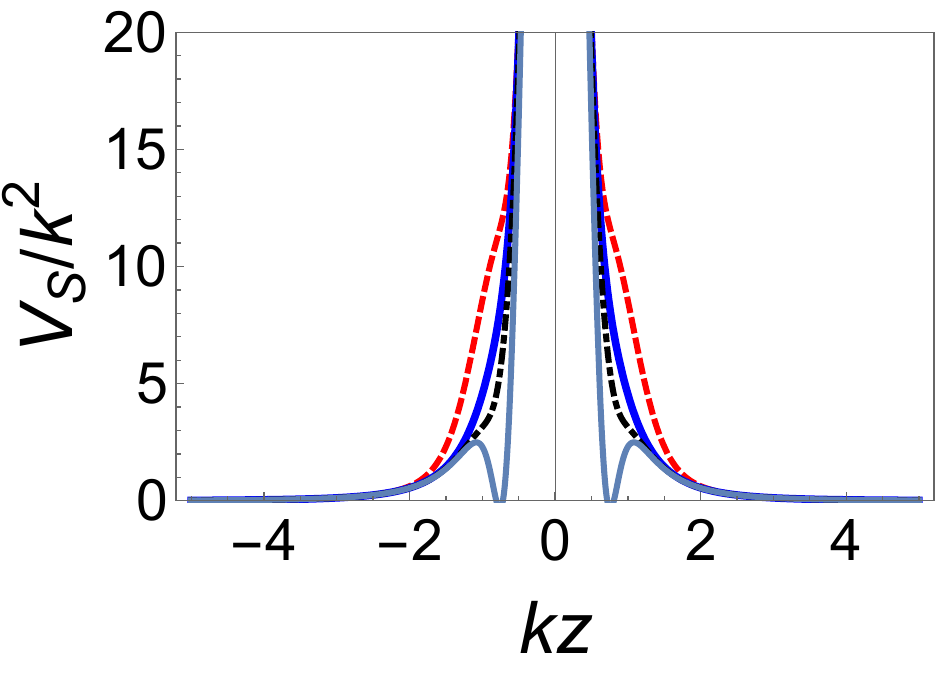}}
	\caption{Profiles of function $V_s$ along the extra dimension for all the thick brane models.}
	\label{sf_per_tachyon_b}
\end{figure}

\begin{table*}[!htb]
	\begin{center}
		\caption{Stabilities of our thick brane models.}
		\begin{tabular}{c|c|c|c|c|c}
		\hline
thick brane solution &~~~~~~~~$K_1$~~~~~~~~   &~~~~~~~~$K_2$~~~~~~~~   &~~~~~~$\Theta$~~~~~~       &~~~~~$V_s$~~~~~        &~~~~~stability~~~~~\\
		\hline
		solution\_a1 & positive definite  &positive definite   & positive definite       & positive definite     & \textbf{stable}     \\
		solution\_a2 & positive definite  &positive definite   & positive definite       & not positive definite & unstable     \\
		solution\_a3 & positive definite  &positive definite   & not positive definite   & positive definite     & unstable     \\
		\hline
		solution\_b1 & positive definite  &positive definite   & not positive definite   & positive definite     & unstable     \\
		solution\_b2 & positive definite  &positive definite   & positive definite       & not positive definite & unstable     \\
		solution\_b3 & positive definite  &positive definite   & positive definite       & positive definite     & \textbf{stable}      \\
		\hline
		solution\_c1 & positive definite  &positive definite   & positive definite       & positive definite     & \textbf{stable}     \\
		solution\_c2 & positive definite  &positive definite   & positive definite       & positive definite     & \textbf{stable}     \\
		solution\_c3 & positive definite  &positive definite   & positive definite       & not positive definite & unstable     \\
		\hline
		solution\_d1 & positive definite  &positive definite   &positive definite        & positive definite     & \textbf{stable}     \\
		solution\_d2 & positive definite  &positive definite   &positive definite        & positive definite     & \textbf{stable}     \\
		solution\_d3 & $< 0$              &positive definite   &positive definite        & positive definite     & unstable     \\
		\hline
		solution\_e1 & positive definite  &positive definite   & not positive definite   & positive definite     & unstable     \\
		solution\_e2 & positive definite  &positive definite   & positive definite       & positive definite     & \textbf{stable}     \\
		solution\_e3 & positive definite  &positive definite   & positive definite       & positive definite     & \textbf{stable}     \\
		solution\_e4 & $< 0$              &positive definite   & positive definite       & positive definite     & unstable     \\
		\hline
		\end{tabular}
		\label{stability}
	\end{center}
\end{table*}

So far, we have studied the properties of thick branes in five-dimensional GB gravity. We propose the numerical method to solve the thick brane solutions, the localization of massless KK graviton and whether the four-dimensional Newtonian potential can be recovered are still studied, the ghost instability of four-dimensional tensor perturbation and tachyon instability of scalar perturbations are also discussed. Here, we briefly summarize how the GB term affects the properties of the thick brane. For simplicity, we only consider the solutions in Table \ref{tab_case5} to discuss this.

When the GB coupling parameter $\tilde{\alpha}$ goes from the negative value to the positive value, the profile of scalar field will gradually transform from the red dashed line to the dark blue line as shown in Fig. \ref{fig_case4_2}. Our further study shows that such change will induce the ghost instability of tensor perturbation and the tachyon instability of scalar perturbation, see the details in Figs. \ref{ghost_e} and subfigure \ref{sf_per_tachyon_a_e}. We find that the thick brane solution will have the ghost stability of tensor perturbation but not have the tachyon instability of scalar perturbations with the negative $\tilde\alpha$. When the coupling parameter $\tilde\alpha$ is vanishing, the thick brane will be stable and there are no any instabilities. When the coupling parameter $\tilde\alpha$ becomes positive, both the tachyon instabilities of scalar perturbations and the ghost instability of tensor perturbation will exist.}}

\section{Conclusions} \label{sec5}

It is well known that the GB term is a topological invariant and does not affect the equations of motion in four-dimensional spacetime. However, when the spacetime dimension satisfies $D\geq 5$, the GB term is no longer a topological invariant and its influence should be considered. Thus, in this paper we investigated the property of a thick brane in the five-dimensional Einstein GB gravity and showed that how the GB term affects the property of a thick brane. We introduced two methods for solving the thick brane solutions. In the first method, all the variables including the warp factor $A$, the scalar potential $V$, and the extra dimensional coordinate $y$ are considered as functions of the scalar field $\phi$. Then we can write the brane solution formally under the help of an auxiliary superpotential $W(\phi)$, which is given by Eqs.~(\ref{ptl}), (\ref{extradim and phi}), and (\ref{wf and phi}). Note that, in this method, we have an assumption, i.e., the scalar field $\phi(y)$ is a monotonic function of the extra dimensional coordinate $y$. It was shown that this novel method works well for solving thick brane solutions with various superpotentials. In the second method, the expressions for the scalar field and scalar potential were obtained in terms of the warp factor, and one can directly obtain the thick brane solution by specifying a warp factor.

By choosing three special kinds of superpotentials, we first derived three types of thick brane solutions. We gave the constraints for the parameters in the superpotentials \eqref{supp1}, \eqref{supp2}, and \eqref{supp3} that could support the existence of a thick brane solution. We showed that one can construct multi-kink thick brane solutions with a suitable superpotential. We then obtained the thick brane solution by specifying a polynomial expanded warp factor.


{We showed that the massless four-dimensional tensor zero modes can be localized on the thick branes and the four-dimensional Newtonian potential can be recovered. The effective potentials in these brane models for the KK gravitons have the rich structures such as multi-well, which would lead to massive resonant KK gravitons~\cite{Guo:2011wr,Xu:2014jda,Chen:2020zzs,Yu:2015wma}. Furthermore, it was found that one can obtain a discontinuous effective potential and nonsmooth tensor zero mode in the smooth thick brane solution, which was not found in general relativity. This singularity is originated from the zero point of the function $F(y)$ \eqref{Fy} or the divergence of the function $H(y)$ given in Eq.~\eqref{Hy} in the effective potential (\ref{Veff(w)}). The singularity condition is given by Eq.~\eqref{singularityCondition}, from which it is clear that this singularity comes from the GB term for some parameter space and there is no singularity in general relativity. Such situation was also found in the $f(R)$ brane model, see Ref.~\cite{Xu:2014jda} for the details. The inner structure of the effective potential with singularities will also support a series of resonant KK gravitons. Moreover, we also studied the ghost instability of four-dimensional tensor perturbation and the tachyon instability of scalar perturbation. We found that these thick brane with rich inner structures might be unstable due to the existence of the ghost and tachyon. Our results indicate that deviation of $\tilde\alpha$ from zero too far in both the positive and negative directions can lead to instability of the thick brane, and only the thick brane solutions with GB coupling parameter $\tilde\alpha$ near zero could be stable.}

\section{Acknowledgements}
We thank the anonymous referee for his/her important comments for the revision of this paper. This work was supported in part by the National Key Research and Development Program of China (Grant No. 2020YFC2201503), the National Natural Science Foundation of China (Grants No. 11875151, No. 12105126, and No. 12047501), the China Postdoctoral Science Foundation (Grant No. 2021M701531), the 111 Project (Grant No. B20063), and the Fundamental Research Funds for the Central Universities (Grant No.~lzujbky-2020-it04, No. lzujbky-2021-pd08). Y.X. Liu was supported by Lanzhou City's scientific research funding subsidy to Lanzhou University.


\begin{thebibliography}{10}




\bibitem{Akama:1982jy}
K. Akama,
{\it An Early Proposal of `Brane World'},
Lect. Notes Phys. {\bfseries 176} (1982) 267.

\bibitem{Rubakov:1983kea}
V. A. Rubakov and M.E. Shaposhnikov,
{\it Do we live inside a domain wall?}
Phys. Lett. B {\bfseries 125} (1983) 136.


\bibitem{Kaluza:1921tu}
Th. Kaluza,
{\it Zum Unit\"atsproblem der Physik},
{Int. J. Mod. Phys.} D {\bfseries 27} (2018) 1870001.

\bibitem{Klein:1991}
O. Klein,
{{Quantum theory and 5-dimensional relativity theory}},
{{Z. Phys.} {\bf 37} (1926) 895.}

\bibitem{Arkani:1998}
N. Arkani-Hamed, S. Dimopoulos, and G.R. Dvali,
{{The hierarchy problem and new dimensions at a millimeter}},
{{Phys. Lett.} B {\bfseries 429} (1998) 263, }					

\bibitem{Randall:1999}
L. Randall and R. Sundrum,
{{A Large mass hierarchy from a small extra dimension}},
{{Phys. Rev. Lett. } {\bfseries 83} (1999) 3370,}.

\bibitem{Randall2:1999}
L. Randall and R. Sundrum,
{{An alternative to compactification}},
{{Phys. Rev. Lett.} {\bfseries 83} (1999) 4690,}

\bibitem{Hsin:2010}
H.-C. Cheng,
{{{Theoretical Advanced Study Institute in Elementary Particle Physics}: {Physics of the Large and the Small}}},
[{{\tt arXiv:1003.1162[hep-ph]}}].

\bibitem{Rizzo:2010}
T.G. Rizzo,
{{Introduction to Extra Dimensions}},
{{AIP Conf. Proc. }{\bfseries 1256} (2010)  27,}
[{{\tt arXiv:1003.1698[hep-ph]}}].

\bibitem{Liu:2017gcn}
Y.-X. Liu,
{{Introduction to Extra Dimensions and Thick Braneworlds}},
{{Memorial Volume for Yi-Shi Duan, pp. 211-275 (2018)}},
[{{\tt arXiv:1707.08541[hep-ph]}}].

\bibitem{Maartens:2010}
R. Maartens and K. Koyama,
{{Brane-World Gravity}},
\href{http://dx.doi.org/10.12942/lrr-2010-5}{{Living Rev. Rel.}
	{\bfseries 13} (2010)   5},
[\href{https://arxiv.org/abs/1004.3962}{{\tt arXiv:1004.3962[hep-th]}}].

\bibitem{Merab1998}
M. Gogberashvili,
{{Hierarchy problem in the shell-Universe model}},
\href {https://www.worldscientific.com/doi/abs/10.1142/S0218271802002992}
{{Int. J. Mod. Phys.} D {\bfseries 11} (2002) 1635,}
[\href{https://arxiv.org/abs/hep-ph/9812296}{{\tt arXiv:hep-ph/9812296}}].

\bibitem{Israel1966}
W. Israel,
{{Singular hypersurfaces and thin shells in general relativity}},
\href{https://doi.org/10.1007/BF02710419}
{{Nuovo Cim.} B {\bfseries44} (1966) 1.}

\bibitem{DeWolfe:1999cp}
O. DeWolfe, D.Z. Freedman, S.S. Gubser, and A. Karch,
{{Modeling the fifth-dimension with scalars and gravity}},
\href {https://doi.org/	10.1103/PhysRevD.62.046008}
{{Phys. Rev.} D {\bfseries62} (2000) 046008,}
[\href{https://arxiv.org/abs/hep-th/9909134}{{\tt arXiv:hep-th/9909134}}].

\bibitem{Bronnikov:2003gg}
K.A. Bronnikov and B.E. Meierovich,
{{A General thick brane supported by a scalar field}},
{{Grav. Cosmol.} {\bfseries 9} (2003) 313,}
[\href{https://arxiv.org/abs/gr-qc/0402030}{{\tt arXiv:gr-qc/0402030}}].

\bibitem{Bazeia:2008zx}
D. Bazeia, A.R. Gomes, L. Losano, and  R. Menezes,
{{Braneworld Models of Scalar Fields with Generalized Dynamics}},
\href {https://doi.org/10.1016/j.physletb.2008.12.039}
{{Phys. Lett.} B {\bfseries 671} (2009) 402,}
[\href{https://arxiv.org/abs/0808.1815}{{\tt arXiv:0808.1815[hep-th]}}].

\bibitem{Toharia:2010ex}
M. Toharia, M. Trodden, and E.J. West,
{{Scalar Kinks in Warped Extra Dimensions}},
\href {https://doi.org/10.1103/PhysRevD.82.025009}
{{Phys. Lett.} D {\bfseries 82} (2010) 025009,}
[\href{https://arxiv.org/abs/1002.0011}{{\tt arXiv:1002.0011[hep-ph]}}].

\bibitem{Bazeia:2003aw}
D. Bazeia, C. Furtado, and A.R. Gomes,
{{Brane structure from scalar field in warped space-time}},
\href {https://doi.org/10.1088/1475-7516/2004/02/002}
{{JCAP} \textbf{02} (2004) 002,}
[\href{https://arxiv.org/abs/hep-th/0308034}{{\tt arXiv:hep-th/0308034}}].

\bibitem{Novikov:2015iph}
Oleg O.	Novikov, A.A. Andrianov, and V.A. Andrianov,
{{Thick branes from self-gravitating scalar fields}},
\href {https://doi.org/10.1063/1.4891147}
{{AIP Conf. Proc. } \textbf{1606} (2015) 313.}

\bibitem{Bazeia:2015owa}
D. Bazeia, A.S. Lob\~ao Jr,  and R. Menezes,
{{Thick brane models in generalized theories of gravity}},
\href {https://doi.org/10.1016/j.physletb.2015.02.037}
{{Phys. Lett.} B \textbf{743} (2015) 98,}
[\href{https://arxiv.org/abs/1502.04757}{{\tt arXiv:hep-th/1502.04757}}].				

\bibitem{Xie:2021ayr}
Q.-Y.~Xie, Q.-M.~Fu, T.-T.~Sui, L.~Zhao, and Y.~Zhong,
{{First-order formalism and thick branes in mimetic gravity}},
\href {https://doi.org/10.3390/sym13081345}
{{Symmetry} \textbf{13~ (13)} (2021) 1345,}
[\href{https://arxiv.org/pdf/2102.10251}{{\tt arXiv:gr-qc/2102.10251}}].	

\bibitem{Wan:2020smy}
J.-J. Wan, Z.-Q. Cui, W.-B. Feng, and Y.-X. Liu,
{{Smooth braneworld in $6$-dimensional asymptotically AdS spacetime}},
\href{http://dx.doi.org/10.1007/JHEP05(2021)017}
{{JHEP} {\bf 05} (2021) 017,}
[\href{http://arxiv.org/abs/hep-th/2010.05016}{{\tt arXiv:hep-th/2010.05016}}].

\bibitem{Nozari:2009zr}
K. Nozari and S.D.~Sadatian,
{{Bouncing Universe with a Nonminimally Coupled Scalar Field on a Moving Domain Wall}},
\href{https://doi.org/10.1016/j.physletb.2009.04.064}
{{Phys. Lett.} B \textbf{676} (2009) 1,}
[\href{https://arxiv.org/abs/0904.4029}{{\tt arXiv:0904.4029[gr-qc]}}].

\bibitem{Liu:2012gv}
Y.-X.~Liu, F.-W.~Chen, H. Guo, and X.-N.~Zhou,
{{Non-minimal Coupling Branes}},
\href{https://doi.org/10.1007/JHEP05(2012)108}
{{JHEP} \textbf{05} (2012) 108,}
[\href{https://arxiv.org/abs/1205.0210}{{\tt arXiv:1205.0210 [hep-th]}}].

\bibitem{Guo:2011wr}
H.~Guo, Y.-X.~Liu, Z.-H.~Zhao, and F.-W.~Chen,
{{Thick branes with a non-minimally coupled bulk-scalar field}},
\href{https://doi.org/10.1103/PhysRevD.85.124033}
{{Phys. Rev.} D \textbf{85} (2012) 124033,}
[\href{https://arxiv.org/abs/1106.5216}{{\tt arXiv:1106.5216 [hep-th]}}].


\bibitem{Xu:2014jda}
Z.-G. Xu, Y. Zhong, H. Yu, and Y.-X. Liu,
{{The structure of $f(R)$-brane model}},
\href{https://doi.org/10.1140/epjc/s10052-015-3597-0}
{{Eur. Phys. J.} C. \textbf{75} (2015) 368,}
[\href{https://arxiv.org/abs/1405.6277}{{\tt arXiv:1405.6277[hep-th]}}].

\bibitem{Zhong:2015pta}
Y. Zhong and Y.-X. Liu,
{{Pure geometric thick $f(R)$-branes: stability and localization of gravity}},
\href{https://doi.org/10.1140/epjc/s10052-016-4163-0}
{{Eur. Phys. J.} C. \textbf{76} (2016) 321,}
[\href{https://arxiv.org/abs/1507.00630}{{\tt arXiv:1507.00630[hep-th]}}].

\bibitem{Gu:2016nyo}
B.-M. Gu, Y.-P. Zhang, H. Yu, and Y.-X. Liu,
{{Full linear perturbations and localization of gravity on $f(R,T)$ brane}},
\href{https://doi.org/10.1140/epjc/s10052-017-4666-3}
{{Eur. Phys. J.} C. \textbf{77} (2017) 115,}
[\href{https://arxiv.org/abs/1606.07169}{{\tt arXiv:1606.07169[hep-th]}}].

\bibitem{Zhou:2017xaq}
X.-N. Zhou, Y.-Z. Du, H. Yu, and Y.-X. Liu,
{{Localization of Gravitino Field on $f(R)$ Thick Branes}},
\href{https://doi.org/10.1007/s11433-018-9246-2}
{{Sci. China Phys. Mech. Astron.} \textbf{61} (2018) 110411,}
[\href{https://arxiv.org/abs/1703.10805}{{\tt arXiv:1703.10805[hep-th]}}].

\bibitem{Zhong:2017ffr}
Y. Zhong, K. Yang, Ke, and Y.-X. Liu,
{{Linearization of a warped $f(R)$ theory in the higher-order frame II: the~Equation of motion approach}},
\href{https://doi.org/10.1103/PhysRevD.97.044032}
{{Phys. Rev.} D \textbf{97} (2018) 044032,}
[\href{https://arxiv.org/abs/1708.03737}{{\tt arXiv:1708.03737 [gr-qc]}}].

\bibitem{Gu:2018lub}
B.-M. Gu, Y.-X. Liu, and Y. Zhong,
{{Stable Palatini $f(\mathcal{R})$ braneworld}},
\href{http://dx.doi.org/10.1103/PhysRevD.98.024027}
{{Phys. Rev.} D \textbf{98} (2018) 024027,}
[\href{https://arxiv.org/pdf/1804.00271}{{\tt	arXiv:1804.00271[hep-th]}}].

\bibitem{Cui:2020fiz}
Z.-Q. Cui, Z.-C. Lin, J.-J. Wan, Y.-X. Liu, and L. Zhao,
{{Tensor Perturbations and Thick Branes in Higher-dimensional $f(R)$ Gravity}},
\href{http://dx.doi.org/10.1007/JHEP12(2020)130}
{{JHEP} \textbf{12} (2020) 130,}
[\href{https://arxiv.org/pdf/2009.00512}{{\tt arXiv:2009.00512[hep-th]}}].

\bibitem{Chen:2020zzs}
J. Chen, W.-D. Guo, and Y.-X. Liu,
{{Thick branes with inner structure in mimetic f(R) gravity}},
\href{http://dx.doi.org/10.1140/epjc/s10052-021-09504-y}
{{Eur. Phys. J.} C \textbf{81} (2021) 709,}
[\href{https://arxiv.org/pdf/2011.03927}{{\tt	arXiv:2011.03927[gr-qc]}}].

\bibitem{Yu:2015wma}
H. Yu, Y. Zhong, B.-M. Gu, and Y.-X. Liu,
{{Gravitational resonances on $f(R)$-brane}},
\href{http://dx.doi.org/10.1140/epjc/s10052-016-4039-3}
{{Eur. Phys. J.} C \textbf{76} (2016) 195,}
[\href{https://arxiv.org/pdf/1506.06458}{{\tt	arXiv:1506.06458[gr-qc]}}].

\bibitem{Rosa:2022fhl}
J.L. Rosa, A.S. Lob\~ao, D. Bazeia,
{{Impact of compactlike and asymmetric configurations of thick branes on the scalar\textendash{}tensor representation of $f\left( R,T\right) $ gravity}},
\href{http://dx.doi.org/10.1140/epjc/s10052-022-10159-6}
{{Eur. Phys. J. } C \textbf{82~(3)} (2022) 191,}
[\href{https://arxiv.org/pdf/2202.10713}{{\tt	arXiv:2202.10713[gr-qc]}}].

\bibitem{Rosa:2021tei}
J.L. Rosa, M.A. Marques, D. Bazeia, S.N. Lobo,
{{Thick branes in the scalar\textendash{}tensor representation of f(R,~T) gravity}},
\href{http://dx.doi.org/10.1140/epjc/s10052-021-09783-5}
{{Eur. Phys. J. } C \textbf{81~(11)} (2021) 981,}
[\href{https://arxiv.org/pdf/2105.06101}{{\tt	arXiv:2105.06101[gr-qc]}}].

\bibitem{Wang:2019igp}
L.-L. Wang, H. Guo, C.-E. Fu, Q.-Y. Xie,
{{Gravity and Matters on a pure geometric thick polynomial $f(R)$ brane}},
\href{http://dx.doi.org/10.1140/epjc/s10052-022-10159-6}
{{Eur. Phys. J. } C \textbf{82~(3)} (2019) 191,}
[\href{https://arxiv.org/pdf/hep-th/1912.01396}{{\tt arXiv:1912.01396[hep-th]}}].

\bibitem{Dzhunushaliev:2019wvv}
V. Dzhunushaliev, V. Folomeev, G. Nurtayeva, S.D. Odintsov,
{{Thick branes in higher-dimensional $f(R)$ gravity}},
\href{http://dx.doi.org/10.1142/S021988782050036X}
{{Int. J. Geom. Meth. Mod. Phys. }  \textbf{17~(3)} (2020) 2050036,}
[\href{https://arxiv.org/pdf/gr-qc/1908.01312}{{\tt arXiv:1908.01312[gr-qc]}}].

\bibitem{Nozari:2019shm}
K. Nozari, N. Sadeghnezhad,
{{Braneworld mimetic $f(R)$ gravity}},
\href{http://dx.doi.org/10.1142/S0219887819500427}
{{Int. J. Geom. Meth. Mod. Phys. }  \textbf{16~(3)} (2019) 1950042,}
[\href{https://arxiv.org/pdf/gr-qc/1908.01312}{{\tt arXiv:1908.01312[gr-qc]}}].

\bibitem{Banerjee:2017lxi}
N. Banerjee, T. Paul,
{{Inflationary scenario from higher curvature warped spacetime}},
\href{http://dx.doi.org/10.1140/epjc/s10052-017-5256-0}
{{Eur. Phys. J.} C \textbf{77~(10)} (2017) 672,}
[\href{https://arxiv.org/pdf/1706.05964}{{\tt	arXiv:1706.05964[hep-th]}}].

\bibitem{Elizalde:2018rmz}
E. Elizalde, S.D. Odintsov, T. Paul, D. S\'aez-Chill\'on G\'omez,
{{Inflationary universe in $F(R)$ gravity with antisymmetric tensor fields and their suppression during its evolution}},
\href{http://dx.doi.org/10.1103/PhysRevD.99.063506}
{{Phys. Rev.} D \textbf{99 (6)} (2019) 063506,}
[\href{https://arxiv.org/pdf/1706.05964}{{\tt	arXiv:1811.02960[gr-qc]}}].

\bibitem{Banerjee:2020uil}
I. Banerjee, T. Paul, S. SenGupta,
{{Bouncing cosmology in a curved braneworld}},
\href{http://dx.doi.org/10.1088/1475-7516/2021/02/041}
{{JCAP}  \textbf{02} (2021) 041,}
[\href{https://arxiv.org/pdf/2011.11886}{{\tt	2011.11886[gr-qc]}}].

\bibitem{Lovelock:1971yv}
D. Lovelock,
{{The Einstein tensor and its generalizations}},
\href{https://doi.org/10.1063/1.1665613}
{{J. Math. Phys.} \textbf{12} (1971) 498.}


\bibitem{Wheeler:1985nh}
J.T.~Wheeler,
{{Symmetric Solutions to the Gauss-Bonnet Extended Einstein~Equations}},
\href{https://doi.org/10.1016/0550-3213(86)90268-3}
{{Nucl. Phys.} B \textbf{268} (1986) 737.}

\bibitem{Boulware:1985wk}
D.G.~Boulware and S.~Deser,
{{String Generated Gravity Models}},
\href{https://doi.org/10.1103/PhysRevLett.55.2656}
{{Phys. Rev. Lett.} \textbf{55} (1985) 2656}.

\bibitem{Zwiebach:1985kea}
B. Zwiebach,
{{Curvature squared terms and string theories}},
\href{https://doi.org/10.1016/0370-2693(85)91616-8}
{{Phys. Lett.} B {\bfseries 156} (1985) 315.}

\bibitem{Gross:1986mw}
D.J.~Gross and J.H. Sloan,
{{The Quartic Effective Action for the Heterotic String}},
\href{https://doi.org/10.1016/0550-3213(87)90465-2}
{{Nucl. Phys.} B {\bf 291} (1987) 41}.


\bibitem{Odintsov:2020sqy}
S.D. Odintsov, V.K. Oikonomou, and F.P. Fronimos,
{{Rectifying Einstein-Gauss-Bonnet Inflation in View of GW170817}},
\href{http://dx.doi.org/10.1016/j.nuclphysb.2020.115135}
{{Nucl. Phys. }B {\bf 958} (2020) 115135,}
[\href{http://arxiv.org/abs/2003.13724}{{\tt arXiv:2003.13724 [gr-qc]}}].

\bibitem{Oikonomou:2021kql}
V.K. Oikonomou,
{{A refined Einstein\textendash{}Gauss\textendash{}Bonnet inflationary theoretical framework}},
\href{https://iopscience.iop.org/article/10.1088/1361-6382/ac2168}
{{Class. Quant. Grav.} {\bf 38 (19)} (2021) 195025,}
[\href{http://arxiv.org/abs/2108.10460}{{\tt arXiv:2108.10460[gr-qc]}}].

\bibitem{Odintsov:2020xji}
S.D. Odintsov,  V.K. Oikonomou, and  F.P. Fronimos,
{{Non-minimally coupled Einstein\textendash{}Gauss\textendash{}Bonnet inflation phenomenology in view of GW170817}},
\href{http://dx.doi.org/10.1016/j.aop.2020.168250}
{{Annals Phys.} {\bf 420} (2020) 168250,}
[\href{http://arxiv.org/abs/2007.02309}{{\tt arXiv:2007.02309[gr-qc]}}].

\bibitem{Oikonomou:2020oil}
V.K. Oikonomou and  F.P. Fronimos,
{{A Nearly Massless Graviton in Einstein-Gauss-Bonnet Inflation with Linear Coupling Implies Constant-roll for the Scalar Field}},
\href{http://dx.doi.org/10.1209/0295-5075/131/30001}
{{EPL} {\bf 131 (3)} (2020) 30001,}
[\href{http://arxiv.org/abs/2007.11915}{{\tt arXiv:2007.11915[gr-qc]}}].

\bibitem{Elizalde:2020zcb}
E. Elizalde, S.D. Odintsov, V.K. Oikonomou, T. Paul,
{{Extended matter bounce scenario in ghost free $f(R,\mathcal{G})$ gravity compatible with GW170817}},
\href{http://dx.doi.org/10.1016/j.nuclphysb.2020.114984}
{{Nucl. Phys.} B {\bf 954} (2020) 114984,}
[\href{http://arxiv.org/abs/2003.04264}{{\tt arXiv:2003.04264[gr-qc]}}].

\bibitem{Chirkov:2021epn}
D. Chirkov, S.A. Pavluchenko,
{{Some aspects of the cosmological dynamics in Einstein\textendash{}Gauss\textendash{}Bonnet gravity}},
\href{http://dx.doi.org/10.1142/S0217732321500929}
{{Mod. Phys. Lett.} A {\bf 36 (13)} (2021) 2150092,}
[\href{http://arxiv.org/abs/2101.12066}{{\tt arXiv:2101.12066[gr-qc]}}].

\bibitem{Odintsov:2021nim}
S.D. Odintsov,  V.K. Oikonomou, and  F.P.  Fronimos,
{{Late-time cosmology of scalar-coupled $f (R, \mathcal{G})$ gravity}},
\href{http://dx.doi.org/10.1088/1361-6382/abe24f}
{{Class. Quant. Grav. }{\bf 38 (7)} (2021) 075009,}
[\href{http://arxiv.org/abs/2102.02239}{{\tt arXiv:2102.02239[gr-qc]}}].

\bibitem{Shamir:2021ptw}
M.F. Shamir,
{{Bouncing universe in f(G,T) gravity}},
\href{http://dx.doi.org/10.1016/j.dark.2021.100794}
{{Phys. Dark Univ.}{\bf 32} (2021) 100794}.

\bibitem{Yerra:2022alz}
P.K. Yerra, C. Bhamidipati,
{{Topology of black hole thermodynamics in Gauss-Bonnet gravity}} (2022),
[\href{http://arxiv.org/abs/2202.10288}{{\tt arXiv:2202.10288[gr-qc]}}].

\bibitem{Li:2021wqa}
G.-Q. Li, J.-X. Mo, Y.-W. Zhuang,
{{Corrections to Hawking radiation and Bekenstein-Hawking entropy of novel four-dimensional black holes in Gauss- Bonnet gravity}},
\href{http://dx.doi.org/10.1007/s10714-021-02875-3}
{{Gen. Rel. Grav.} {\bf 11} (2021) 107}.

\bibitem{Chen:2018nbh}
B. Chen, P.-C. Li, Y. Tian, C.-Y. Zhang,
{{Holographic Turbulence in Einstein-Gauss-Bonnet Gravity at Large $D$}},
\href{http://dx.doi.org/10.1007/JHEP01(2019)156}
{{JHEP} {\bf 01} (2019) 156,}
[\href{http://arxiv.org/abs/1804.05182}{{\tt arXiv:hep-th/1804.05182}}].

\bibitem{Hu:2013cia}
C. Hu, X.-X. Zeng and X.-M. Liu,
{{Phase transition and critical phenomenon of AdS black holes in Einstein-Gauss-Bonnet gravity}},
\href{http://dx.doi.org/10.1007/s11433-013-5107-4}
{{Sci. China Phys. Mech. Astron.} {\bf 56} (2013) 1652-1663.}

\bibitem{Li:2021rff}
Y.-Z. Li, C.-Y. Zhang, X.-M. Kuang,
{{Entanglement wedge cross-section with Gauss-Bonnet corrections and thermal quench}},
\href{http://dx.doi.org/10.1007/s11433-021-1791-1}
{{Sci. China Phys. Mech. Astron.} {\bf 64 (12)} (2021) 120413,}
[\href{http://arxiv.org/abs/2102.12171}{{\tt arXiv:2102.12171[hep-th]}}].

\bibitem{Kim:2000ym}
J.E. Kim and H.M. Lee,
{{Gravity in the Einstein-Gauss-Bonnet theory with the Randall-Sundrum background}},
\href{https://doi.org/10.1016/S0550-3213(01)00119-5}
{{Nucl. Phys.} B \textbf{602} (2001) 346,}
[\href{https://arxiv.org/abs/hep-th/0010093}{{\tt arXiv:0010093[hep-th]}}].

\bibitem{Kim:2000pz}
J.E.~Kim, B. Kyae, and H.M. Lee,
{{Various modified solutions of the Randall-Sundrum model with the Gauss-Bonnet interaction}},
\href{https://doi.org/10.1016/S0550-3213(00)00318-7}
{{Nucl. Phys.} B \textbf{582} (2000) 296,}
[\href{https://arxiv.org/abs/hep-th/0004005}{{\tt arXiv:hep-th/0004005}}].


\bibitem{Neupane:2001kd}
I.P. Neupane,
{{Completely localized gravity with higher curvature terms}},
\href{https://doi.org/10.1088/0264-9381/19/21/315}
{{Class. Quant. Grav.} \textbf{19} (2002) 5507,}
[\href{https://arxiv.org/abs/hep-th/0106100}{{\tt arXiv:hep-th/0106100}}].

\bibitem{Meissner:2001xg}
K.A. Meissner and M. Olechowski,
{{Brane localization of gravity in higher derivative theory}},
\href{https://doi.org/10.1103/PhysRevD.65.064017}
{{Phys. Rev.} D \textbf{65} (2002) 064017,}
[\href{https://arxiv.org/abs/hep-th/0106203}{{\tt arXiv:hep-th/0106203}}].

\bibitem{Aoyanagi:2004zz}
K. Aoyanagi and K. Maeda,
{{Creation of a Brane World with Gauss-Bonnet Term}},
\href{https://doi.org/10.1103/PhysRevD.70.123506}
{{Phys. Rev.} D \textbf{70} (2004) 123506,}
[\href{https://arxiv.org/abs/hep-th/0408008}{{\tt arXiv:hep-th/0408008}}].

\bibitem{Corradini:2000sw}
O. Corradini and Z. Kakushadze,
{{Localized gravity and higher curvature terms}},
\href{https://doi.org/10.1016/S0370-2693(00)01196-5}
{{Phys. Lett.} B \textbf{494} (2000) 302,}
[\href{https://arxiv.org/abs/hep-th/0009022}{{\tt arXiv:hep-th/0009022}}].

\bibitem{Neupane:2000wt}
I.P. Neupane,
{{Consistency of higher derivative gravity in the brane background}},
\href{https://doi.org/10.1088/1126-6708/2000/09/040}
{{JHEP} \textbf{09} (2000) 040,}
[\href{https://arxiv.org/abs/hep-th/0008190}{{\tt arXiv:hep-th/0008190}}].

\bibitem{Andrianov:2013vqa}
A.A. Andrianov, V.A. Andrianov, and O.O. Novikov,
{{Gravity effects on thick brane formation from scalar field dynamics}},
\href{http://dx.doi.org/10.1140/epjc/s10052-013-2675-4}
{{Eur. Phys. J.} C \textbf{73} (2013) 2675,}
[\href{http://arxiv.org/abs/1306.0723}{{\tt arXiv:1306.0723[hep-th]}}].	


\bibitem{German:2013sk}
G.~Germ\'an, A.~Herrera--Aguilar, D.~Malag\'on--Morej\'on, I.~Quiros and R.~da Rocha,
{{Study of field fluctuations and their localization in a thick braneworld generated by gravity nonminimally coupled to a scalar field with the Gauss-Bonnet term}},
\href{https://doi.org/10.1103/PhysRevD.89.026004}
{{Phys. Rev.} D {\bfseries89} (2014) 026004,}
[\href{https://arxiv.org/abs/1301.6444}{{\tt arXiv:1301.6444 [hep-th]}}].



\bibitem{Giovannini:2001ta}
M.~Giovannini,
{{Thick branes and Gauss-Bonnet self-interactions}},
\href{http://dx.doi.org/10.1103/PhysRevD.64.124004}
{{ Phys. Rev.} D \textbf{64} (2001) 124004,}
[\href{http://arxiv.org/abs/hep-th/0107233}{{\tt arXiv:hep-th/0107233}}].

\bibitem{Farakos:2006sr}
K. Farakos and P. Pasipoularides,
{{Gauss-Bonnet gravity, brane world models, and non-minimal coupling}},
\href{https://doi.org/10.1103/PhysRevD.75.024018}
{{Phys. Rev.} D \textbf{75} (2007) 024018,}
[\href{https://arxiv.org/abs/hep-th/0610010}{{\tt arXiv:hep-th/0610010}}].

\bibitem{HerreraAguilar:2011jm}
H.A. Alfredo, M.M. Dagoberto, R.M.L. Refugio and I. Quiros,
{{Thick braneworlds generated by a non-minimally coupled scalar field and a Gauss-Bonnet term: conditions for localization of gravity}},
\href{http://dx.doi.org/10.1088/0264-9381/29/3/035012}
{{Class. Quant. Grav.} {\bfseries29} (2012)  035012,}
[\href{http://arxiv.org/abs/1105.5479}{{\tt arXiv:1105.5479[hep-th]}}].

\bibitem{Dias:2015gga}
M. Dias,  J.M. Hoff da Silva and R. da Rocha,
{{Thick Braneworlds and the Gibbons-Kallosh-Linde No-go Theorem in the Gauss-Bonnet Framework}},
\href{http://dx.doi.org/10.1209/0295-5075/110/20004}
{{EPL } {\bfseries 110 (2)} (2015)  20004,}
[\href{http://arxiv.org/abs/1504.04243}{{\tt arXiv:1504.04243[gr-qc]}}].

\bibitem{Charmousis:2002rc}
C. Charmousis and J.F. Dufaux,
{{General Gauss-Bonnet brane cosmology}},
\href{http://dx.doi.org/10.1088/0264-9381/19/18/304}
{{Class. Quant. Grav. }  \textbf{19} (2002) 4671,}
[\href{http://arxiv.org/abs/hep-th/0202107}{{\tt arXiv:hep-th/0202107}}].

\bibitem{Brown:2006mh}
R.A. Brown,
{{Brane universes with Gauss-Bonnet-induced-gravity}},
\href{http://dx.doi.org/10.1007/s10714-007-0398-2}
{{Gen. Rel. Grav. }  \textbf{39} (2007) 477,}
[\href{http://arxiv.org/abs/gr-qc/0602050}{{\tt arXiv:gr-qc/0602050}}].

\bibitem{abdesselam2002brane}
B. Abdesselam and N. Mohammedi,
{{Brane world cosmology with Gauss-Bonnet interaction}},
\href{http://dx.doi.org/10.1103/PhysRevD.65.084018}
{{Phys. Rev.} D \textbf{65} (2002) 084018,}
[\href{http://arxiv.org/abs/hep-th/0110143}{{\tt arXiv:hep-th/0110143}}].

\bibitem{Alberghi:2005vq}
G.L. Alberghi and  A. Tronconi,
{{Gauss-Bonnet brane cosmology with radion stabilization}},
\href{https://doi.org/10.1103/PhysRevD.73.027702}
{{Phys. Rev.} D \textbf{73} (2006) 027702,}
[\href{https://arxiv.org/abs/hep-ph/0510267}{{\tt arXiv:hep-ph/0510267}}].

\bibitem{Nozari:2013hra}
K. Nozari, F. Kiani, and N. Rashidi,
{{Gauss-Bonnet Braneworld Cosmology with Modified Induced Gravity on the Brane}},
\href{https://doi.org/10.1155/2013/968016}
{{Adv. High Energy Phys.} {\bf 2013} (2013) 968016,}
[\href{https://arxiv.org/abs/1308.5770}{{\tt arXiv:1308.5770[gr-qc]}}].

\bibitem{Konya:2006wr}
K. Konya,
{{Gauss-Bonnet brane-world cosmology without Z(2)-symmetry}},
\href{https://doi.org/10.1088/0264-9381/24/10/019}
{{Class. Quant. Grav.} \textbf{24} (2007) 2761,}
[\href{https://arxiv.org/abs/gr-qc/0605119}{{\tt arXiv:gr-qc/0605119}}].

\bibitem{Okada:2014eva}
N. Okada and S. Okada,
{{Simple inflationary models in Gauss\textendash{}Bonnet brane-world cosmology}},
\href{https://doi.org/10.1088/0264-9381/33/12/125034}
{{Class. Quant. Grav.} \textbf{33} (2016) 125034,}
[\href{https://arxiv.org/abs/1412.8466}{{\tt arXiv:1412.8466[hep-ph]}}].

\bibitem{Herrera:2010vv}
R. Herrera and N. Videla,
{{Intermediate inflation in Gauss-Bonnet braneworld}},
\href{https://doi.org/10.1140/epjc/s10052-010-1264-z}
{{Eur. Phys. J.} C \textbf{67} (2010) 499,}
[\href{https://arxiv.org/abs/1003.5645}{{\tt arXiv:astro-ph.CO/1003.5645}}].

\bibitem{Fomin:2018typ}
I.V. Fomin,
{{Cosmological Inflation with Einstein\textendash{}Gauss\textendash{}Bonnet Gravity}},
\href{https://doi.org/10.1134/S1063779618040226}
{{Phys. Part. Nucl.}  \textbf{49} (2018) 525.}



\bibitem{Nojiri:2001ae}
S. Nojiri, S.D. Odintsov, and S. Ogushi,
{{Cosmological and black hole brane world universes in higher derivative gravity}},
\href{http://dx.doi.org/10.1103/PhysRevD.65.023521}
{{Phys. Rev. } D {\bf 65} (2002) 023521,}
[\href{http://arxiv.org/abs/hep-th/0108172}{{\tt arXiv:hep-th/0108172}}].




\bibitem{Nojiri:2002hz}
S. Nojiri, S.D. Odintsov, and S. Ogushi,
{{Friedmann-Robertson-Walker brane cosmological equations from the five-dimensional bulk (A)dS black hole}},
\href{http://dx.doi.org/10.1142/S0217751X02012156}
{{Int. J. Mod. Phys. }A {\bf 17} (2002) 4809--4870,}
[\href{http://arxiv.org/abs/hep-th/0205187}{{\tt arXiv:hep-th/0205187}}].


\bibitem{Lidsey:2002zw}
J.E. Lidsey, S. Nojiri, and S.D. Odintsov,
{{Brane world cosmology in (anti)-de Sitter Einstein-Gauss-Bonnet-Maxwell gravity}},
\href{http://dx.doi.org/10.1088/1126-6708/2002/06/026}
{{JHEP } {\bf 06} (2002) 026,}
[\href{http://arxiv.org/abs/hep-th/0202198}{{\tt arXiv:hep-th/0202198}}].

\bibitem{Diaz:2020iwx}
R. D\'\i{}az, F. G\'omez, M. Pinilla,
{{De Sitter brane-world solution in 5-dimensional Einstein\textendash{}Gauss\textendash{}Bonnet gravity}},
\href{http://dx.doi.org/10.1007/s10714-020-02742-7}
{{Gen. Rel. Grav.} {\bf 52 (9)} (2020) 86.}

\bibitem{Giovannini:2006rj}
M. Giovannini,
{{Kink-anti-kink, trapping bags and five-dimensional Gauss-Bonnet gravity}},
\href{http://dx.doi.org/10.1103/PhysRevD.74.087505}
{{Phys. Rev.} D {\bf 74} (2006) 087505,}
[\href{http://arxiv.org/abs/hep-th/0609136}{{\tt arXiv:hep-th/0609136}}].

\bibitem{Giovannini:2006ye}
M. Giovannini,
{{Gravitating multidefects from higher dimensions}},
\href{http://dx.doi.org/10.1103/PhysRevD.75.064023}
{{Phys. Rev.} D {\bf 75} (2007) 064023,}
[\href{http://arxiv.org/abs/hep-th/0612104}{{\tt arXiv:hep-th/0612104}}].

\bibitem{Giovannini:2006sw}
M. Giovannini,
{{Non-topological gravitating defects in five-dimensional anti-de Sitter space}},
\href{http://dx.doi.org/10.1088/0264-9381/23/23/L01}
{{Class. Quant. Grav.}  {\bf 23} (2006) L73,}
[\href{http://arxiv.org/abs/hep-th/0607229}{{\tt arXiv:hep-th/0607229}}].

\bibitem{Zhong:2018fdq}
Y.~Zhong, Y.~P.~Zhang, W.~D.~Guo and Y.~X.~Liu,
JHEP \textbf{04}, 154 (2019).

\bibitem{Tan:2022uex}
Q.~Tan, Y.~P.~Zhang, W.~D.~Guo, J.~Chen, C.~C.~Zhu and Y.~X.~Liu,
Eur. Phys. J. C \textbf{83}, 84 (2023).

\bibitem{Lovelock:1972vz}
D. Lovelock,
{{The 4-dimensionality of space and the Einstein tensor}},
\href{https://doi.org/10.1063/1.1666069}
{{J. Math. Phys.} {\bf 13} (1972) 874.}


\bibitem{Liu:2009ve}
Y.-X.~Liu, J.~Yang, Z.-H.~Zhao, C.-E.~Fu, and Y.-S.~Duan,
{{Fermion Localization and Resonances on A de Sitter Thick Brane}},
\href{http://dx.doi.org/10.1103/PhysRevD.80.065019}
{{Phys. Rev. D} \textbf{80} (2009) 065019, }
[\href{https://arxiv.org/abs/0904.1785}{{\tt arXiv:0904.1785 [hep-th]}}].

\bibitem{Bazeia:2003cv}
D.~Bazeia, F.A.~Brito, and J.R.S.~Nascimento,
{{Supergravity brane worlds and tachyon potentials}},
\href{https://doi.org/10.1103/PhysRevD.68.085007}
{{Phys. Rev.} D \textbf{68} (2003) 085007,}
[\href{https://arxiv.org/pdf/hep-th/0306284}{{\tt	arXiv:hep-th/0306284}}].

\bibitem{Brito:2001hd}
F.~Brito, M.~Cvetic, and S.~Yoon,
{{From a thick to a thin supergravity domain wall}},
\href{https://doi.org/10.1103/PhysRevD.64.064021}
{{Phys. Rev.} D \textbf{64} (2001) 064021,}
[\href{https://arxiv.org/pdf/hep-ph/0105010}{{\tt	arXiv:hep-ph/0105010}}].

\bibitem{Afonso:2006gi}
V.I. Afonso, D.~Bazeia, and L.~Losano,
{{First-order formalism for bent brane}},
\href{http://dx.doi.org/10.1016/j.physletb.2006.02.017}
{{Phys. Lett.} B {\bfseries634} (2006) 526,}
[\href{http://arxiv.org/abs/hep-th/0601069}{{\tt arXiv:hep-th/0601069}}].

\bibitem{1999Gravitational}
K. Skenderis and P.K. Townsend,
{{Gravitational stability and renormalization-group flow}},
{{Phys. Lett.} B {\bfseries 468} (1999) 46.}

\bibitem{2004Fake}
D.Z. Freedman, C. N\'u\~nez,  M.  Schnabl, and K. Skenderis,
{{Fake supergravity and domain wall stability}},
{{Phys. Lett.} D {\bfseries 69} (2004) 104027.}

\bibitem{1993GReGr}
V.H. Hamity and D.E. Barraco,
{{First order formalism of f(R) gravity}},
\href{http://dx.doi.org/10.1007/BF00756965}
{{Gen. Relativ. Gravit.} {\bf 25} (1993) 461}.

\bibitem{Julia:1999tk}
B.~Julia and S.~Silva,
{{On first order formulations of supergravities}},
{{JHEP} \textbf{01} (2000) 026},
[\href{http://arxiv.org/abs/hep-th/9911035}{{\tt arXiv:hep-th/9911035}}].

\bibitem{Low:2000pq}
I.~Low and A.~Zee,
{{Naked singularity and Gauss-Bonnet term in brane world scenarios}},
{{Nucl. Phys. B} \textbf{585} (2000) 395}.
[\href{http://arxiv.org/abs/hep-th/0004124}{{\tt arXiv:hep-th/0004124}}].

\bibitem{Koley:2004at}
R.~Koley and S.~Kar,
{{Scalar kinks and fermion localisation in warped spacetimes}},
{{Class. Quant. Grav.} \textbf{22} (2005) 753},
[\href{http://arxiv.org/abs/hep-th/0407158}{{\tt arXiv:hep-th/0407158}}].

\bibitem{Liu:2008pi}
Y.-X.~Liu, L.-D.~Zhang, L.-J.~Zhang, and Y.-S.~Duan,
{{Fermions on Thick Branes in Background of Sine-Gordon Kinks}},
{{Phys. Rev. D} \textbf{78} (2008) 065025},
[\href{http://arxiv.org/abs/0804.4553}{{\tt arXiv:0804.4553 [hep-th]}}].

\bibitem{Zhong:2012nt}
Y.~Zhong and Y.~X.~Liu,
Phys. Rev. D \textbf{88} (2013) no.2, 024017
doi:10.1103/PhysRevD.88.024017
[arXiv:1212.1871 [hep-th]].

\bibitem{Zhong:2015pta}
Y.~Zhong and Y.~X.~Liu,
Eur. Phys. J. C \textbf{76} (2016) no.6, 321
doi:10.1140/epjc/s10052-016-4163-0
[arXiv:1507.00630 [hep-th]].

\bibitem{Giovannini:2001fh}
M.~Giovannini,
Phys. Rev. D \textbf{64} (2001), 064023
doi:10.1103/PhysRevD.64.064023
[arXiv:hep-th/0106041 [hep-th]].

\bibitem{Kobayashi:2001jd}
S.~Kobayashi, K.~Koyama and J.~Soda,
Phys. Rev. D \textbf{65} (2002), 064014
doi:10.1103/PhysRevD.65.064014
[arXiv:hep-th/0107025 [hep-th]].

\bibitem{Giovannini:2001xg}
M.~Giovannini,
Phys. Rev. D \textbf{65} (2002), 064008
doi:10.1103/PhysRevD.65.064008
[arXiv:hep-th/0106131 [hep-th]].




\end{thebibliography}
\end{document}